\documentclass[aps,prd,twocolumn,superscriptaddress,preprintnumbers,showpacs,amsmath,amssymb]{revtex4-1}

\usepackage{graphicx} 
\usepackage{dcolumn}  
\usepackage[colorlinks,citecolor=blue]{hyperref}
\graphicspath{{ps}}

\usepackage{float}
\usepackage{indentfirst}
\usepackage[abs]{overpic}
\usepackage{subfigure}
\usepackage{multirow}

\usepackage{amsmath} 
\usepackage{amssymb}

\usepackage{epsfig}
\usepackage{color}
\usepackage{dcolumn}
\usepackage{bm}
\usepackage{appendix}

\begin{document}



\title{ \quad\\[1.0cm] Measurement of $\eta_{c}(1S)$, $\eta_{c}(2S)$ and non-resonant $\eta'\pi^{+}\pi^{-}$ production via two-photon collisions
}

\noaffiliation
\affiliation{University of the Basque Country UPV/EHU, 48080 Bilbao}
\affiliation{Beihang University, Beijing 100191}
\affiliation{Budker Institute of Nuclear Physics SB RAS, Novosibirsk 630090}
\affiliation{Faculty of Mathematics and Physics, Charles University, 121 16 Prague}
\affiliation{Chonnam National University, Kwangju 660-701}
\affiliation{University of Cincinnati, Cincinnati, Ohio 45221}
\affiliation{Deutsches Elektronen--Synchrotron, 22607 Hamburg}
\affiliation{University of Florida, Gainesville, Florida 32611}
\affiliation{SOKENDAI (The Graduate University for Advanced Studies), Hayama 240-0193}
\affiliation{Hanyang University, Seoul 133-791}
\affiliation{University of Hawaii, Honolulu, Hawaii 96822}
\affiliation{High Energy Accelerator Research Organization (KEK), Tsukuba 305-0801}
\affiliation{J-PARC Branch, KEK Theory Center, High Energy Accelerator Research Organization (KEK), Tsukuba 305-0801}
\affiliation{IKERBASQUE, Basque Foundation for Science, 48013 Bilbao}
\affiliation{Indian Institute of Science Education and Research Mohali, SAS Nagar, 140306}
\affiliation{Indian Institute of Technology Bhubaneswar, Satya Nagar 751007}
\affiliation{Indian Institute of Technology Guwahati, Assam 781039}
\affiliation{Indian Institute of Technology Madras, Chennai 600036}
\affiliation{Indiana University, Bloomington, Indiana 47408}
\affiliation{Institute of High Energy Physics, Chinese Academy of Sciences, Beijing 100049}
\affiliation{Institute of High Energy Physics, Vienna 1050}
\affiliation{Institute for High Energy Physics, Protvino 142281}
\affiliation{INFN - Sezione di Napoli, 80126 Napoli}
\affiliation{INFN - Sezione di Torino, 10125 Torino}
\affiliation{Advanced Science Research Center, Japan Atomic Energy Agency, Naka 319-1195}
\affiliation{J. Stefan Institute, 1000 Ljubljana}
\affiliation{Kanagawa University, Yokohama 221-8686}
\affiliation{Institut f\"ur Experimentelle Kernphysik, Karlsruher Institut f\"ur Technologie, 76131 Karlsruhe}
\affiliation{Kennesaw State University, Kennesaw, Georgia 30144}
\affiliation{King Abdulaziz City for Science and Technology, Riyadh 11442}
\affiliation{Department of Physics, Faculty of Science, King Abdulaziz University, Jeddah 21589}
\affiliation{Korea Institute of Science and Technology Information, Daejeon 305-806}
\affiliation{Korea University, Seoul 136-713}
\affiliation{Kyungpook National University, Daegu 702-701}
\affiliation{\'Ecole Polytechnique F\'ed\'erale de Lausanne (EPFL), Lausanne 1015}
\affiliation{P.N. Lebedev Physical Institute of the Russian Academy of Sciences, Moscow 119991}
\affiliation{Faculty of Mathematics and Physics, University of Ljubljana, 1000 Ljubljana}
\affiliation{Ludwig Maximilians University, 80539 Munich}
\affiliation{Luther College, Decorah, Iowa 52101}
\affiliation{University of Malaya, 50603 Kuala Lumpur}
\affiliation{University of Maribor, 2000 Maribor}
\affiliation{Max-Planck-Institut f\"ur Physik, 80805 M\"unchen}
\affiliation{School of Physics, University of Melbourne, Victoria 3010}
\affiliation{University of Mississippi, University, Mississippi 38677}
\affiliation{University of Miyazaki, Miyazaki 889-2192}
\affiliation{Moscow Physical Engineering Institute, Moscow 115409}
\affiliation{Moscow Institute of Physics and Technology, Moscow Region 141700}
\affiliation{Graduate School of Science, Nagoya University, Nagoya 464-8602}
\affiliation{Nara Women's University, Nara 630-8506}
\affiliation{National Central University, Chung-li 32054}
\affiliation{National United University, Miao Li 36003}
\affiliation{Department of Physics, National Taiwan University, Taipei 10617}
\affiliation{H. Niewodniczanski Institute of Nuclear Physics, Krakow 31-342}
\affiliation{Nippon Dental University, Niigata 951-8580}
\affiliation{Niigata University, Niigata 950-2181}
\affiliation{Novosibirsk State University, Novosibirsk 630090}
\affiliation{Osaka City University, Osaka 558-8585}
\affiliation{Pacific Northwest National Laboratory, Richland, Washington 99352}
\affiliation{Panjab University, Chandigarh 160014}
\affiliation{University of Pittsburgh, Pittsburgh, Pennsylvania 15260}
\affiliation{Theoretical Research Division, Nishina Center, RIKEN, Saitama 351-0198}
\affiliation{University of Science and Technology of China, Hefei 230026}
\affiliation{Showa Pharmaceutical University, Tokyo 194-8543}
\affiliation{Soongsil University, Seoul 156-743}
\affiliation{University of South Carolina, Columbia, South Carolina 29208}
\affiliation{Stefan Meyer Institute for Subatomic Physics, Vienna 1090}
\affiliation{Sungkyunkwan University, Suwon 440-746}
\affiliation{School of Physics, University of Sydney, New South Wales 2006}
\affiliation{Department of Physics, Faculty of Science, University of Tabuk, Tabuk 71451}
\affiliation{Tata Institute of Fundamental Research, Mumbai 400005}
\affiliation{Department of Physics, Technische Universit\"at M\"unchen, 85748 Garching}
\affiliation{Department of Physics, Tohoku University, Sendai 980-8578}
\affiliation{Earthquake Research Institute, University of Tokyo, Tokyo 113-0032}
\affiliation{Department of Physics, University of Tokyo, Tokyo 113-0033}
\affiliation{Tokyo Institute of Technology, Tokyo 152-8550}
\affiliation{Tokyo Metropolitan University, Tokyo 192-0397}
\affiliation{University of Torino, 10124 Torino}
\affiliation{Virginia Polytechnic Institute and State University, Blacksburg, Virginia 24061}
\affiliation{Wayne State University, Detroit, Michigan 48202}
\affiliation{Yamagata University, Yamagata 990-8560}
\affiliation{Yonsei University, Seoul 120-749}
 \author{Q.~N.~Xu}\thanks{ Also at University of Chinese Academy of Sciences.} \affiliation{Institute of High Energy Physics, Chinese Academy of Sciences, Beijing 100049} 
  \author{I.~Adachi}\affiliation{High Energy Accelerator Research Organization (KEK), Tsukuba 305-0801}\affiliation{SOKENDAI (The Graduate University for Advanced Studies), Hayama 240-0193} 
  \author{H.~Aihara}\affiliation{Department of Physics, University of Tokyo, Tokyo 113-0033} 
  \author{S.~Al~Said}\affiliation{Department of Physics, Faculty of Science, University of Tabuk, Tabuk 71451}\affiliation{Department of Physics, Faculty of Science, King Abdulaziz University, Jeddah 21589} 
  \author{D.~M.~Asner}\affiliation{Pacific Northwest National Laboratory, Richland, Washington 99352} 
  \author{H.~Atmacan}\affiliation{University of South Carolina, Columbia, South Carolina 29208} 
  \author{V.~Aulchenko}\affiliation{Budker Institute of Nuclear Physics SB RAS, Novosibirsk 630090}\affiliation{Novosibirsk State University, Novosibirsk 630090} 
  \author{T.~Aushev}\affiliation{Moscow Institute of Physics and Technology, Moscow Region 141700} 
  \author{R.~Ayad}\affiliation{Department of Physics, Faculty of Science, University of Tabuk, Tabuk 71451} 
  \author{V.~Babu}\affiliation{Tata Institute of Fundamental Research, Mumbai 400005} 
% ...
\collaboration{The Belle Collaboration}

\begin{abstract}
We report the measurement of $\gamma\gamma\rightarrow\eta_{c}(1S), \eta_{c}(2S)\rightarrow\eta'\pi^{+}\pi^{-}$ with $\eta'$ decays to $\gamma\rho$ and $\eta\pi^{+}\pi^{-}$  using 941 $\text{fb}^{-1}$ of data collected with the Belle detector at the KEKB asymmetric-energy $e^{+}e^{-}$ collider. The $\eta_{c}(1S)$ mass and width are measured to be 
$M$ = [2984.6 $\pm$ 0.7 (stat) $\pm$ 2.2 (syst) $\pm$ 0.3 (model)] MeV/$c^2$ 
and $\Gamma$ = $[30.8^{+2.3}_{-2.2}$ (stat) $\pm$ 2.5 (syst) $\pm$ 1.4 (model)] MeV, 
respectively. 
First observation of $\eta_{c}(2S)\rightarrow\eta'\pi^{+}\pi^{-}$ with a significance of 5.5$\sigma$ including systematic error is obtained, and the $\eta_c(2S)$  mass is measured to 
be $M$ = [3635.1 $\pm$ 3.7 (stat) $\pm$ 2.9 (syst) $\pm$ 0.4 (model)] MeV/$c^2$. 
The products of the two-photon decay width  and branching fraction ($\cal B$) of decays to $\eta'\pi^{+}\pi^{-}$ are determined to be $\mathrm{\Gamma_{\gamma\gamma}{\cal{B}} = [65.4\pm2.6~(stat)\pm7.8~(syst)]}$ eV for $\eta_{c}(1S)$ and $\mathrm{[5.6^{+1.2}_{-1.1}~(stat)\pm1.1~(syst)]}$ eV for $\eta_{c}(2S)$. 
The cross sections for $\gamma\gamma\rightarrow\eta'\pi^{+}\pi^{-}$ and $\eta'f_{2}(1270)$  are measured for the first time.  
\end{abstract}

\pacs{12.38.Qk, 13.25.Gv, 12.40.Yx, 13.66.Bc}


\maketitle

\tighten

\section{Introduction}
The charmonium states $\eta_c(1S)$ and $\eta_c(2S)$ play important role in tests of quantum chromodynamics (QCD) \cite{QCD_char_eta}. Precise measurement of their two-photon decay widths may provide sensitive tests for QCD models \cite{QCD_char_tp_width}. The lowest heavy-quarkonium state $\eta_c(1S)$, together with the $J/\psi$, $\eta_b(1S)$, and $\Upsilon(1S)$, serve as benchmarks for the fine tuning of input parameters for QCD calculations \cite{QCD_char_input_par}. The $\eta_c(1S)$ and $\eta_c(2S)$   resonance parameters were measured in $\psi(2S)$ radiative decay by BESIII, and in $B$ decay and two-photon production by BaBar, Belle and CLEO~\cite{etac_bes3,etac_bes3_1,etac_babar,etac_belle,etac_belle_1,etac2s_epp}.
CLEO made the first measurement of the $\eta_c(2S)$ two-photon decay width $\Gamma_{\gamma\gamma}$ via $K^0_SK^+\pi^-$ but observed no signal for the $\eta_{c}(2S)\rightarrow\eta'\pi^{+}\pi^{-}$ decay \cite{etac2s_epp}. They measured the ratio of the product of $\Gamma_{\gamma\gamma}$ and ${\cal B}(K^0_SK^+\pi^-)$ for $\eta_c(2S)$ to that for $\eta_c(1S)$, as well as  $\Gamma_{\gamma\gamma}$ for $\eta_c(1S)$. Assuming equal $\cal B$ for the $\eta_c(1S)$ and $\eta_c(2S)$ decays, the two-photon width $\Gamma_{\gamma\gamma}$ for $\eta_c(2S)$ is estimated to be (1.3 $\pm$ 0.6) keV. 
On the other hand, the assumption of equal $\cal B$ for $\eta_c(1S)$ and $\eta_c(2S)$ seems implausible since the value of  ${\cal B}(\eta_c(2S)\rightarrow K\bar{K}\pi)$  = $\mathrm{(1.9 \pm 0.4 \pm 1.1)}$\%
measured by BaBar \cite{babar_etac2s} is far from the world-average value 
of ${\cal B}(\eta_c(1S)\rightarrow K\bar{K}\pi)$ = $\mathrm{(7.3 \pm 0.5)}$\%.

Using 637 fb$^{-1}$ of data, Belle reported the measurement of the $\eta_c(1S)$ resonance parameters in two-photon fusion based on its decays to $\eta^{\prime}\pi^+\pi^-$ with $\eta^{\prime}\to\eta\pi^+\pi^-$ \cite{etac1S_zcc}. The above considerations motivate an updated measurement of $\eta_c(1S)$ parameters using the  941 fb$^{-1}$ Belle data set, and, additionally, an attempt to measure $\Gamma_{\gamma\gamma}$ for $\eta_c(2S)$ in order to address the discrepancy between experimental data and QCD predictions  for this parameter, most of which lie in the range of 1.8--5.7 keV \cite{tp_wid_QCD,tp_wid_QCD_1,tp_wid_QCD_2,tp_wid_QCD_3,tp_wid_QCD_4,tp_wid_QCD_5}.

The cross sections for two-photon production of meson pairs have been calculated in perturbative QCD and measured in experiments
in a $W$ region  near or above 3 GeV, where $W$ is the invariant mass of the two-photon system. The leading term in the QCD calculation \cite{qcd_pre_1,qcd_pre_2,qcd_pre_2_1} of the cross section  predicts a $1/(W^{6}\rm{sin}^{4}\theta)$
dependence for a charged-meson pair, and a $1/W^{10}$ dependence and model-dependent angular distribution for a neutral-meson pair. Here, $\theta$ is the scattering angle of a final-state particle in the two-photon CM frame. 
The handbag model \cite{handbag} gives the transition amplitude describing energy dependence and predicts a $1/\rm{sin}^{4}\theta$ angular distribution for both charged- and neutral-meson pairs for large $W$.
The Belle results for the cross sections \cite{belle_result_cr} show that the angular distributions for the charged-meson pairs,
$\gamma\gamma\to\pi^{+}\pi^{-},K^{+}K^{-}$, agree well with the $1/\rm{sin}^{4}\theta$ expectation, while those for the neutral-meson pairs,
$\gamma\gamma\to\pi^{0}\pi^{0},K^{0}_{S}K^{0}_{S},\eta\pi^{0}$ and $\eta\eta$, exhibit more complicated angular  behavior.
The measured exponent $\mathit{n}$  in the energy dependence $1/W^{n}$ for both charged- and neutral-meson pairs is found to lie between 7.3 and 11 with a relative
error of 7--20\%.  Further study with improved precision in both experiment and QCD predictions at higher $W$ mass would provide more sensitive comparisons.
There is no specific QCD prediction for the two-photon production of either the  pseudoscalar-tensor meson pair $\eta'f_{2}(1270)$ or the three-body final state $\eta'\pi^{+}\pi^{-}$.
Our results for the production of these two- and three-body final states would thus provide new information to validate QCD models.

In this paper, we report the updated measurement of the $\eta_c(1S)$ parameters with the  most Belle data sample of 941 fb$^{-1}$, the observation of an $\eta_c(2S)$ signal with its decays to $\eta'\pi^{+}\pi^{-}$ for the first time, the measurement of the product of the two-photon width of $\eta_c(2S)$ and its branching fraction to $\eta'\pi^{+}\pi^{-}$, and the measurement of  non-resonant production of $\eta'\pi^{+}\pi^{-}$ 
with $\eta'\rightarrow\eta\pi^{+}\pi^{-}$ decay
via two-photon collisions.

\section{Detector and MONTE CARLO SIMULATION}

The Belle detector is a large-solid-angle magnetic
spectrometer that consists of a silicon vertex detector,
a 50-layer central drift chamber, an array of aerogel threshold Cherenkov counters, 
a barrel-like arrangement of time-of-flight scintillation counters, and an electromagnetic calorimeter (ECL) comprised of CsI(Tl) crystals  
located inside a superconducting solenoid coil that provides a 1.5~T
magnetic field.  An iron flux-return located outside of
the coil is instrumented to detect $K_L^0$ mesons and to identify muons. 
The detector is described in detail elsewhere~\cite{Belle}.

We generate the two-photon process $\gamma\gamma\rightarrow \eta^{\prime}\pi^{+}\pi^{-}$ using the TREPS code~\cite{TREPS}, where the $\eta^{\prime}$ decays generically according to JETSET7.3 \cite{jet_set}. A distribution uniform in phase space is assumed for the $\eta_{c}(1S)$ and $\eta_{c}(2S)$ decays to the $\eta^{\prime}\pi^{+}\pi^{-}$ final state. 
The GEANT3-based \cite{geant3} simulation package that incorporates the trigger conditions is employed for the propagation of the generated particles through the Belle detector.

\section{Data and Event Selection}
We use two data samples. The first is collected at the $\Upsilon(4S)$ resonance ($\sqrt{s} = 10.58$ GeV) and  
60 MeV below it with integrated luminosity $L_{\rm{int,4S}}$ = 792 fb$^{-1}$, while the other is recorded near the $\Upsilon(5S)$ resonance 
($\sqrt{s} = 10.88$ GeV) with $L_{\rm{int,5S}}$ = 149 fb$^{-1}$. When combining the data in this analysis, a slight dependence of 
the two-photon cross section on $e^+ e^-$ center-of-mass energy is taken into account,  as described in Sec.~\ref{fit_etac}.

Two $\eta^\prime$ decay modes, $\eta^\prime\rightarrow\eta\pi^{+}\pi^{-}$ with $\eta\rightarrow\gamma\gamma$ and $\eta^\prime\rightarrow\gamma\rho$ including non-resonant $\pi^{+}\pi^{-}$ (denoted as $\eta\pi\pi$ and $\gamma\rho$, respectively), are included in the reconstruction of the $\eta^\prime$ meson in the $\eta^\prime \pi^+\pi^-$ final state. 

\subsection{Selection criteria}

At least one neutral cluster and exactly four charged tracks with zero net charge are required in each event. The candidate photons are neutral clusters in the ECL that have an energy deposit greater than 100 MeV and are unmatched with any charged tracks. To suppress background photons from $\pi^0$ ($\pi^0$  or $\eta$) decays for the $\eta \pi\pi$ ($\gamma\rho$) mode, any photon that, in combination with another photon in the event has an invariant mass within the $\pi^0$ ($\pi^0$ or $\eta$) window $|M_{\gamma\gamma} - m_{\pi^0}|<$ 0.018 GeV/$c^2$ ($|M_{\gamma\gamma} - m_{\pi^0}|<$  0.020 GeV/$c^2$ or $|M_{\gamma\gamma} - m_{\eta}|<$  0.024 GeV/$c^2$)  is excluded. Events with an identified kaon ($K^{\pm}$ or $K^0_S\rightarrow \pi^+\pi^-$) or proton are vetoed. Charged pion, kaon and proton identification  strategies and criteria  for the both  $\eta\pi\pi$ and $\gamma\rho$ modes, as well as the event selection criteria for the $\eta \pi\pi$ mode, are the same as those used in Ref. ~\cite{etac1S_zcc} except for the requirement on the transverse momentum $|\Sigma p^{*}_{t}|$  (see Sec.~\ref{sum pt cut}). 
Here, $|\Sigma p^{*}_{t}|$ is the absolute value of the vector sum of the transverse momenta of the $\eta'$,  $\pi^+$, and $\pi^-$ in the $e^+e^-$ center-of-mass system. 
To improve the momentum resolution of the $\eta^\prime$, two separate fits  to the $\eta'$ are applied, one with a constrained vertex and the other with a constrained mass.
  
For the $\eta \pi\pi$ mode, the $\eta$ is reconstructed via its two-photon decay mode, where the two-photon invariant mass is in the window $M_{\gamma\gamma}\in$ [0.524, 0.572] GeV/$c^2$ ($\pm 2\sigma$ of the nominal $\eta$ mass). The $\eta^\prime$ candidate is reconstructed from the $\eta$ candidate and the $\pi^+\pi^-$ track pair that has an invariant mass within $M_{\eta\pi^+\pi^-} \in$ [0.951, 0.963] GeV/$c^2$ ($\pm 2\sigma$ of the nominal $\eta^\prime$ mass).

For the $\gamma\rho$ mode, the event contains  one photon and two $\pi^+ \pi^-$  pairs. 
The $\eta'$ candidates are reconstructed with one photon candidate and a $\rho^0$ candidate comprised of a $\pi^+\pi^-$ pair whose invariant mass lies within the $\rho^0$ signal region  $|M_{\pi^{+}\pi^{-}} - m_{\rho^0}| <$ 0.18 GeV/$c^{2}$. Finally, the photon and $\rho^0$ candidate must satisfy 
$M_{\gamma\rho} \in$ [0.942, 0.974] GeV/$c^2$ ($\pm 2\sigma$ of the nominal $\eta^\prime$ mass).

For both the $\eta\pi\pi$ and $\gamma\rho$ modes, we reconstruct $\eta^{\prime}\pi^{+}\pi^{-}$ candidates by combining the $\eta'$ with the remaining $\pi^{+}\pi^{-}$ pair, which must satisfy a vertex-constrained fit. 
For multicandidate events, the candidate with the smallest  $\chi^{2}$ from the $\eta'$ 
mass-constrained fit is selected. For $\eta'\pi^{+}\pi^{-}$ combinations with an invariant 
mass of $W$ = 2.98 (3.64) GeV/$c^{2}$, we find that 8.2\% (7.3\%) of the signal Monte Carlo (MC) events have more than one candidate per event for the $\eta\pi\pi$ mode and  15\% (9.8\%) for the $\gamma\rho$ mode, from which the correct candidate is selected 94\% (98\%) for the $\eta\pi\pi$ mode and 88\% (89\%) for the $\gamma\rho$ mode. The sum of the ECL cluster energies in the laboratory system  
and the scalar sum of the absolute momenta for all charged and neutral tracks in the 
laboratory system for the $\eta'\pi^+\pi^-$ system 
must satisfy $E_{\rm sum} < 4.5$ GeV and $P_{\rm sum} < 5.5$ GeV/$c$ to further suppress background events produced via $e^+e^-\rightarrow q\bar{q}$  with or without radiative photons.

\subsection{Optimization for the $|\Sigma p^{*}_{t}|$ requirement}
\label{sum pt cut}
\begin{figure*}[htb]
\centering
  \begin{minipage}[t]{0.45\textwidth}
  \centering
  \begin{overpic}[width=1.0\linewidth]{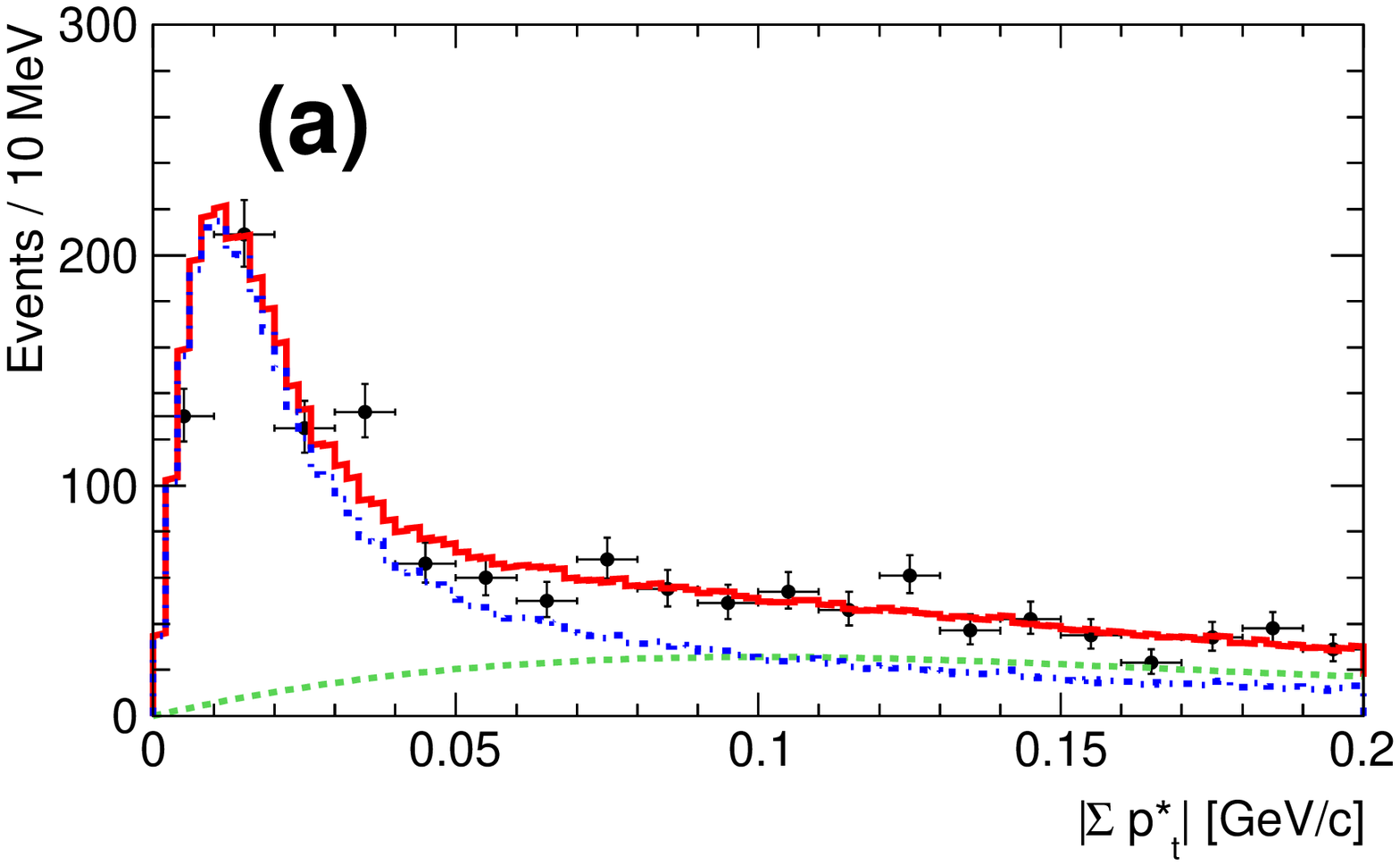}
  \end{overpic}
  \end{minipage}
\begin{minipage}[t]{0.45\textwidth}
\centering
\begin{overpic}[width=1.0\linewidth]{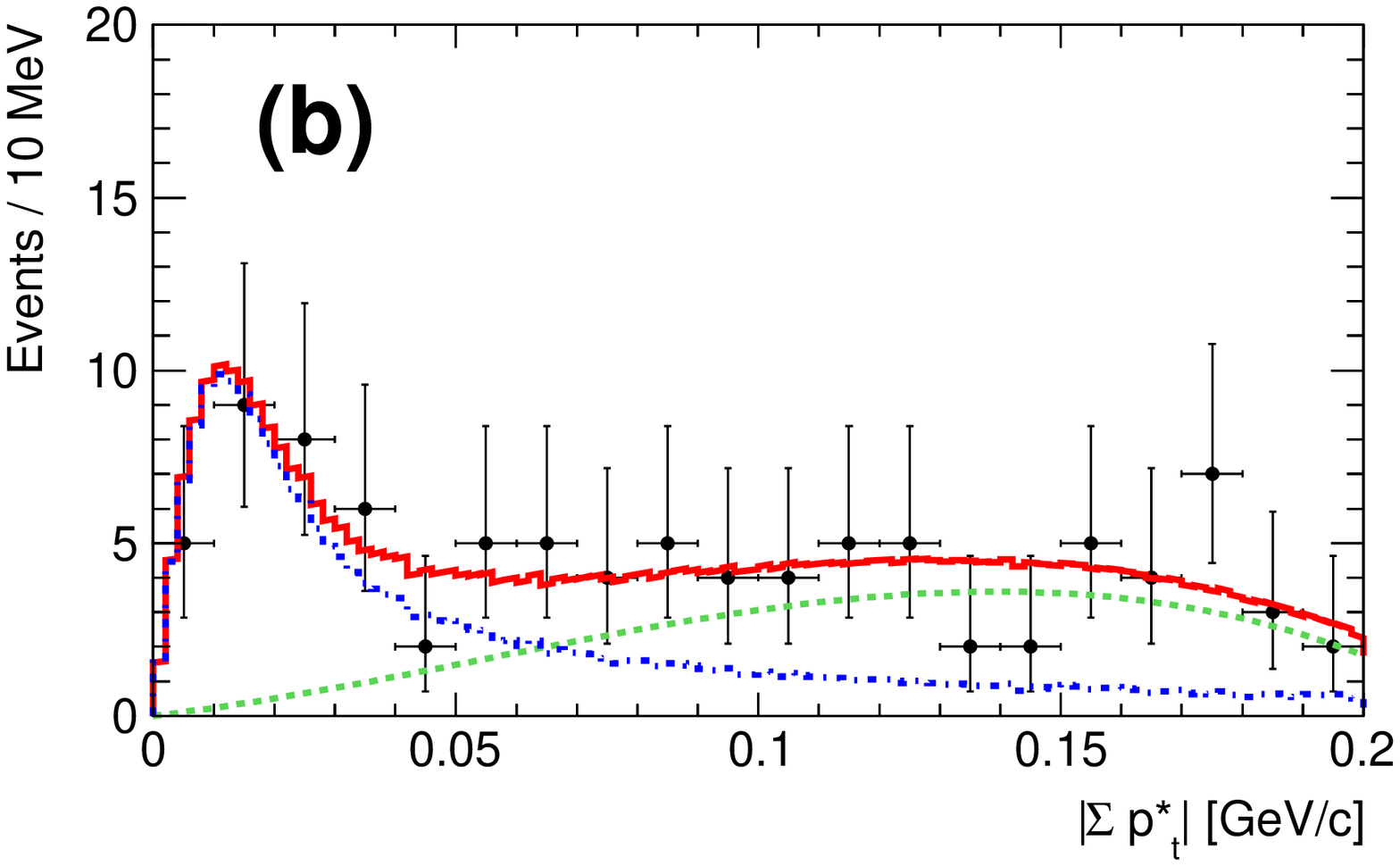}
\end{overpic}
\end{minipage}

 \begin{minipage}[t]{0.45\textwidth}
  \centering
  \begin{overpic}[width=1.0\linewidth]{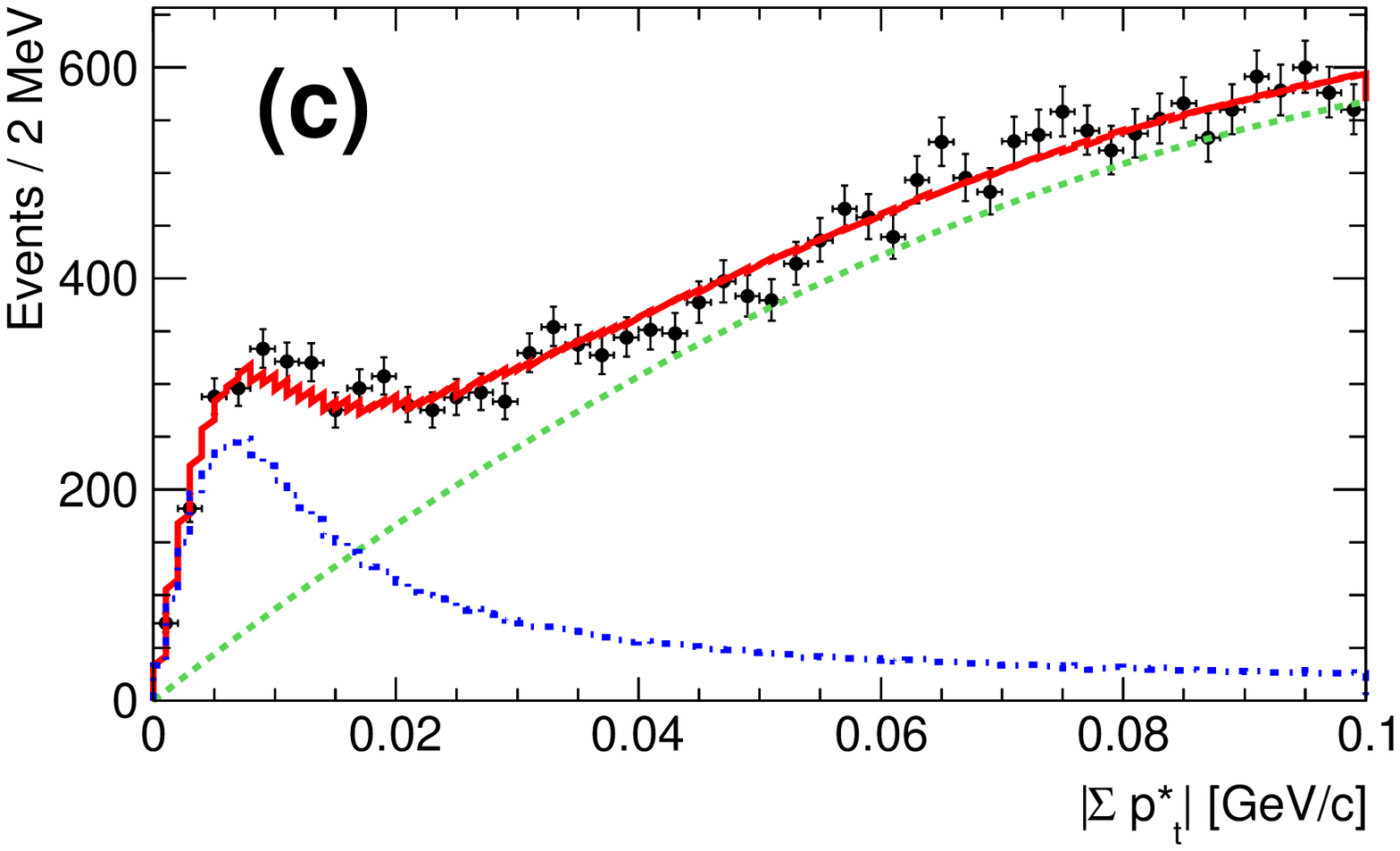}
  \end{overpic}
  \end{minipage}
\begin{minipage}[t]{0.45\textwidth}
\centering
\begin{overpic}[width=1.0\linewidth]{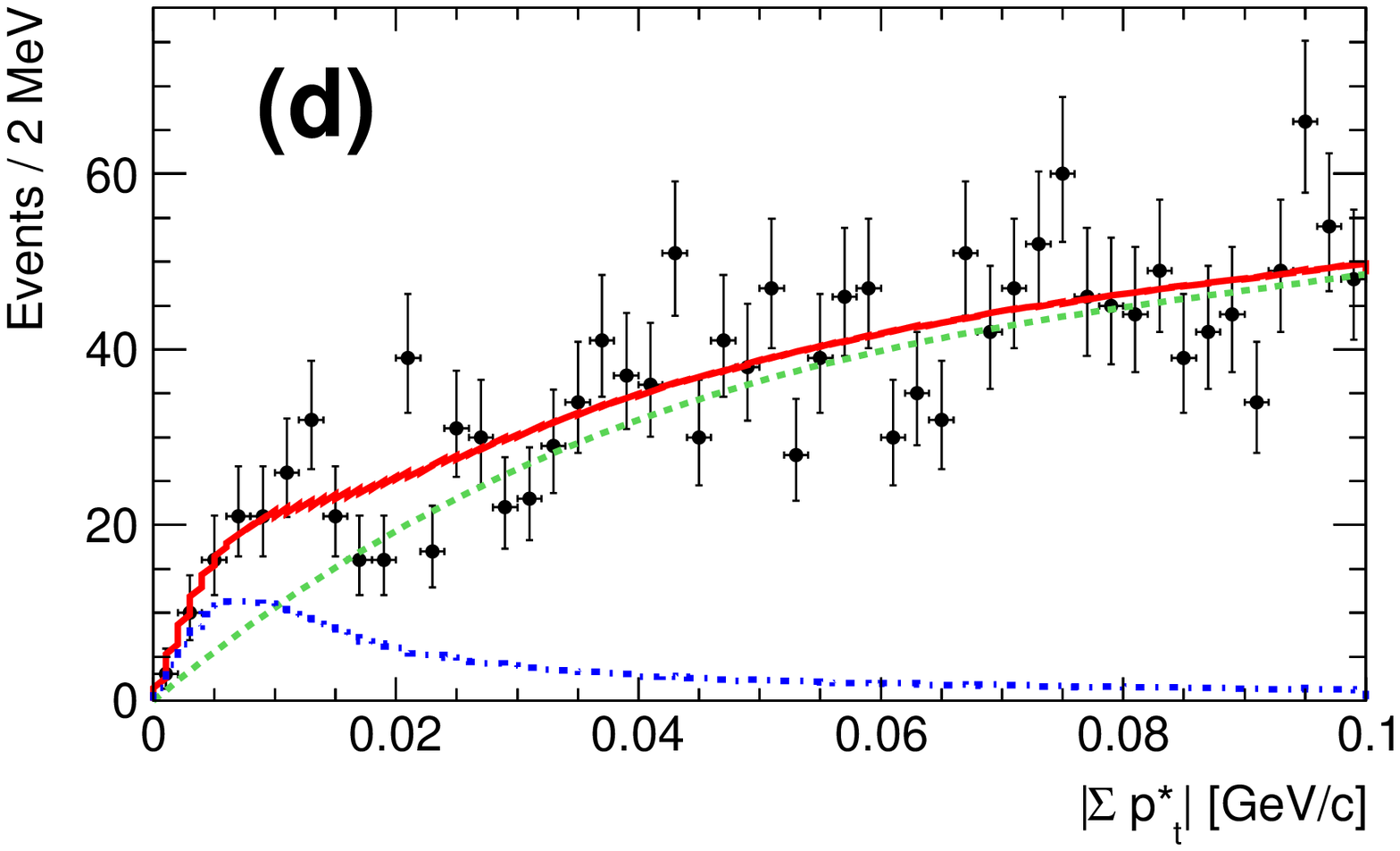}
\end{overpic}
\end{minipage}
\caption{(Color online) The $|\Sigma p^{*}_{t}|$ distributions in the $\eta_{c}(1S)$ $[\eta_{c}(2S)]$ signal region for (a) [(b)] the $\eta\pi\pi$ mode and (c) [(d)] the $\gamma\rho$ mode. The solid points with error bars are data. The solid red line is the fit; the blue dashed-dot and green dashed lines, respectively, show the  signal in MC and the background in data.
}
\label{pt_gr_epp}
\end{figure*}

The prominent feature for the events from an untagged two-photon process in $e^{+}e^{-}$ collisions is that they tend to carry small transverse momentum. Therefore,  a $|\Sigma p^{*}_{t}|$ requirement allows significant background reduction. 
The $|\Sigma p^{*}_{t}|$ distributions for the $\eta\pi\pi$ and $\gamma\rho$  modes  in the signal regions of $W\in$ [2.90, 3.06] GeV for $\eta_{c}(1S)$ and $W\in$ [3.60, 3.68] GeV for $\eta_{c}(2S)$  are shown in Fig.~\ref{pt_gr_epp}. 

The $|\Sigma p^{*}_{t}|$ requirement for selection of the $\eta'\pi^+\pi^-$ candidates from both the $\eta_{c}(1S)$ and $\eta_{c}(2S)$ decays is optimized using signal and background MC samples. The $\eta_{c}$ signal and the background are described by a relativistic Breit-Wigner function [see Eq.~\eqref{R signal pdf} in section ~\ref{fit_etac}] and the exponential of a third-order polynomial, respectively. The background shape in the $\eta_{c}$ signal region is determined  from the fit to the sideband data and normalized.
The requirement on $|\Sigma p^{*}_{t}|$ is determined by maximizing the value of $s/\sqrt{s+b}$ for both $\eta\pi\pi$ and $\gamma\rho$ modes, where $s$ is the $\eta_{c}$ signal yield and $b$ is background yield in the $\eta_{c}$ signal region. 
We find the best $|\Sigma p^{*}_{t}|$ requirements, which are close to each other in the two $\eta_{c}$ mass regions, to be $|\Sigma p^{*}_{t}| <$ 0.15 GeV/$c$ for the $\eta\pi\pi$ mode and $|\Sigma p^{*}_{t}| < $ 0.03 GeV/$c$ for the $\gamma\rho$ mode. 
We find that these values are stable in the range of the expected signal yield based on the previous measurement ~\cite{etac1S_zcc} for $\eta_{c}(1S)$ and an assumption of theoretical expectation for $\eta_{c}(2S)$~\cite{etac2S_br}. We employ the $|\Sigma p^{*}_{t}|$ requirement values optimized for $\eta_{c}(1S)$ to look also for the $\eta_{c}(2S)$ in both $\eta\pi\pi$ and $\gamma\rho$ modes.

\begin{figure*}[htb]
\begin{center}
\begin{minipage}[t]{0.45\textwidth}
\centering
\begin{overpic}[width=1.0\linewidth]{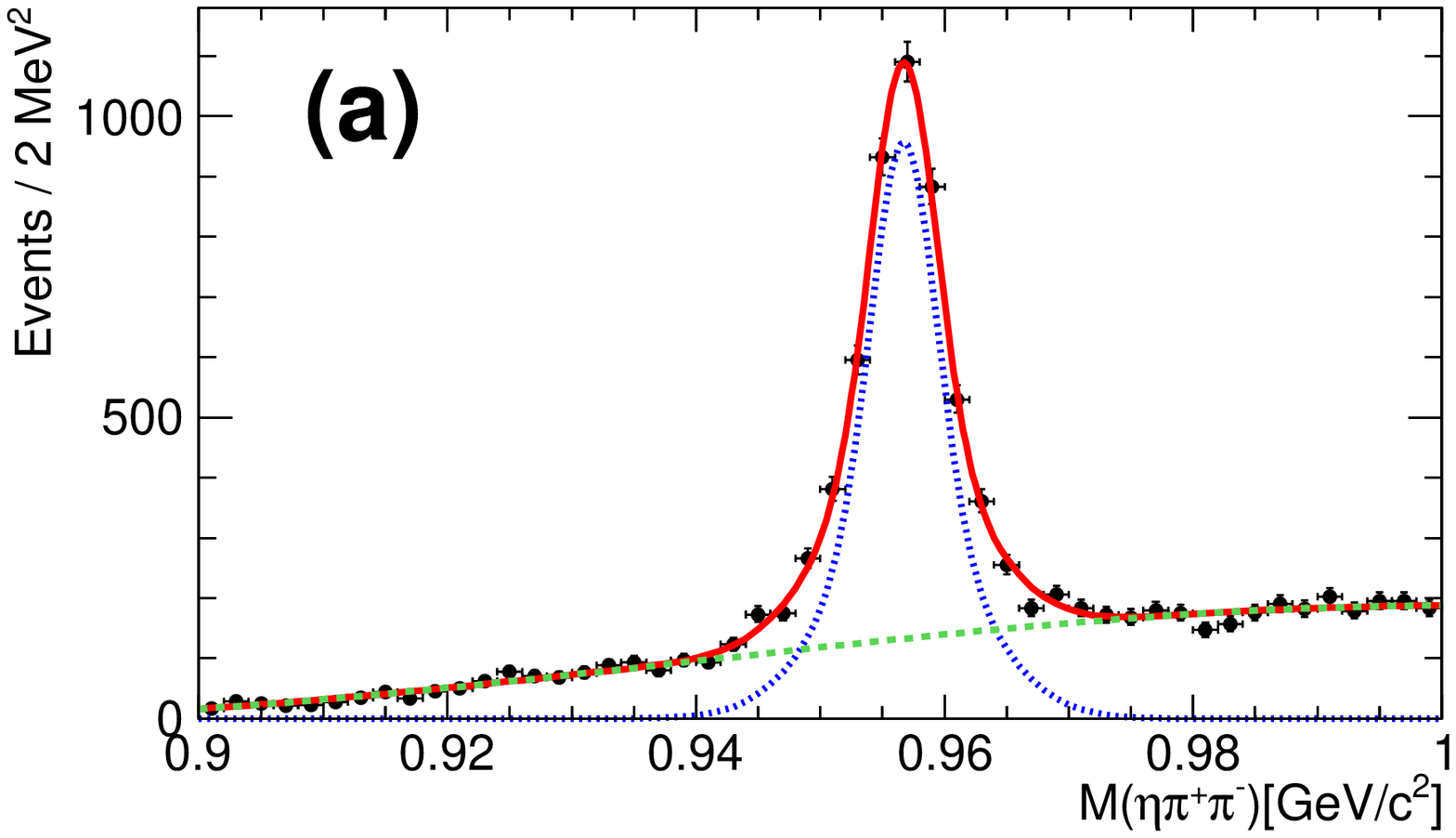}
\end{overpic}
\end{minipage}
\begin{minipage}[t]{0.45\textwidth}
\centering
\begin{overpic}[width=1.0\linewidth]{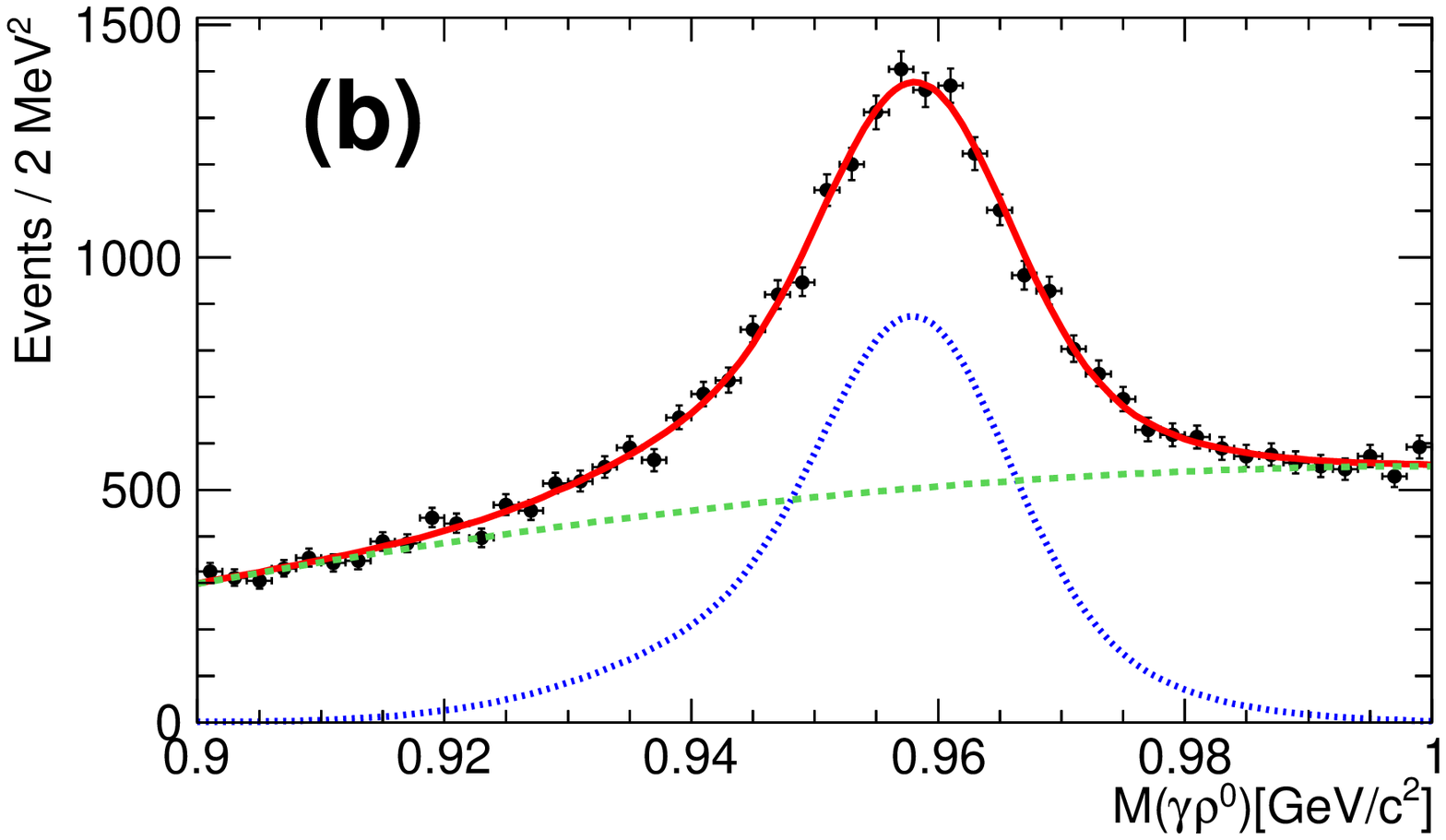}
\end{overpic}
\end{minipage}
\end{center}
\renewcommand{\figurename}{Fig.}
\caption{The invariant mass distributions of (a) $\eta\pi^{+}\pi^{-}$  and $\gamma\rho^{0}$ (b)  for the $\eta'\pi^+\pi^-$ candidate events. Solid red line is the fit. The blue dashed-dot and green dashed lines are the signal and background, respectively.}
\label{metp_twomodes}
\end{figure*}

\begin{figure*}[htb]
\centering
  \begin{minipage}[t]{0.45\textwidth}
  \centering
  \begin{overpic}[width=1.0\linewidth]{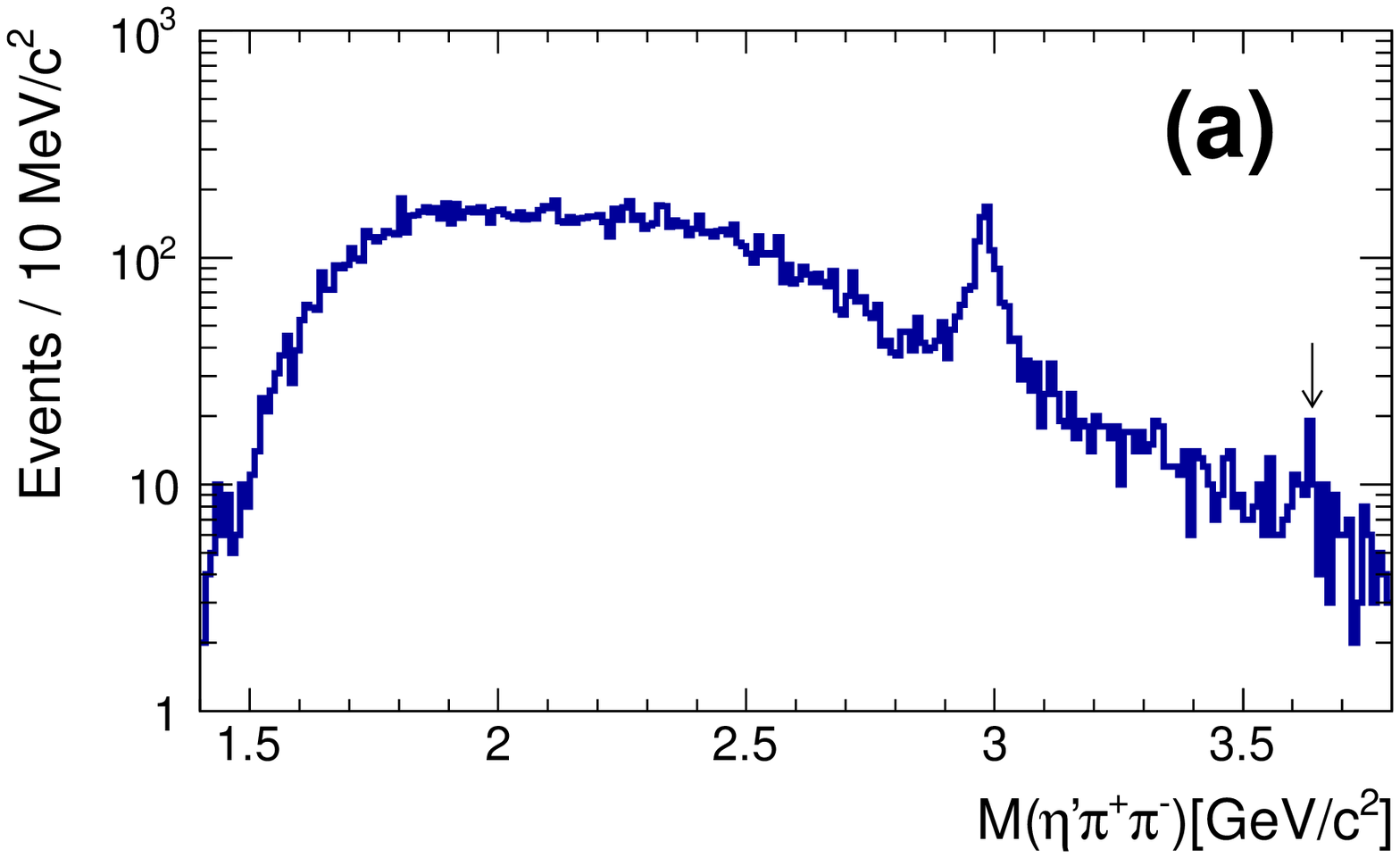}
  \end{overpic}
  \end{minipage}
\begin{minipage}[t]{0.45\textwidth}
\centering
\begin{overpic}[width=1.0\linewidth]{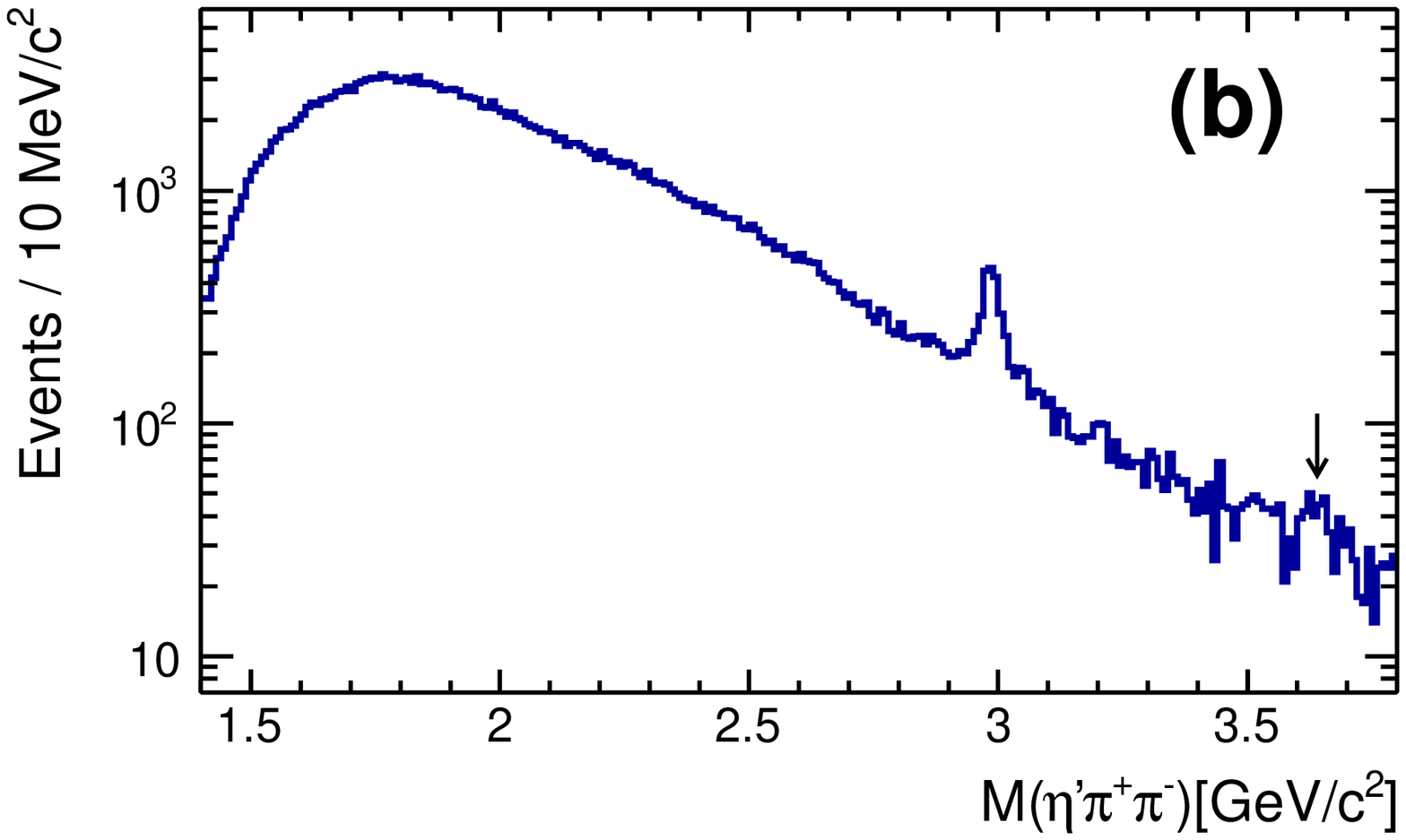}
\end{overpic}
\end{minipage}
\caption{The $\eta'\pi^+\pi^-$ invariant mass distribution for the candidate events with $\eta^\prime$ decays to (a) $\eta\pi^{+}\pi^{-}$ and (b) $\gamma\rho$. Large  $\eta_c(1S)$ signal and evident excess in the  $\eta_c(2S)$ region (as arrow pointed) are seen.}
\label{m_etap2pi}
\end{figure*}

The invariant mass distributions for the candidates of the $\eta'$ and that of  the $\eta^{\prime}\pi^{+}\pi^{-}$ in the $\eta\pi\pi$ and $\gamma\rho$ modes are shown in Fig.~\ref{metp_twomodes} and Fig.~\ref{m_etap2pi}, respectively. In addition to the prominent $\eta_{c}(1S)$ signal, an evident enhancement in the mass region near 3.64 GeV/$c^2$ is seen in both modes.

\section{Fitting for  $\eta_{c}(1S)$ and $\eta_{c}(2S)$}\label{fit_etac}

The probability density function $f_s(W)$ for the resonance $R$ is a Breit-Wigner function~\cite{BW_f,BW_f_1} $f_{\rm BW}(W)$ convolved with a mass-resolution function $R_{\rm ICB}$ after corrections for the detection efficiency $\epsilon_{i}(W)$ 
and the two-photon luminosity function $dL_{\gamma\gamma}/dW$:
\begin{eqnarray}\label{R signal pdf}
&& f_{\rm s}(W) =  f_{\rm BW}(W)\frac{dL_{\gamma\gamma}(W)}{dW}\epsilon_{i}(W)\otimes  
R_{\rm ICB}(W).\quad
\end{eqnarray}
Here, $R_{\rm ICB}$ is an improved Crystal Ball (ICB) function~\cite{icb fun}. 
The efficiency factor $\epsilon_{i}(W)$ includes the branching fractions of $\eta^\prime \rightarrow \eta\pi^{+}\pi^{-}$ with $\eta\rightarrow\gamma\gamma$ for the $\eta\pi\pi$  mode ($i$ = 1) and $\eta^\prime\rightarrow\gamma\rho$ with $\rho\rightarrow\pi^{+}\pi^{-}$ for the $\gamma\rho$ mode ($i$ = 2). 
The number of the $\eta_c(1S)$ mesons produced via the two-photon process is constrained to be equal for both modes in the simultaneous fit.
The luminosity function is evaluated in the Equivalent Photon Approximation (EPA)~\cite{BW_f,BW_f_1} using TREPS~\cite{TREPS}.
The efficiency for each  $\eta^\prime$ decay mode is corrected for the dependence on beam energy in the $\Upsilon(4S)$ and $\Upsilon(5S)$  regions \cite{eff_cor,eff_cor_1}:

\begin{eqnarray}\label{eff_corr}
&& \epsilon = \frac{\epsilon_{\rm{4S}}L_{\rm{int,4S}} + \epsilon_{\rm{5S}}L_{\rm{int,5S}}\cdot\frac{dL_{\gamma\gamma,\rm{5S}}}{dW}/\frac{dL_{\gamma\gamma,\rm 4S}}{dW}}{L_{\rm{int,4S}} + L_{\rm{int,5S}}}, \nonumber\\[1mm]
\end{eqnarray}
where $\epsilon_{\rm 4S}$ ($\epsilon_{\rm 5S}$) and $dL_{\gamma\gamma,\rm 5S}/dW$   ($dL_{\gamma\gamma,\rm 5S}/dW$) are the efficiency and two-photon luminosity  functions, respectively, at the $\Upsilon(4S)$ $[\Upsilon(5S)]$ energy.

The product of the two-photon decay width and the branching fraction for the $R\rightarrow\eta'\pi^{+}\pi^{-}$ decay is determined as
\begin{eqnarray}\label{crs_def_cal}
&& \Gamma_{\gamma\gamma}{\cal B}(R\rightarrow\eta'\pi^{+}\pi^{-}) \nonumber\\[1mm]
&& = \frac{n_{{\rm obs},i}}{L_{\rm int}\cdot\int f_{{\rm BW}}(W)\frac{dL_{\gamma\gamma}(W)}{dW}\epsilon_{i}(W)dW},
\end{eqnarray}
where $n_{{\rm obs},i}$ is the yield of decay mode $i$ of the resonance $R$ in the  simultaneous fit,
while $L_{\rm int}$ is the integrated luminosity. 
Identical $W$ regions of [2.60, 3.4] GeV/$c^2$ for $\eta_c(1S)$ and [3.3, 3.8] GeV/$c^2$ for $\eta_c(2S)$ 
are chosen in the simultaneous  fit for the yield and as  the integral interval in the calculation of $\Gamma_{\gamma\gamma}{\cal B}$.

\subsection{Background estimation} \label{etac_bk}

The background in the $\eta^{\prime} \pi^{+}\pi^{-}$ mass spectrum for the $R$ measurement is dominated by three components: 
(1) non-resonant ($NR$) events produced via two-photon collisions, which have the same  $|\Sigma p^{*}_{t}|$ distribution as that of the $R$ signal;
(2) the $\eta'$ sideband ($\eta^\prime$-$sdb$) arises from wrong combinations of 
$\gamma\gamma\pi^+\pi^-$ ($\gamma\pi^+\pi^-$) for the $\eta\pi\pi$ ($\gamma\rho$) mode 
that survive the $\eta^\prime$ selection criteria, estimated using the events in the margins of the  $\eta^\prime$ signal in the $\eta\pi\pi$ ($\gamma\rho$) invariant-mass distribution; 
(3) $\eta'\pi^{+}\pi^{-}+X$ ($b_{\rm any}$) events having additional particles in the event beyond the $R$ candidate. 
Other nonexclusive events, including those arising from initial-state radiation, are found to be 
negligible~\cite{etac1S_zcc}.

For the determination of the background components, two data subsamples, one with $|\Sigma p^{*}_{t}| <$ 0.15 GeV/$c$ (0.03 GeV/$c$), denoted as $p_t$-balanced, and the other with $|\Sigma p^{*}_{t}|\in [0.17,0.2]$ GeV/$c$ ($\mathrm{[0.15,0.2]}$ GeV/$c$), denoted as $p_t$-unbalanced, for the $\eta\pi\pi$ ($\gamma\rho$) mode, are selected. (See Ref. ~\cite{etac1S_zcc} for the details.) 
The $R$ signal and $NR$ component peak in the $p_t$-balanced sample while the $\eta^\prime$-$sdb$ and $b_{\rm any}$
backgrounds dominate over the signal plus $NR$ in the $p_t$-unbalanced sample. 
For the $\eta\pi\pi$ mode, the $\eta'$-$sdb$ component is well estimated using the $\eta^\prime$ sideband, defined by  $\mathrm{\mathit{M}_{\eta\pi^{+}\pi^{-}}\in [0.914,0.934]}$ GeV/$c^2$ and $\mathrm{\in [0.98,1.00]}$ GeV/$c^2$. The $b_{\rm any}$ component is determined using the events in the $p_t$-unbalanced sample
with the $\eta'$-$sdb$ contribution subtracted.
Here, the assumption of the same shape in the invariant mass distribution for the $b_{\rm any}$ component in the $p_t$-balanced and  $p_t$-unbalanced samples is implied.
For the $\gamma\rho$ mode, the sum of $\eta'$-$sdb$ and $b_{\rm any}$ is determined from the events in the $p_t$-unbalanced sample. These two components are hard to distinguish because of peaking background in the $\gamma\rho^{0}$ invariant mass distribution, caused by the  large width of the $\rho$ meson and the $\eta^\prime$ mass-constraint fit. 

The yield and shape for the two components, $\eta'$-$sdb$ and $b_{\rm any}$, separated (combined) for the $\eta\pi\pi$ ($\gamma\rho$) mode, are fixed in the simultaneous  fit. The exponential of a second-order polynomial is used to describe the $NR$ component with the yield and shape floating in the fit for both the $\eta\pi\pi$ and $\gamma\rho$  modes.

\subsection{Results of the $\eta_{c}(1S)$ and $\eta_{c}(2S)$ fits}
\begin{figure*}[htb]
\centering
\includegraphics[width=7cm]{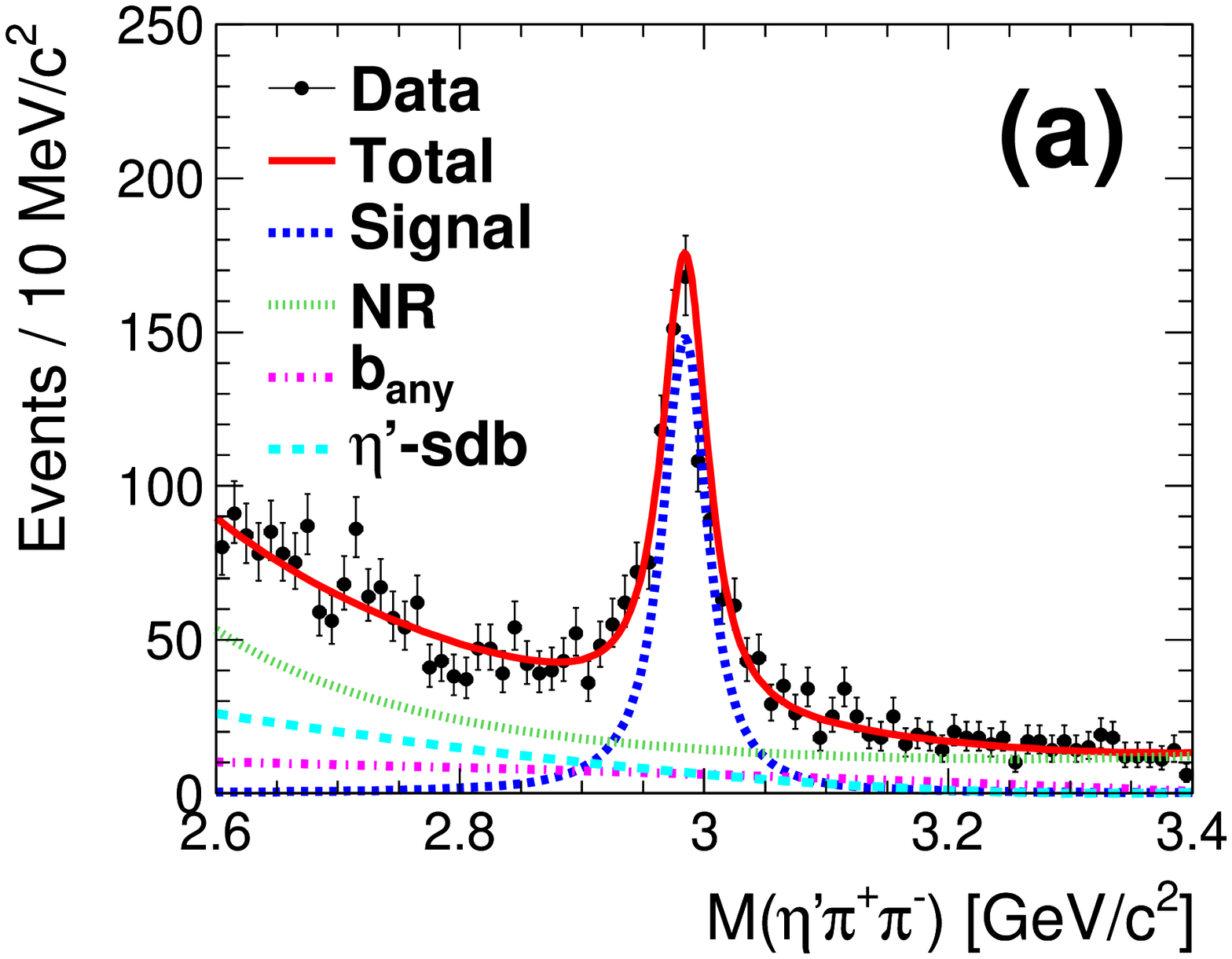}
\includegraphics[width=7cm]{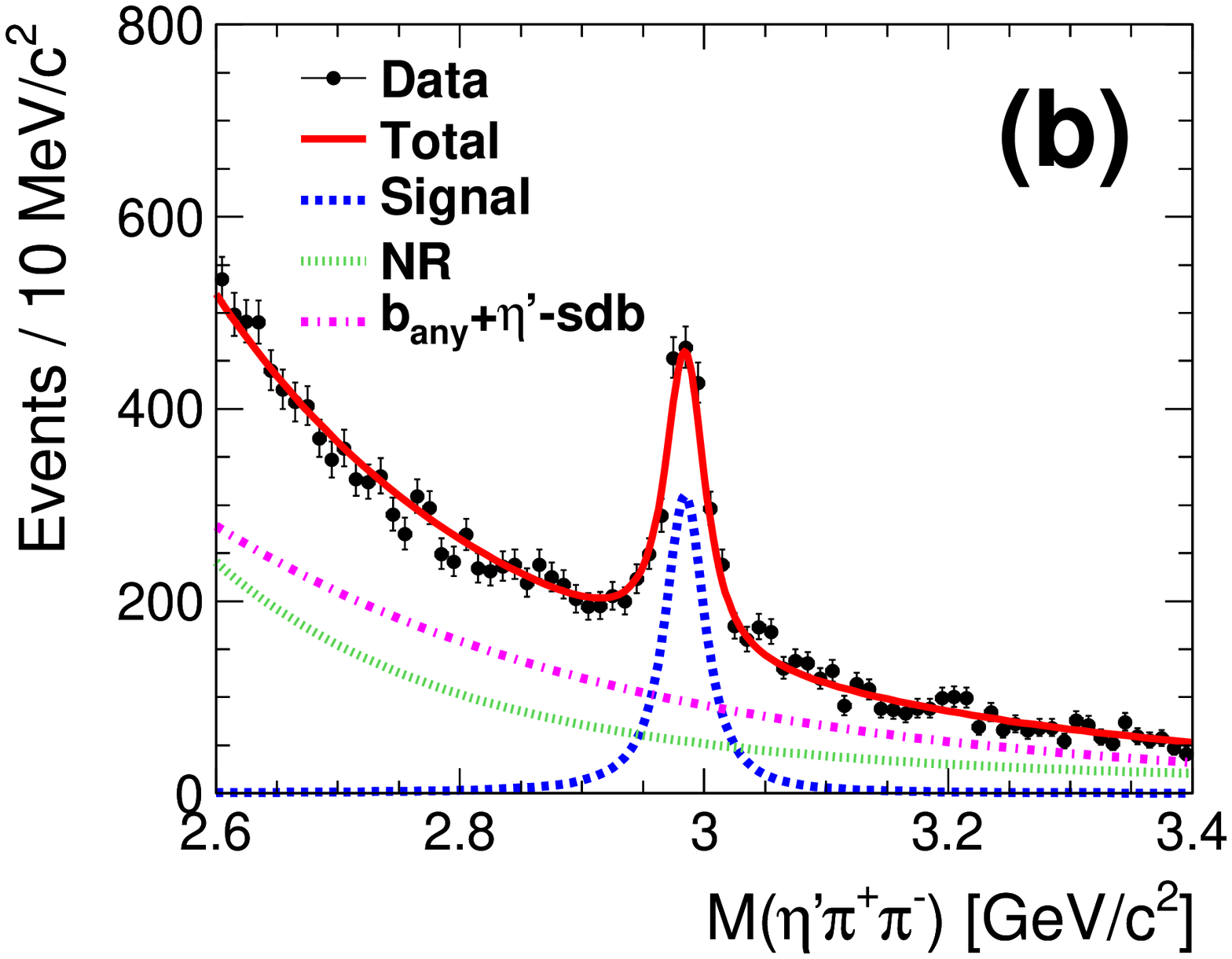}
\caption{(Color online)  The invariant mass distribution for the $\eta'\pi^+\pi^-$ candidates for (a) the $\eta\pi\pi$ mode and (b) the $\gamma\rho$ mode, in the $\eta_c(1S)$ region. The dots with error bars are data. The red solid line is the fit; the blue dashed line is fitted signal for $\eta_{c}(1S)$. The green dot, cyan long-dashed, and magenta dashed-dot lines are the $NR$, $\eta'$-$sdb$ and $b_{\rm any}$ ($b_{\rm any}$ + $\eta'$-$sdb$ merged into the magenta dashed-dot line for the $\gamma\rho$ mode) background components, respectively. }
\label{mx_1s_sumfit}
\end{figure*}

Simultaneous fits to the $\eta^{\prime}\pi^{+}\pi^{-}$ mass spectra with the $\eta\pi\pi$ and $\gamma\rho$ modes combined are performed for both $\eta_{c}(1S)$ and $\eta_{c}(2S)$. The result of the fit for the $\eta_{c}(1S)$ signal and background contributions are shown in Fig.~\ref{mx_1s_sumfit}. The $\eta_{c}(1S)$ mass and width are determined to be $\mathrm{\mathit{M}=2984.6\pm0.7}$ MeV/$c^{2}$ and $\mathrm{\Gamma = 30.8^{+2.3}_{-2.2}}$ MeV, with yields of  $\mathrm{\mathit{n}_{1} =945^{+38}_{-37}}$ for the $\eta\pi\pi$ mode and  $\mathrm{\mathit{n}_{2} =1728^{+69}_{-68}}$ for the $\gamma\rho$ mode.

\begin{figure*}[htb]
\centering
\includegraphics[width=7cm]{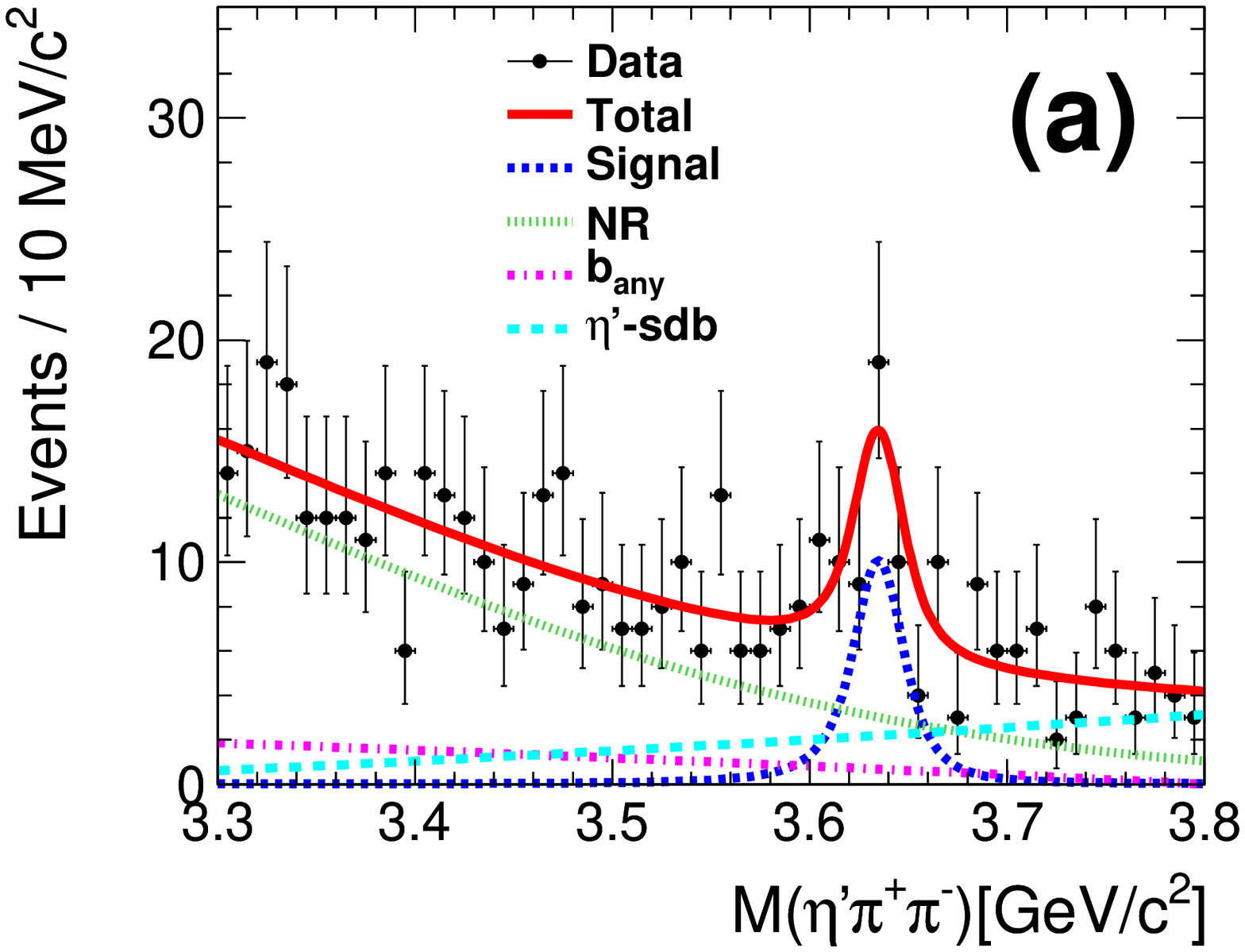}
\includegraphics[width=7cm]{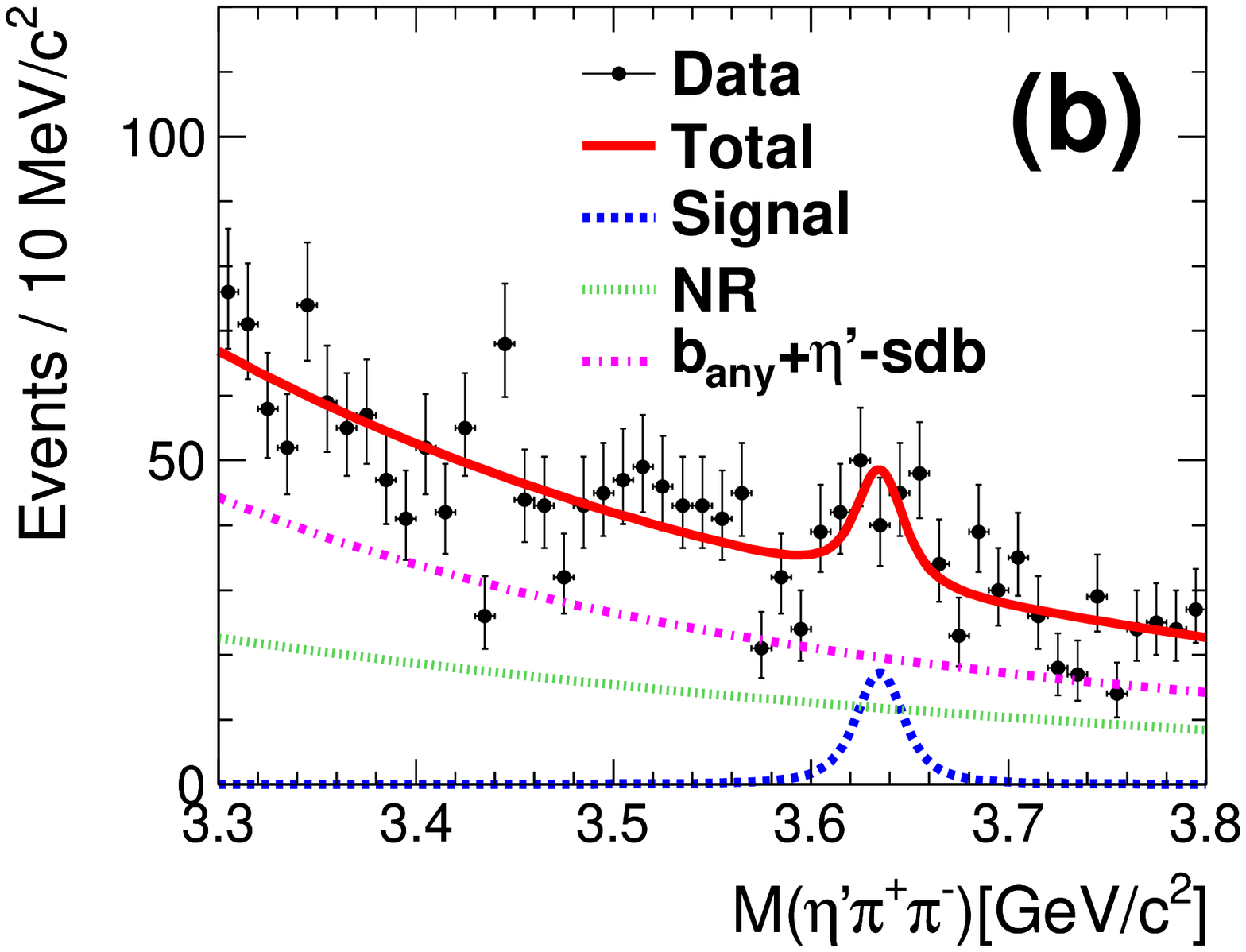}
\caption{(Color online) The invariant mass distribution for the $\eta'\pi^+\pi^-$ candidates for (a) the $\eta\pi\pi$ mode and (b) the $\gamma\rho$ mode,  in the $\eta_c(2S)$ region. The dots with error bars are data. The red solid line is the fit; the blue dashed line is fitted signal for $\eta_{c}(2S)$. The green dot, cyan long-dashed, and magenta dashed-dot lines are the $NR$, $\eta'$-$sdb$ and $b_{\rm any}$($b_{\rm any}$ + $\eta'$-$sdb$ merged into the magenta dashed-dot line for the $\gamma\rho$ mode) background components, respectively. }
\label{mx_2s_sumfit}
\end{figure*}

Figure~\ref{mx_2s_sumfit} shows the result of the fit for the  $\eta_{c}(2S)$ region, which results in a signal with a statistical significance of $\mathrm{5.5\sigma}$, and yields of  $\mathrm{\mathit{n}_{1} =41^{+9}_{-8}}$ for the  $\eta\pi\pi$ mode and $\mathrm{\mathit{n}_{2} =65^{+14}_{-13}}$ for the $\gamma\rho$ mode. The $\eta_{c}(2S)$ mass is determined to be $\mathrm{\mathit{M}=(3635.1\pm3.7)}$ MeV/$c^{2}$; its width is fixed to the world-average value of 11.3 MeV \cite{PDG_2016} in the fit.
The statistical significance for the $\eta_{c}(2S)$ signal is calculated with the $\chi^2$ distribution $-2\rm{ln}({\cal L}_{0}/{\cal L}_{\rm max})$ for $N_{\rm dof}$ degrees of freedom. Here, ${\cal L}_{\rm max}$ and ${\cal L}_{0}$ are the maximum likelihoods of the fits with the signal yield floating and fixed to zero, respectively, and $N_{\rm dof}$ = 2 is the difference in the number of floating parameters between the nominal fit and the latter fit.

From Eq.~\eqref{crs_def_cal}, with the fitted signal yields as input, the product of the two-photon decay width and the branching fraction for the $\eta_{c}(1S)$ and $\eta_{c}(2S)$ are calculated to be $\mathrm{\Gamma_{\gamma\gamma}{\cal B}(\eta^\prime\pi^{+}\pi^{-}) = (65.4\pm2.6)}$ eV and 
$\mathrm{(5.6^{+1.2}_{-1.1})}$ eV, respectively. 
The fit results for the $\eta_{c}(1S)$ and $\eta_{c}(2S)$ are summarized in Table~\ref{etac_resl}.
 
\begin{table*}[!hbpt]
\caption{Summary of the results for the $\eta_{c}(1S)$ and $\eta_{c}(2S)$: $n_{s}$ is the yield; $M$ and $\Gamma$ are the mass and width; $\Gamma_{\gamma\gamma}{\cal B}$ is the product of the two-photon decay width and the branching fraction for $\eta_{c}\rightarrow\eta'\pi^{+}\pi^{-}$. The first error is statistical and the second is systematic.}
\begin{center}
\begin{tabular}{ccccc}
\hline
\hline
\multicolumn{1}{c}{~} & \multicolumn{2}{c}{$\eta_{c}(1S)$} &  \multicolumn{2}{c}{$\eta_{c}(2S)$} \\ 
\hline
                 & $\gamma\rho$  &  $\eta\pi^{+}\pi^{-}$ & $\gamma\rho$ &  $\eta\pi^{+}\pi^{-}$  \\
\hline
$n_{s}$   & $1728^{+69}_{-68}$   &  $945^{+38}_{-37}$   & $65^{+14}_{-13}$    &  $41^{+9}_{-8}$ \\
\hline
\multicolumn{1}{c}{$M$ (MeV/$c^{2}$)}      & \multicolumn{2}{c}{$2984.6\pm0.7\pm2.2$} &  \multicolumn{2}{c}{$3635.1\pm3.7\pm2.9$} \\
\multicolumn{1}{c}{$\Gamma$ (MeV)}    & \multicolumn{2}{c}{$30.8^{+2.3}_{-2.2}\pm2.5$} &  \multicolumn{2}{c}{11.3 [fixed] }      \\
$\Gamma_{\gamma\gamma}{\cal B}$ (eV) & \multicolumn{2}{c}{$65.4\pm2.6\pm7.8$}    &  \multicolumn{2}{c}{$5.6^{+1.2}_{-1.1}\pm1.1$}    \\
\hline
\hline
\end{tabular}
\end{center}
\label{etac_resl}
\end{table*}

\subsection{Systematic uncertainties}
The systematic uncertainties are summarized in Table~\ref{serr_MW_twomodes}.
We estimate the uncertainty in the trigger efficiency using signal MC events.  The differences between the two efficiencies with and without simulation of the trigger conditions are evaluated to be $0.5\%$ ($0.6\%$) for $\eta_{c}(1S)$ 
($\eta_{c}(2S)$) in the $\gamma\rho$ mode, and  1.4\% for both $\eta_{c}$ mesons in the $\eta\pi\pi$ mode.  
The contribution to the systematic uncertainty arising from pion identification is studied using an inclusive $D^{*}$ sample. 
The uncertainties of pion identification are found to be $1.8\%$ $(2.3\%)$
in the $\gamma\rho$ mode and $1.5\%$ $(1.8\%)$ in the $\eta\pi\pi$ mode for $\eta_{c}(1S)$ $[\eta_{c}(2S)]$. 
The averaged values of deviations in the yield, mass, and width between the two simultaneous fits, with the $|\Sigma p^*_{t}|$ requirement changed by $\pm 0.01$ GeV/$c$ in the $\gamma\rho$ mode and by $\pm 0.02$ GeV/$c$ in the $\eta\pi\pi$ mode, are treated as systematic uncertainties.

Two methods are applied to evaluate the systematic uncertainty related to the uncertainty in the $NR$ background shape: (1) changing the mass window size in the fit; (2) altering the fit function for the background-shape description.  
The difference between the average values of the two fit yields calculated by changing the mass window width by  $\pm 100$ MeV/$c^{2}$  
is regarded as systematic uncertainty:  we find $2.3\%$ $(9.0\%)$ in the $\gamma\rho$ mode and 
$2.2\%$ $(9.5\%)$ in the $\eta\pi\pi$ mode for $\eta_{c}(1S)$ $(\eta_{c}(2S))$.
The contribution to the uncertainty in the fit yield estimated by varying the order of the polynomial function is found to be minor and thus is neglected. 

The uncertainty in the determination of the $\eta'$-$sdb$ and $b_{\rm any}$ backgrounds is estimated with changes in the $\eta'$-$sdb$ window size by $\pm 0.01$ GeV/$c^2$. The resulting difference in yields is evaluated to be 2.5\% for $\eta_{c}(1S)$ and 4.8\% for $\eta_{c}(2S)$  and is treated as the uncertainty.

The uncertainty from the $\pi^{0}$-veto is estimated as the difference in efficiency with and without the $\pi^{0}$-veto. The uncertainties for the $\eta$ reconstruction efficiency are studied using an inclusive $\eta$ sample, and its deviation from the MC simulation plus its error in quadrature is 4.9\%.
The systematic uncertainties related to  charged track reconstruction efficiency, luminosity function calculation, and experimental-conditions dependence are studied via charmonium decay to four charged mesons \cite{etac_belle,etac_belle_1}. 
The evolution of the background conditions over time adds  an additional uncertainty of $3\%$ in the yield determination. 
The accuracy of the two-photon luminosity is estimated to be $5\%$  including the uncertainties from radiative corrections $(2\%)$, the uncertainty from the form-factor effect $(2\%)$, and the error of the integrated luminosity $(1.36\%)$.

\begin{figure*}[!hbpt]
\centering
\includegraphics[width=7cm]{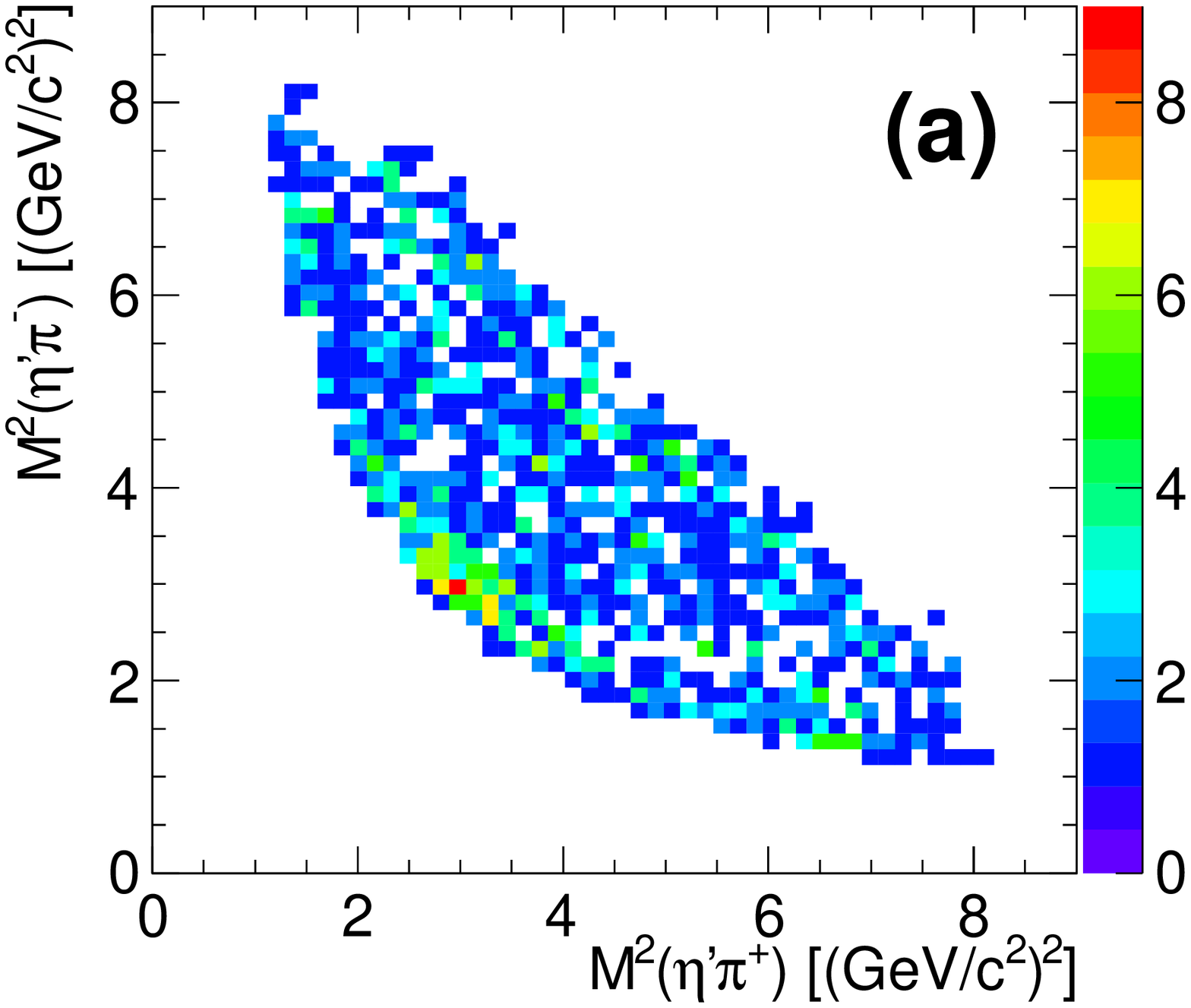}
\includegraphics[width=7cm]{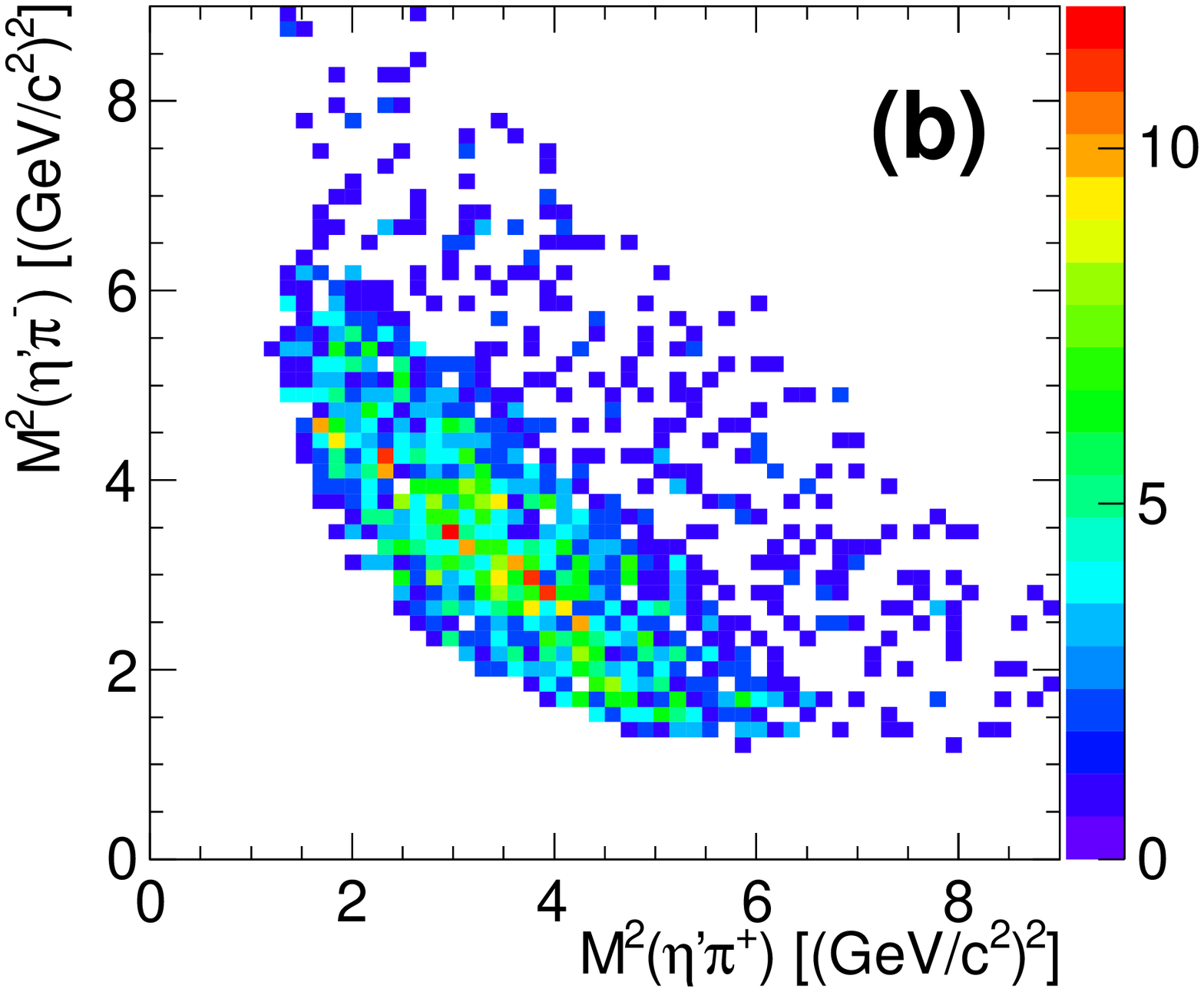}
\caption{The Dalitz plots for events selected in the $\eta_c(1S)$ signal (a) and $sdb$ (b) regions.}
\label{dalitz_sig_vs_sdb}
\end{figure*}

\begin{figure*}[!hbpt]
\begin{center}
\begin{minipage}[t]{0.45\textwidth}
\centering
\begin{overpic}[width=1.0\linewidth]{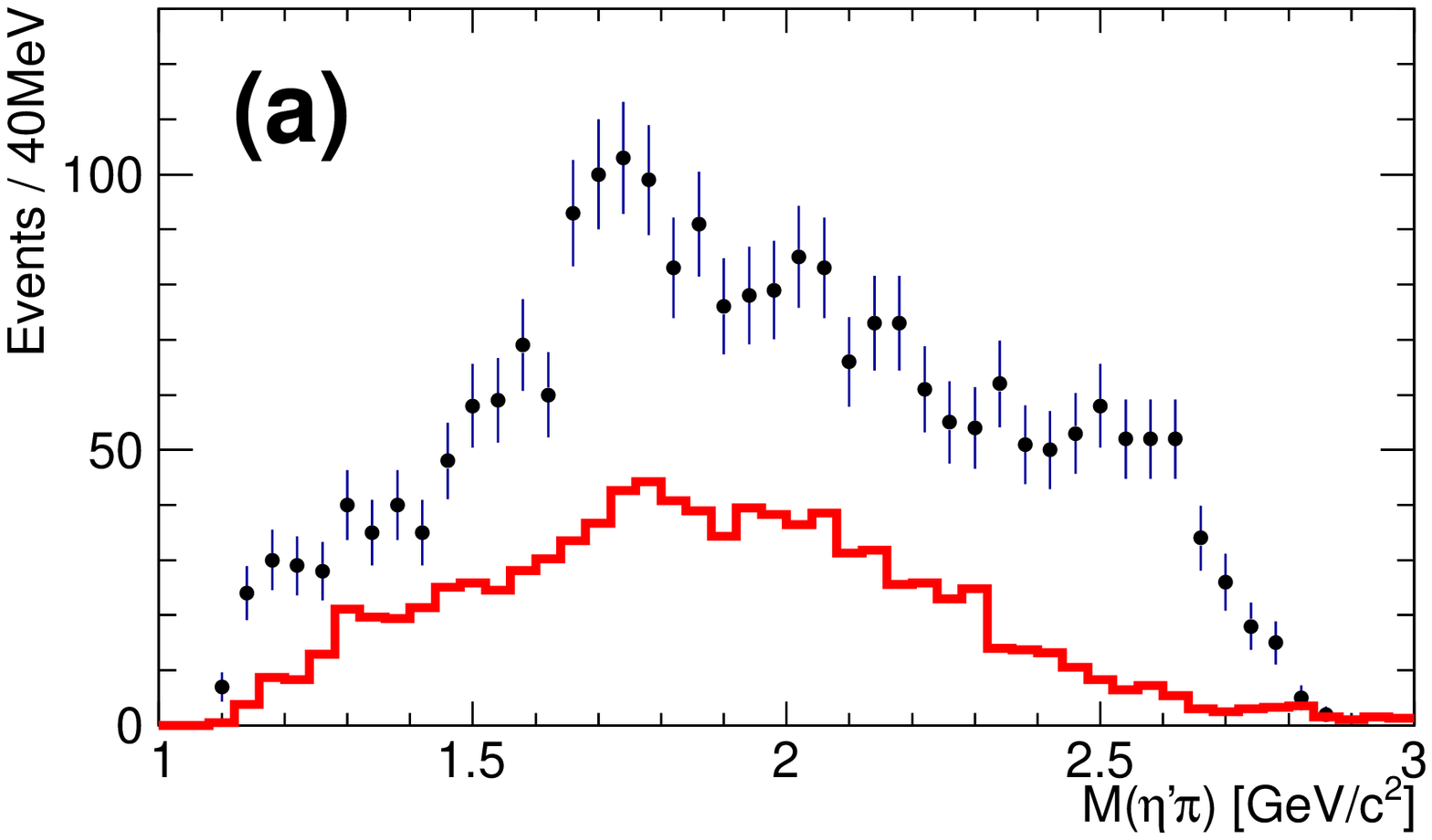}
\end{overpic}
\end{minipage}
\begin{minipage}[t]{0.45\textwidth}
\centering
\begin{overpic}[width=1.0\linewidth]{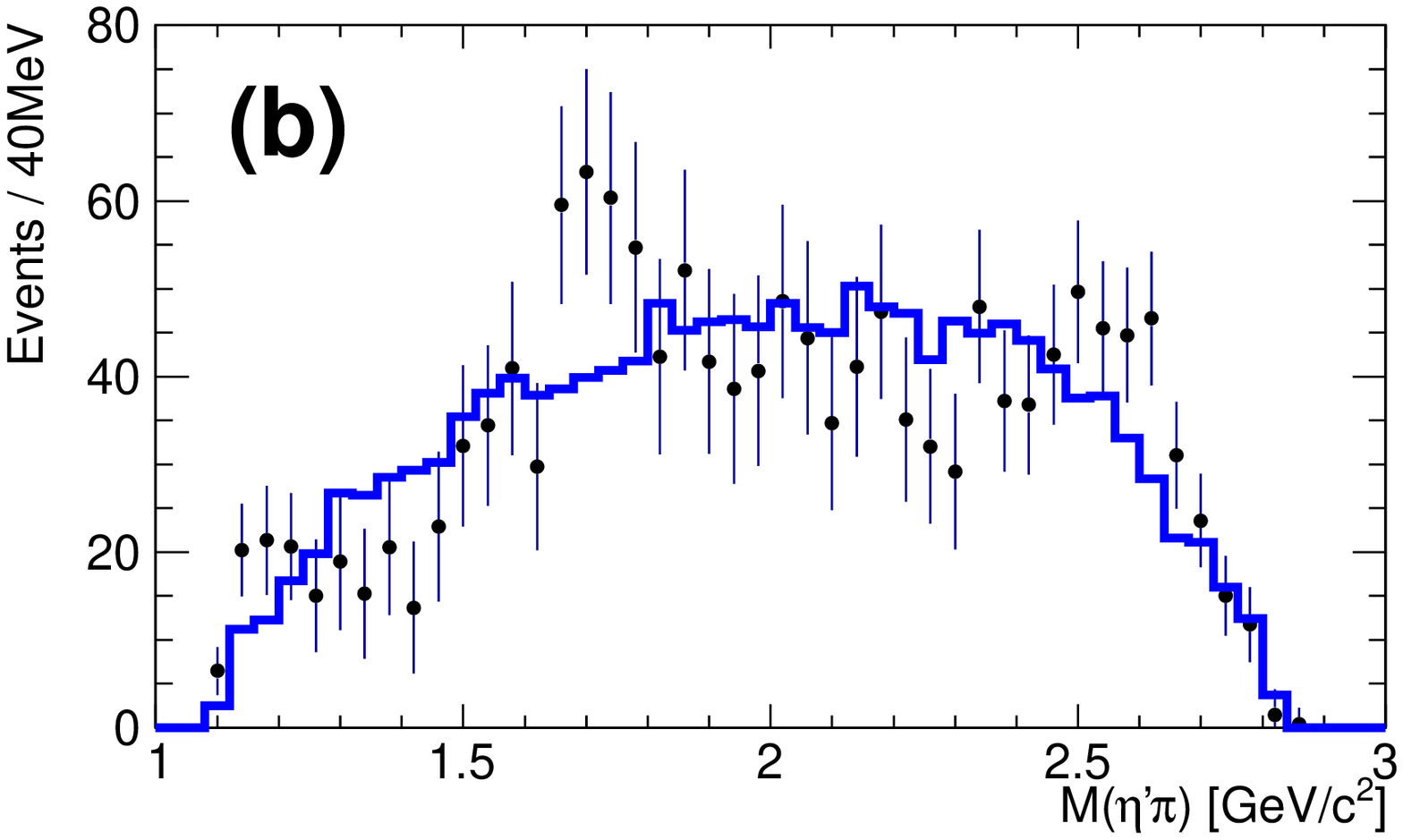}
\end{overpic}
\end{minipage}

\begin{minipage}[t]{0.45\textwidth}
\centering
\begin{overpic}[width=1.0\linewidth]{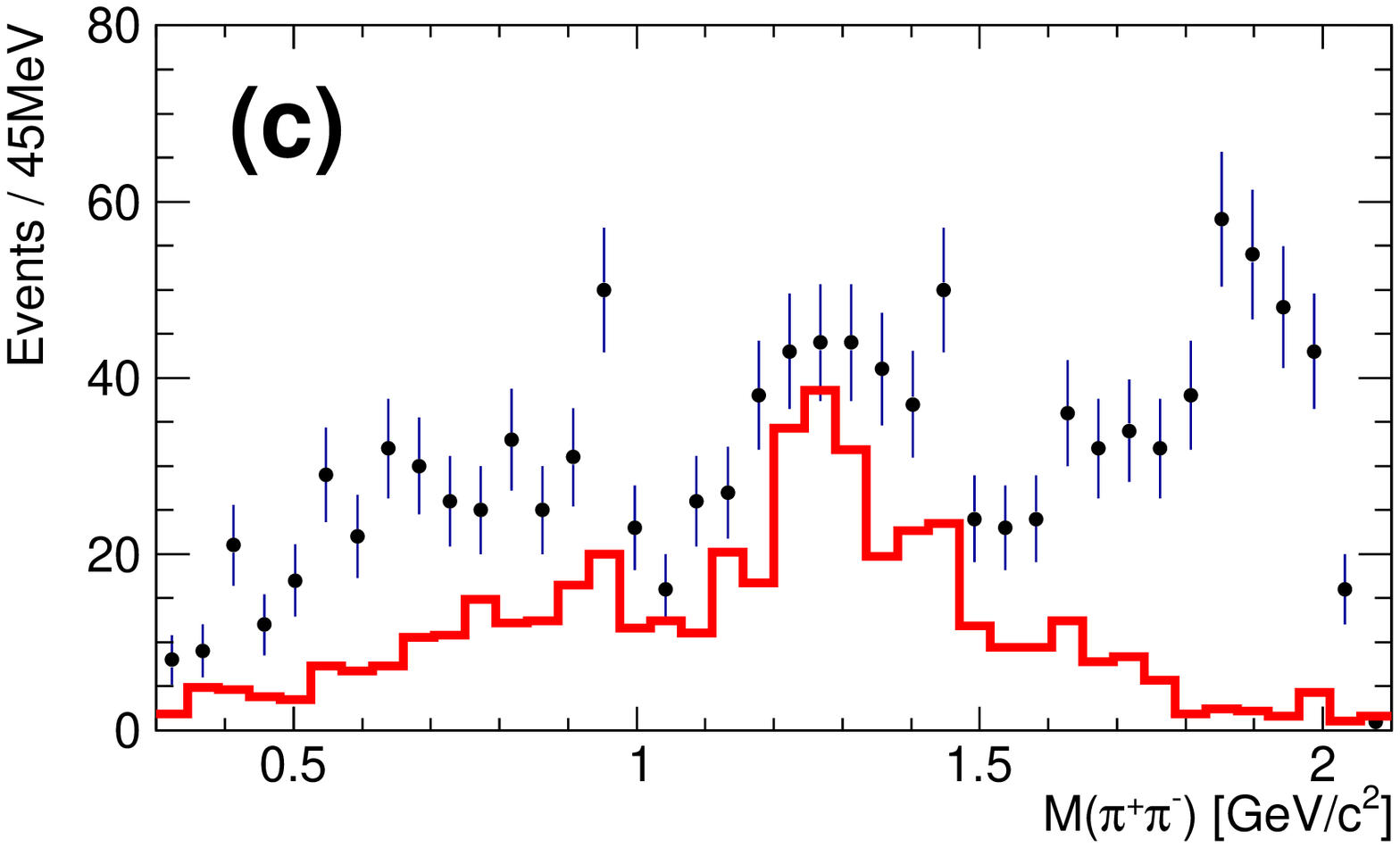}
\end{overpic}
\end{minipage}
\begin{minipage}[t]{0.45\textwidth}
\centering
\begin{overpic}[width=1.0\linewidth]{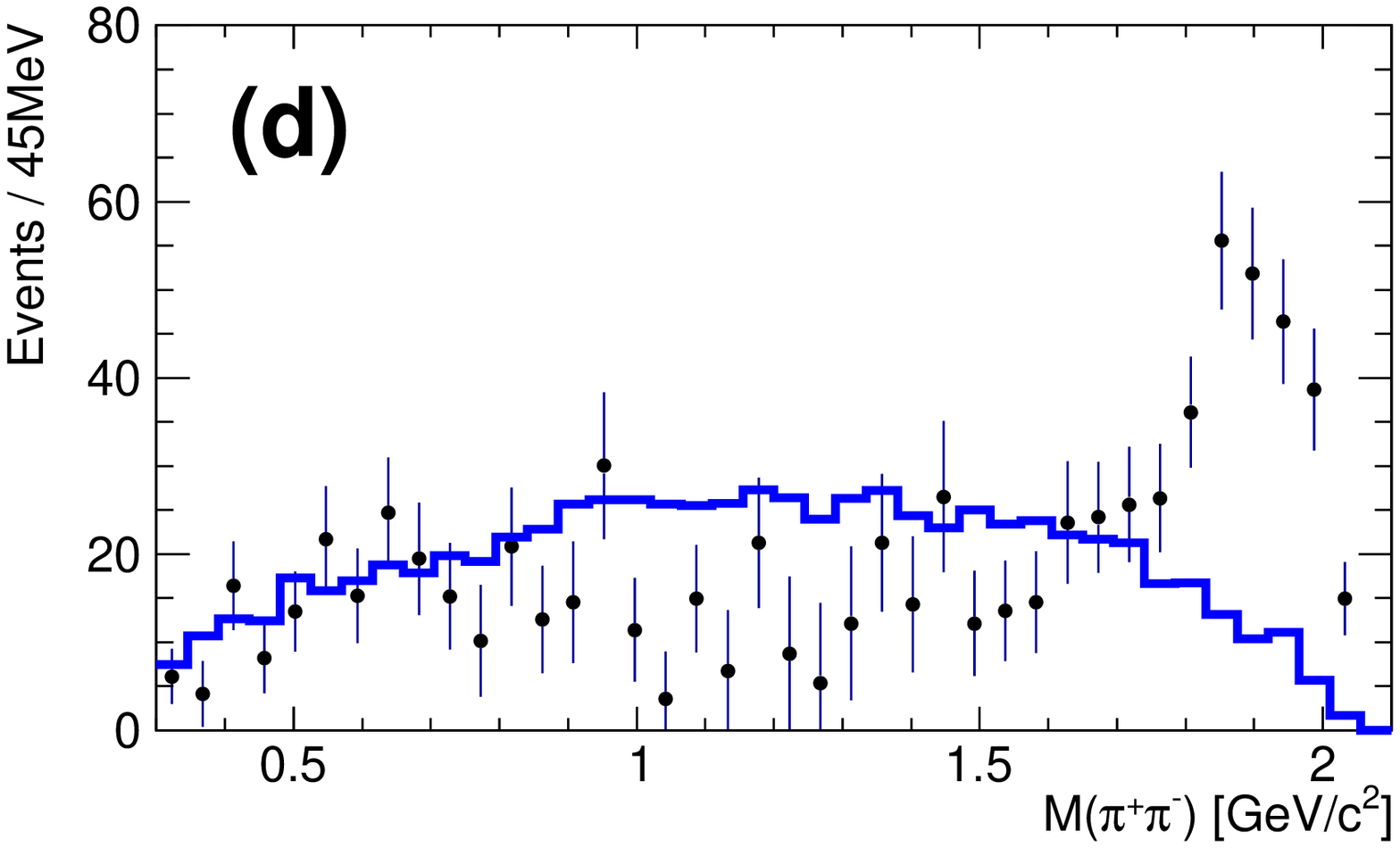}
\end{overpic}
\end{minipage}
\end{center}
\renewcommand{\figurename}{Fig.}
\caption{(a) [(c)] The invariant mass M($\eta'\pi^+$) distributions (two entries per event) [M($\pi^+\pi^-$) distributions] in data for the events selected in the $\eta_c(1S)$ signal region is drawn as the black solid dots with error bars. The red histogram is for the normalized $sdb$ background events. (b) [(d)] The black solid dots with error bars is the M($\eta'\pi^+$) [M($\pi^+\pi^-$)] distribution in the $\eta_c(1S)$ signal region in data after subtraction of the $sdb$ background and the blue histogram normalized to data is for MC events of the $\eta_c(1S)$ decays to three-body final state  according to PHSP distribution.}
\label{metp_m2pix_sub_sdb}
\end{figure*}

The efficiency for the $\eta^\prime\pi^+\pi^-$ events is determined with the MC sample generated with $\eta_c(1S)$ decays to three-body $\eta^\prime\pi^+\pi^-$ according to phase space distribution.
Possible intermediate states in $\eta_c(1S)$ decays  are checked in data.
Figure~\ref{dalitz_sig_vs_sdb} shows the Dalitz plots for the $\eta'\pi^+\pi^-$ events selected in the $\eta_c(1S)$ signal window of $\mathrm{ [2.90,3.06]}$ GeV/$c^2$ and sideband region of $\mathrm{ [2.60,2.81]\cup[3.15,3.36]}$ GeV/$c^2$ (denoted as $sdb$) in the $\eta\pi\pi$ mode.
Figures~\ref{metp_m2pix_sub_sdb}(a) and (c) show the $\eta'\pi^+$ (charge conjugate implied, two entries per event) and $\pi^+\pi^-$ invariant mass distributions  for the events selected in the $\eta_c(1S)$ signal and $sdb$ regions. The corresponding mass distributions after subtraction of the normalized $sdb$ background are shown in Figs.~\ref{metp_m2pix_sub_sdb}(b) and (d).  
 Broad structures are seen in distributions of both $M(\eta'\pi^+)$ near 1.7 GeV/$c^2$ and $M(\pi^+\pi^-)$ near 2 GeV/$c^2$. 
To estimate the effect on the efficiency due to the two-body intermediate states in $\eta_c(1S)$ decays,  a possible two-body intermediate state $\eta_c(1S)\rightarrow\eta'f_0(2100)$ is assumed and simulated, and the averaged efficiency of this mode and the three-body phase space sample is calculated.  
 Here, an approximately equal ratio of two yields $n_{\rm s,three-body}/n_{\rm s,two-body}$  is assumed in averaging the two modes. 
The relative difference in efficiencies between the phase space (PHSP) MC sample and the average efficiency is estimated to be $\Delta\epsilon_{{\rm avr},\eta \pi\pi}$ = 8.8\% ($\Delta\epsilon_{{\rm avr},\gamma\rho}$ = 3.6\%) for the $\eta\pi\pi$ ($\gamma\rho$) mode. Taking the yield-weighted mean of $\Delta\epsilon_{{\rm avr},\eta \pi\pi}$ and $\Delta\epsilon_{{\rm avr},\gamma\rho}$ for the $\eta\pi\pi$ and $\gamma\rho$ modes combined in the fits, the uncertainty in efficiency related to the assumption of the uniform distribution in PHSP is found to be 6\%, which is added to the systematic error.

To examine the systematic uncertainty in the mass measurement for the $R\rightarrow\eta^\prime\pi^+\pi^-$ decay,
an inclusive control sample of the decay $D^{0}\rightarrow\eta^\prime K^0_S$ with $K^0_S\rightarrow\pi^{+}\pi^{-}$ is selected with a tight mass window for $\eta'$. 
The $D^0$ mass resulting from fits to the invariant mass spectra of $\eta^\prime K^0_S$ is shifted from its nominal value by 1.26 MeV/$c^2$ (0.93 MeV/$c^2$) in the $\eta\pi\pi$ ($\gamma\rho$) mode. The sum of the shift and statistical error in quadrature, scaled linearly to the $\eta_c$ mass, is taken as the contribution of the uncertainty for the mass scale. 
The uncertainty in the width determination 
is estimated by changing the mass resolution by $\pm 1$ MeV/$c^2$, and is found to be 1.2 MeV/$c^2$ 
for the $\eta_{c}(1S)$. 
The uncertainties for the resonance mass and width coming from $|\Sigma p^*_{t}|$ and background shape are determined with the same method as that for the  $\Gamma_{\gamma\gamma}\cal B$ measurement. 

Taking the yield-weighted mean of squared uncertainty for the $\gamma\rho$ and $\eta\pi^{+}\pi^{-}$  modes combined in the fits, the total systematic uncertainties in the measurements of $\Gamma_{\gamma\gamma}\cal B$, mass and width for $\eta_{c}(1S)$ $[\eta_{c}(2S)]$ are calculated by adding the individual mean uncertainties in quadrature.

\begin{table}[!hbpt]
\caption{Summary of systematic uncertainty contributions to the $\Gamma_{\gamma\gamma}{\cal B}$, mass and width for $\eta_{c}(1S)$, $\eta_{c}(2S)$ in the fit with $\gamma\rho$ and $\eta\pi^{+}\pi^{-}$ modes combined.}
\begin{center}
\begin{tabular}{c|cc}
\hline
\hline
\multicolumn{3}{c}{$\Delta(\Gamma_{\gamma\gamma} {\cal B})$/$(\Gamma_{\gamma\gamma}{\cal B})(\%)$}  \\
\hline
Source                     & $\eta_{c}(1S)$ & $\eta_{c}(2S)$  \\
\hline
Trigger efficiency         & 0.9 & 1.0 \\
$\pi^{\pm}$ identification efficiency         & 1.7 & 2.1    \\
$|\Sigma p^*_{t}|$         & 1.5 & 9.8    \\
Background shape           & 2.3 & 9.2    \\
$\eta$-$sdb$ and $b_{\rm any}$   & 2.5 & 4.8    \\
$\pi^{0}$-veto             & 2.4 & 2.2    \\
$\eta_c(2S)$ width error  & --  &  8.8 \\
$\eta$ reconstruction efficiency     & \multicolumn{2}{c}{4.9}   \\
Track reconstruction efficiency      & \multicolumn{2}{c}{5.5}   \\
Run dependence          & \multicolumn{2}{c}{3}    \\
Two-photon luminosity      & \multicolumn{2}{c}{5}   \\
PHSP assumption & \multicolumn{2}{c}{6}   \\
\hline
Total                      & 12  & 20    \\
\hline
\hline
 \multicolumn{3}{c}{$\Delta M$ (MeV/$c^{2})$}  \\
\hline
Mass scale                 & 2.1 & 2.6  \\
$|\Sigma p^*_{t}|$             & 0.1 & 1.1    \\
Background shape           & 0.7 & 0.4   \\
$\eta_c(2S)$ width error  & --  &  0.1 \\
Total                      & 2.2   & 2.9    \\
\hline
\multicolumn{3}{c}{$\Delta \Gamma$(MeV)}  \\
\hline
Mass resolution            & 1.2   & --  \\
$|\Sigma p^*_{t}|$         & 0.7 & --    \\
Background shape           & 2.1 & --    \\
Total                      & 2.5 & --   \\
\hline
\hline
\end{tabular}
\end{center}
\label{serr_MW_twomodes}
\end{table}



\section{measurements of the cross sections}

We utilize the data sample selected in the $\eta'\rightarrow\eta\pi\pi$ mode to 
measure the  non-resonant production of $\eta'\pi^+\pi^-$ final states via 
two-photon collisions.
The cross section of $e^{+}e^{-}\rightarrow e^{+}e^{-}h$  production is expressed as

\begin{eqnarray}\label{crs_def1}
&&\sigma_{e^{+}e^{-}\rightarrow e^{+}e^{-}h} = \int\sigma_{\gamma\gamma
\rightarrow h}(W,|\rm cos\theta^{*}|)\nonumber\\[1mm]
&&\times\frac{dL_{\gamma\gamma}(W)}{dW}dW d|\rm cos\theta^{*}|,
\end{eqnarray}
where $h$ denotes one of two  hadronic final states:  $\eta'\pi^{+}\pi^{-}$ or $\eta' f_2(1270)$. Here, $\theta^{*}$ is the angle between the $\eta'$ momentum and the beam direction in the $\gamma\gamma$ rest frame.

The differential cross section in the measurement of the $W$ and $|$cos$\theta^*|$ two-dimensional (2D) distribution for the final-state particles is calculated with the formula below, accounting for the efficiencies as a function of the measured variables.

\begin{eqnarray}\label{crs_def2}
&&\frac{d\sigma_{\gamma\gamma\rightarrow h}(W,\rm cos\theta^{*})}{d|\rm cos\theta^{*}|} = \nonumber\\[1mm]
&&\frac{\Delta N(W,\rm cos\theta^{*})/\epsilon(\mathit{W},\rm cos\theta^{*})}{L_{\rm int}\frac{dL_{\gamma\gamma}(W)}{dW}\Delta W\Delta |\rm cos\theta^{*}|},
\end{eqnarray}
where the yield $\Delta N$ is extracted by fitting the $|\Sigma p^*_{t}|$ [$M(\pi^{+}\pi^{-})$] distribution in a data subsample sliced in each 2D bin for the $\gamma\gamma\rightarrow\eta'\pi^{+}\pi^{-}$ $[\gamma\gamma\rightarrow\eta' f_{2}(1270)]$ production.
The efficiency $\epsilon(W,\rm cos\theta^{*})$ is evaluated using MC events for each 2D bin.  
$L_{\rm int}$ is the total integrated luminosity of the data and $dL_{\gamma\gamma}/dW$ is the two-photon luminosity function.

The $W$-dependent cross sections of $\gamma\gamma\rightarrow h$ are obtained by a summation over $|$cos$\theta^{*}|$ bins as

\begin{eqnarray}\label{crs_def3} 
&&\sigma_{\gamma\gamma\rightarrow h}(W) = \nonumber\\[1mm] 
&& \sum_{\Delta|\rm{cos}\theta^{*}|}\frac{d\sigma_{\gamma\gamma\rightarrow h}(W,\rm{cos}\theta^{*})}{d|\rm{cos}\theta^{*}|}\Delta|\rm{cos}\theta^{*}|.
\end{eqnarray}

\subsection{\texorpdfstring{Cross sections of $\gamma\gamma\rightarrow\eta'\pi^{+}\pi^{-}$ (including $\eta^\prime f_{2}(1270)$)}{sigam(gammagamma->eta'pi+pi-)} }

 \label{cr_fitmode}

\begin{figure*}[!hbpt]
  \begin{overpic}[width=7cm]{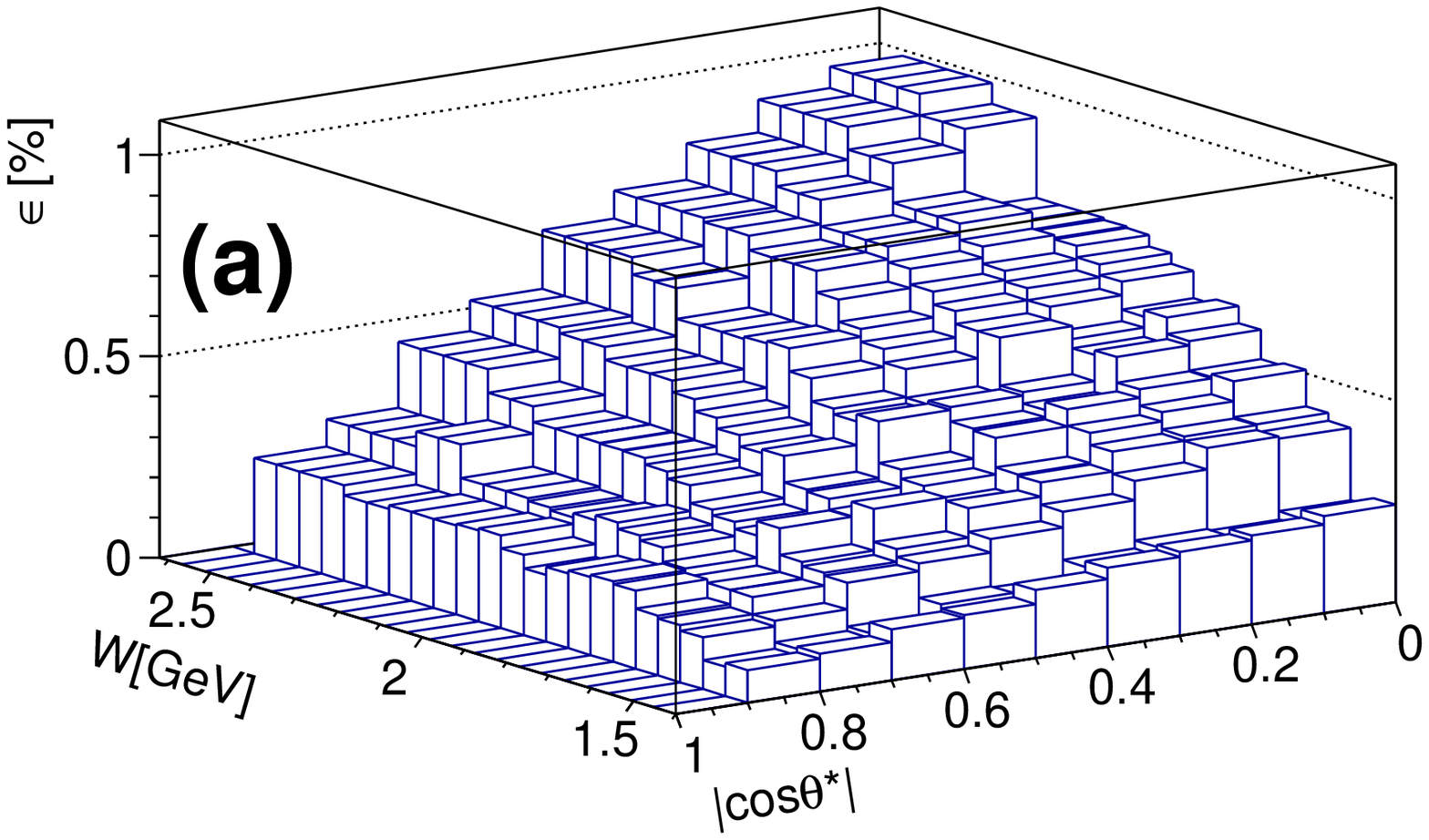}
  \end{overpic}
\begin{overpic}[width=7cm]{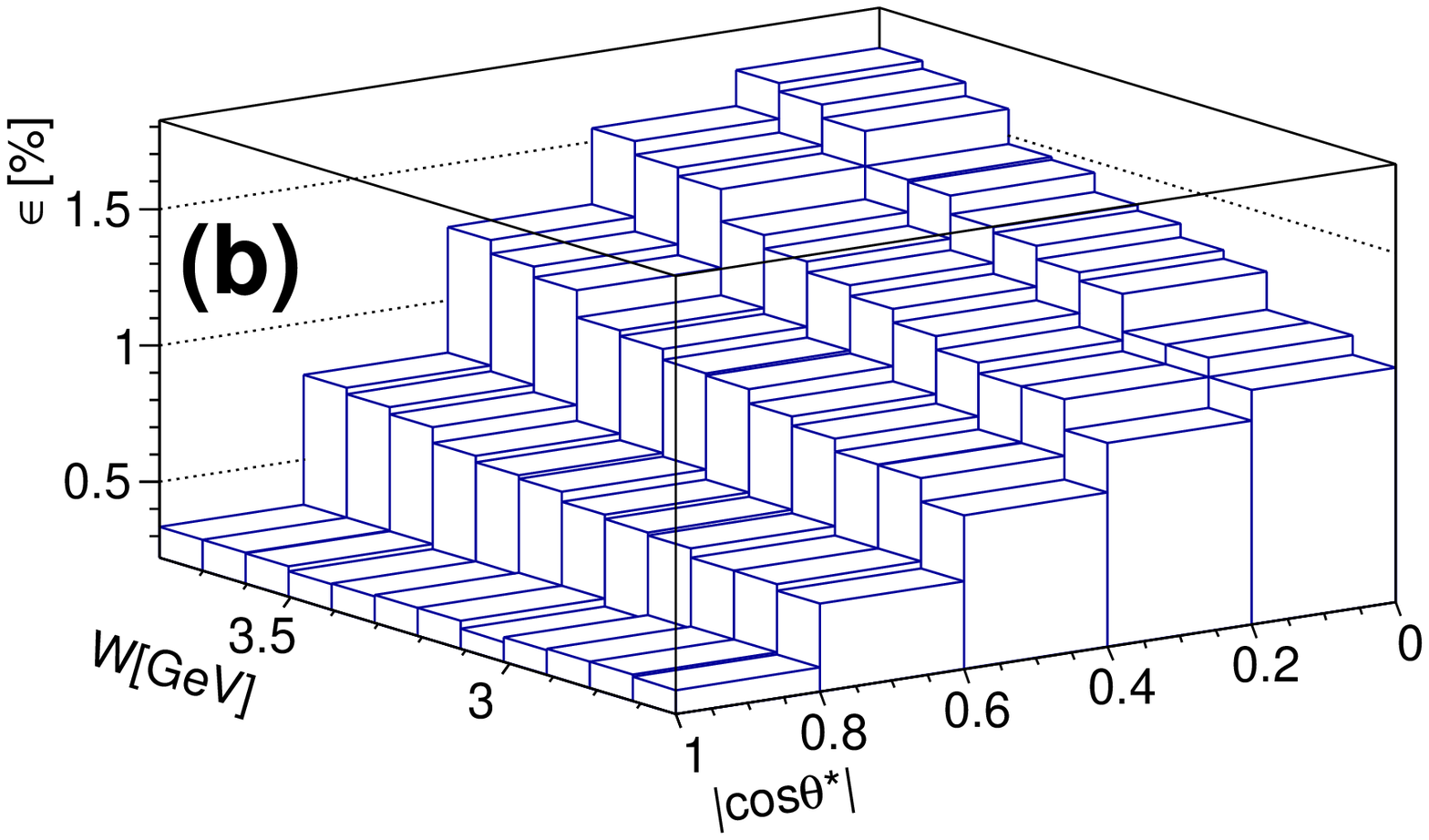}
\end{overpic}

\caption{Detection efficiency $\epsilon$ as a function of $W$ and $|$cos$\theta^{*}|$ for $\gamma\gamma\rightarrow\eta'\pi^{+}\pi^{-}$ with the $\eta\pi^{+}\pi^{-}$ mode in the regions of (a) $W\in [1.40,2.66)$ GeV and (b) $W\in [2.66,3.80]$ GeV.}
\label{eff_epp_gr}
\end{figure*}

We divide the $W$ distribution between 1.40 and 3.80 GeV into 35 bins and the  $|$cos$\theta^{*}|$ distribution into 10 and 5 bins for the $W$ regions of 1.40 to 2.66 GeV and 2.66 to 3.80 GeV,
respectively. The defined bin size and total number of bins in $W$ and $|$cos$\theta^{*}|$ are listed in the Table~\ref{W_bin}. Detection efficiencies as a function of $W$ and $|$cos$\theta^{*}|$ are shown in Fig.~\ref{eff_epp_gr}. The yield $\Delta N $ in Eq.~\eqref{crs_def2} is extracted by fitting the $|\Sigma p^*_{t}|$ distribution in data for each 2D bin. For the fit, the signal shape in MC is fixed, the $\eta'$-$sdb$ background in data is normalized and fixed, and the $b_{\rm any}$ background is
described by a third-order polynomial with its constant term fixed at 0 and the other parameters floating.

\begin{table}[!hbpt]
\caption{Defined bin size and total number of bins in $W$ and $|$cos$\theta^{*}|$ in individual $W$ ranges.}
\begin{center}
\begin{tabular}{c|c|c}
\hline
\hline
\multicolumn{1}{c|}{ \multirow{1}{*}{ $W$ [GeV] }} & \multicolumn{1}{c|}{\multirow{1}{*}{$\Delta W \times N_{\rm bins}$   [GeV]}} &  \multicolumn{1}{c}{$\Delta |$cos$\theta^{*}| \times N_{\rm bins}$} \\ 
\hline
1.40 -- 1.66 & $ 0.26\times1$     & $0.1\times10$ \\
1.66 -- 1.82 & $\ 0.08\times2$     & $0.1\times10$ \\
1.82 -- 2.66 & $\ 0.04\times21$    & $0.1\times10$ \\
2.66 -- 3.08 & $\ 0.06\times7$     & $0.2\times5$  \\
3.08 -- 3.40 & $ 0.16\times2$     & $0.2\times5$  \\
3.40 -- 3.80 & $ 0.20\times2$     & $0.2\times5$  \\
\hline
\hline
\end{tabular}
\end{center}
\label{W_bin}
\end{table}

\begin{figure}[htb]
\centering
\includegraphics[width=7cm]{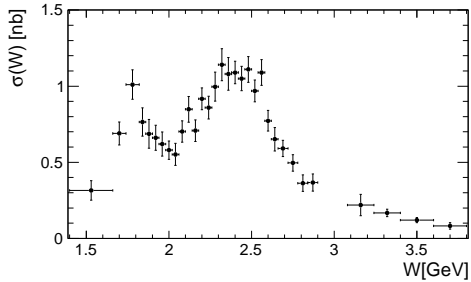}
\caption{Measured cross section of $\gamma\gamma\rightarrow\eta'\pi^{+}\pi^{-}$ (including $\eta'f_{2}(1270)$) for the $\eta\pi\pi$ mode.}
\label{cr_epp_gr}
\end{figure}

A background arising from $\eta'\rightarrow\gamma\rho$ decays in the candidate events of the $\eta\pi\pi$ mode is studied using the MC sample. 
One photon and four charged-pion tracks in the MC event, produced for the $\gamma\rho$ mode, plus a fake photon, is wrongly chosen as an $\eta^{\prime} \pi^+\pi^-$ combinatorial candidate for the $\eta\pi\pi$ mode. Here, the fake photon with low momentum is a neutral track composed of background hits or hit clusters split from charged pion tracks in the ECL.
This appears as a background component because of the additional fake photon in the event; it is  estimated using the pre-measured cross section for $\gamma\gamma\rightarrow\eta^{\prime} \pi^+\pi^-$ in data for the $\eta\pi\pi$ mode and is found to be small. The measured cross section for $\gamma\gamma\rightarrow\eta'\pi^{+}\pi^{-}$ for the $\eta\pi\pi$ mode after subtraction of this small contamination is shown in Fig.~\ref{cr_epp_gr}.

\subsection{\texorpdfstring{Result for the $\gamma\gamma\rightarrow\eta'f_{2}(1270)$ cross section measurement}{sigma(gammagamma->eta'f2(1270))}}

\begin{figure*}[htb]
\centering
\includegraphics[width=7cm]{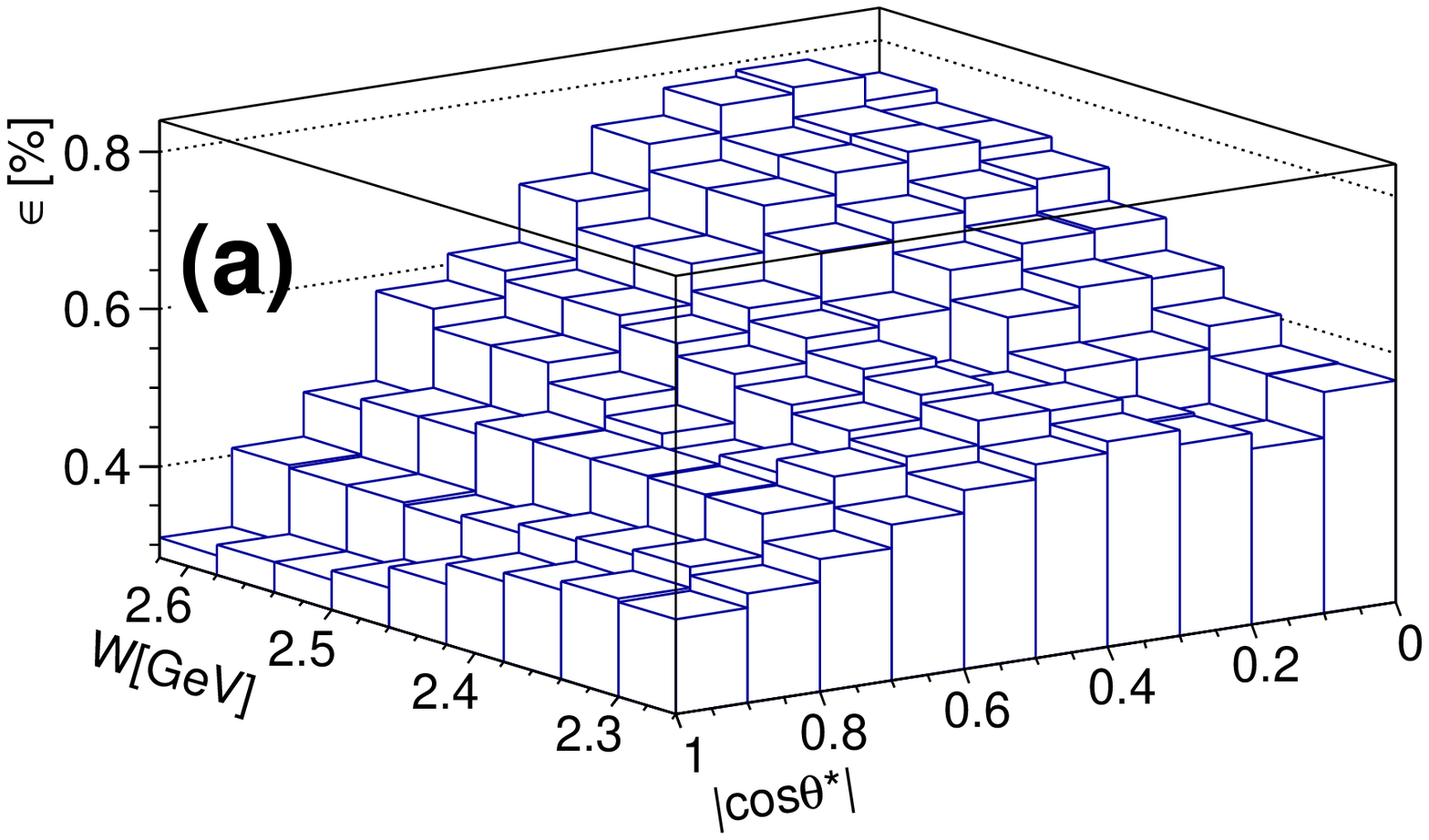}
\includegraphics[width=7cm]{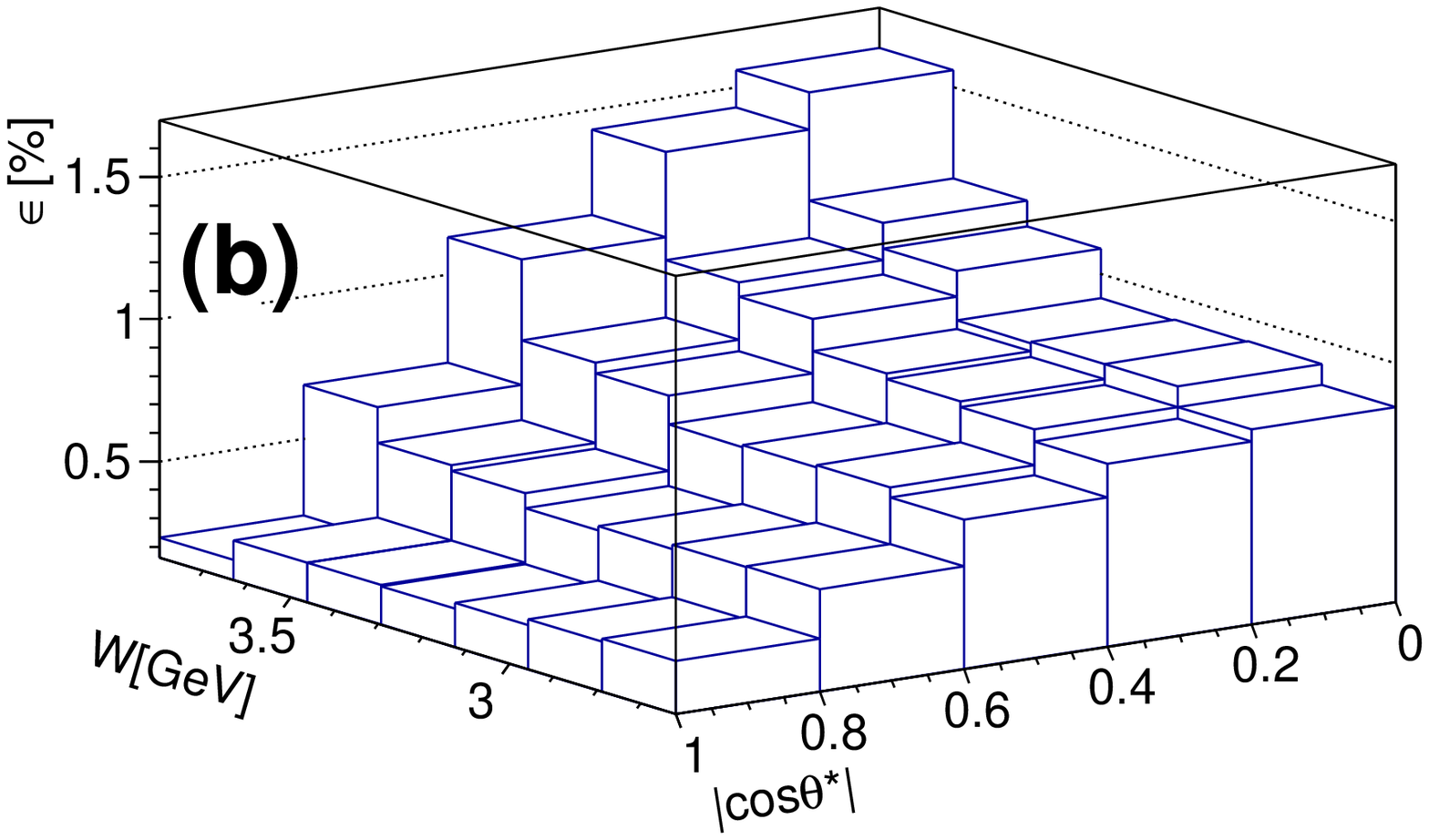}
\caption{Detection efficiency $\epsilon$ as function of $W$ and $|$cos$\theta^{*}|$ for $\gamma\gamma\rightarrow\eta'f_{2}(1270)$ in the  $\eta\pi\pi$ mode in the $W$ ranges of (a) [2.26, 2.62) GeV and (b) [2.62, 3.80] GeV.}
\label{eff_epp_f1270}
\end{figure*}

To calculate the cross section for the $\gamma\gamma\rightarrow\eta'f_{2}(1270)$ production, we divide $W$ into 16 bins from 2.26 to 3.80 GeV, and $|$cos$\theta^{*}|$ into 10 and 5 bins ($0<|$cos$\theta^{*}|< 1$) for the regions of $W \in [2.26, 2.62)$ GeV and $[2.62, 3.80]$ GeV, respectively. The efficiency $\epsilon$ in each 2D bin, evaluated using signal MC events for $\gamma\gamma\rightarrow\eta'f_{2}(1270)$ with the phase-space distribution, is shown in Fig.~\ref{eff_epp_f1270}.

The yield $\Delta N $ of $f_{2}(1270)$ in Eq.~\eqref{crs_def2} is extracted by fitting the invariant mass spectrum of $\pi^+\pi^-$ for the $f_{2}(1270)$ signal using the data subsample in each 2D bin. A broad $f_{2}(1270)$ signal in the $W$ region from 2.26 to 2.62 GeV near threshold is described by a $D$-wave Breit-Wigner function

\begin{eqnarray}
f_{{\rm BW}} = \frac{1}{(W^{2} - M^{2})^{2} + M^{2}\Gamma^{2}} q p^{5}, 
\end{eqnarray} 
where $M$ and $\Gamma$ are the $f_{2}(1270)$ mass and width. 
The $q$ and $p$ momentum variables are, respectively, of the $f_{2}(1270)$ in the $\gamma\gamma$ rest frame and of the $\pi$ meson from the $f_{2}(1270)$ decay in the $f_{2}(1270)$ rest frame. In the fits, $\Gamma$ is fixed to the world-average value, and $M$ is fixed to the value extracted from fitting the $\pi^+\pi^-$ invariant mass spectrum for the  $f_{2}(1270)$ 
using events in the full range of $W$ ($|$cos$\theta^{*}| <$ 1). 
The $f_{2}(1270)$ signal in the $W$ region above 2.62 GeV is described by a  normal Breit-Wigner function with both $M$ and $\Gamma$ fixed to the world-average values. We fix the fraction of the $\eta'$-$sdb$ background in the  fits. The combinatorial background, including non-$f_{2}(1270)$ and $b_{\rm any}$ events, is described by a fourth-order polynomial with its parameters fixed to the values extracted from the $f_{2}(1270)$ fit for each $W$ bin.

\begin{figure*}[!htp]
\centering
\includegraphics[width=7cm]{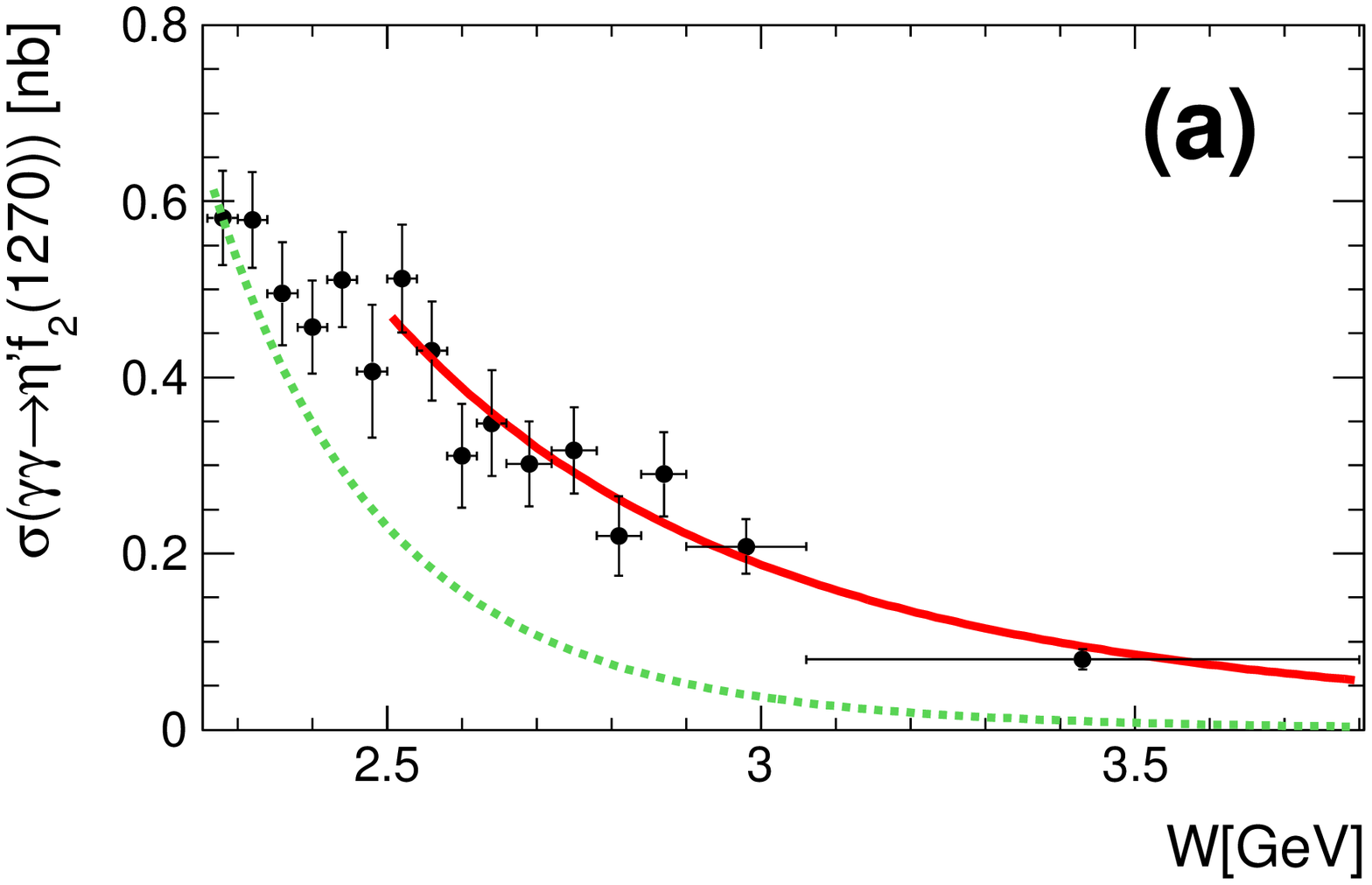}
\includegraphics[width=7cm]{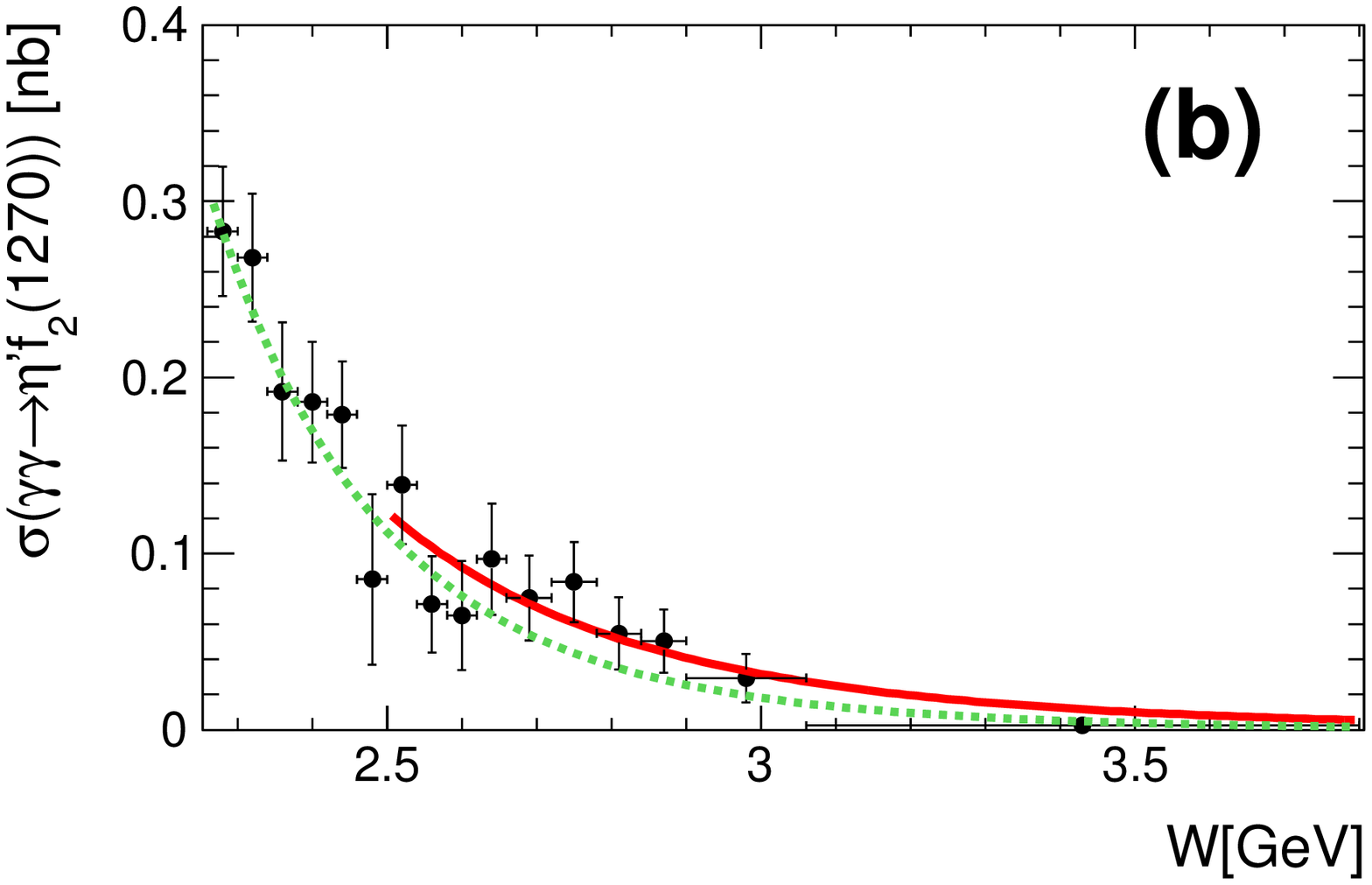}
\caption{(Color online) Measured cross sections for $\gamma\gamma\rightarrow\eta'f_{2}(1270)$.
The black dots with statistical error bars are the data within (a) $|\rm{cos}\theta^{*}|<1$ and (b) $|\rm{cos}\theta^{*}|<0.6$. The red solid lines are fitted curves with the $W$-power index  $n=5.1\pm1.0$ and $n=7.5\pm2.0$, respectively, assuming a $W$ dependence of $1/W^n$. The green dashed line corresponds to the  leading-term QCD prediction for neutral meson pairs ($n = 10$).}
\label{cr_f1270}
\end{figure*}

\begin{table}[!hbpt]
\caption{Measured cross sections as a function of $W$ within $|\rm{cos}\theta^{*}|<1$ for $\gamma\gamma\rightarrow\eta'f_{2}(1270)$ in the $\eta\pi\pi$ mode. The first error is statistical and the second is systematic.}
\begin{center}
\begin{tabular}{cc}
\hline
\hline
\multicolumn{1}{c}{ $W$(GeV)} & \multicolumn{1}{c}{$\sigma(\gamma\gamma\rightarrow\eta'f_{2}(1270))$ $(\rm nb)$}\\
\hline
2.26 -- 2.30 & $\mathrm{0.58\pm0.05\pm0.11}$\\
2.30 -- 2.34 & $\mathrm{0.58\pm0.05\pm0.11}$\\
2.34 -- 2.38 & $\mathrm{0.495\pm0.059\pm0.091}$\\
2.38 -- 2.42 & $\mathrm{0.457\pm0.053\pm0.087}$\\
2.42 -- 2.46 & $\mathrm{0.511\pm0.054\pm0.098}$\\
2.46 -- 2.50 & $\mathrm{0.407\pm0.075\pm0.086}$\\
2.50 -- 2.54 & $\mathrm{0.512\pm0.061\pm0.091}$\\
2.54 -- 2.58 & $\mathrm{0.430\pm0.056\pm0.078}$\\
2.58 -- 2.62 & $\mathrm{0.311\pm0.059\pm0.063}$\\
2.62 -- 2.66 & $\mathrm{0.348\pm0.060\pm0.063}$\\
2.66 -- 2.72 & $\mathrm{0.302\pm0.048\pm0.058}$\\
2.72 -- 2.78 & $\mathrm{0.317\pm0.049\pm0.053}$\\
2.78 -- 2.84 & $\mathrm{0.220\pm0.045\pm0.037}$\\
2.84 -- 2.90 & $\mathrm{0.290\pm0.048\pm0.051}$\\
2.90 -- 3.06 & $\mathrm{0.208\pm0.031\pm0.043}$\\
3.06 -- 3.80 & $\mathrm{0.080\pm0.011\pm0.019}$\\
\hline
\hline
\end{tabular}
\end{center}
\label{cr_f1270_sta_sys_epp}
\end{table}

\begin{figure*}[!hbpt]
\begin{overpic}[width=7cm]{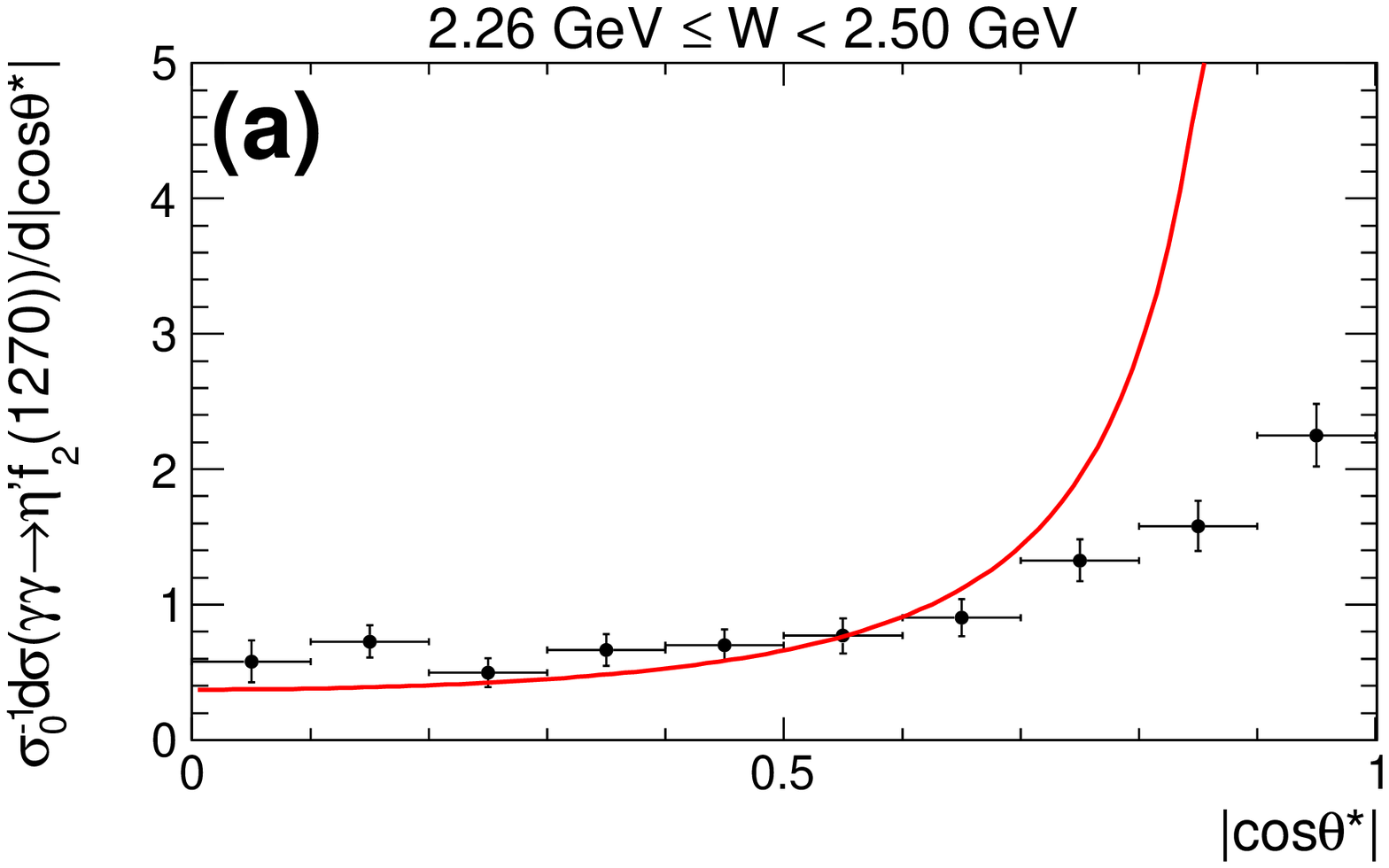}
\end{overpic}
\begin{overpic}[width=7cm]{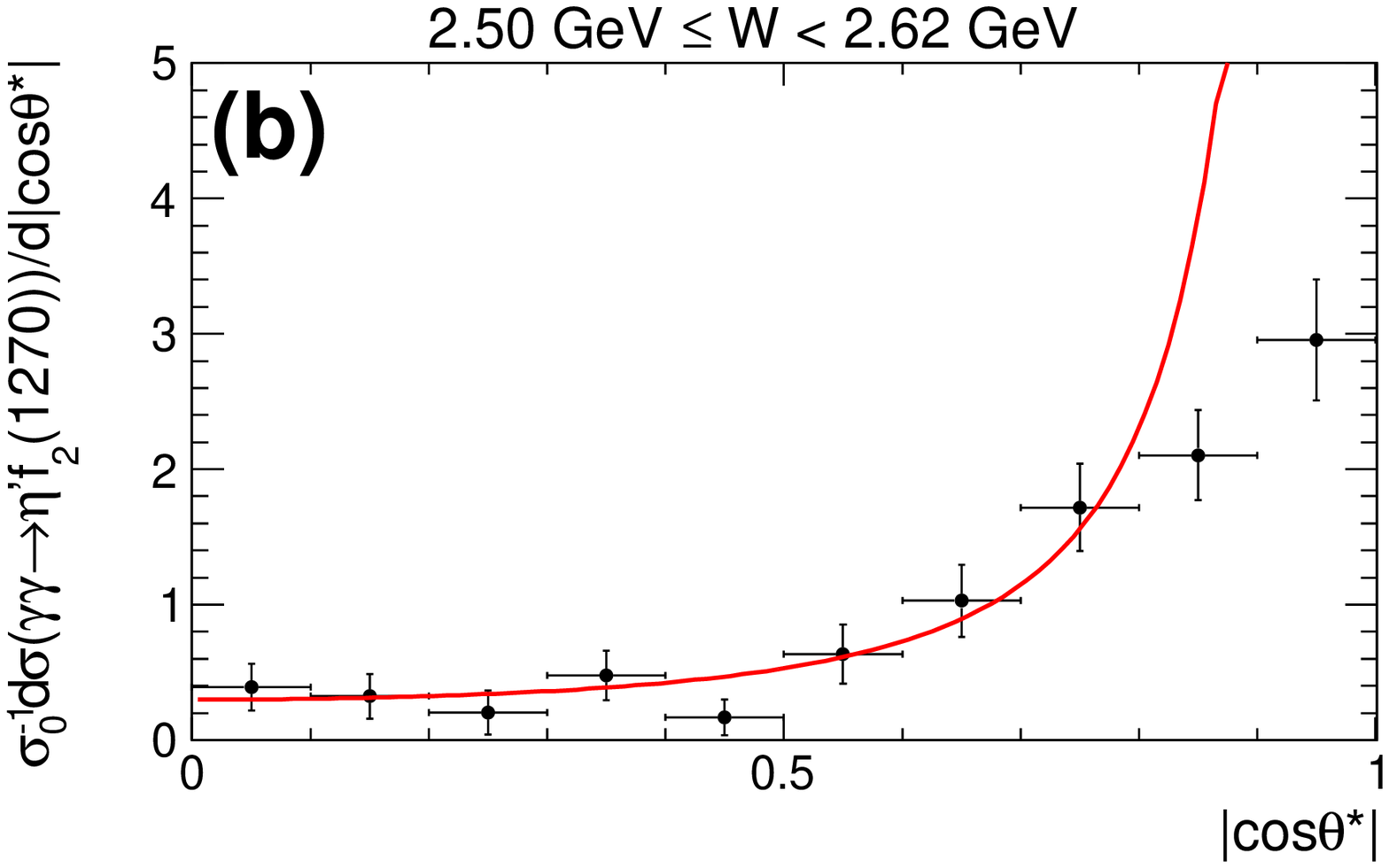}
\end{overpic}

\begin{overpic}[width=7cm]{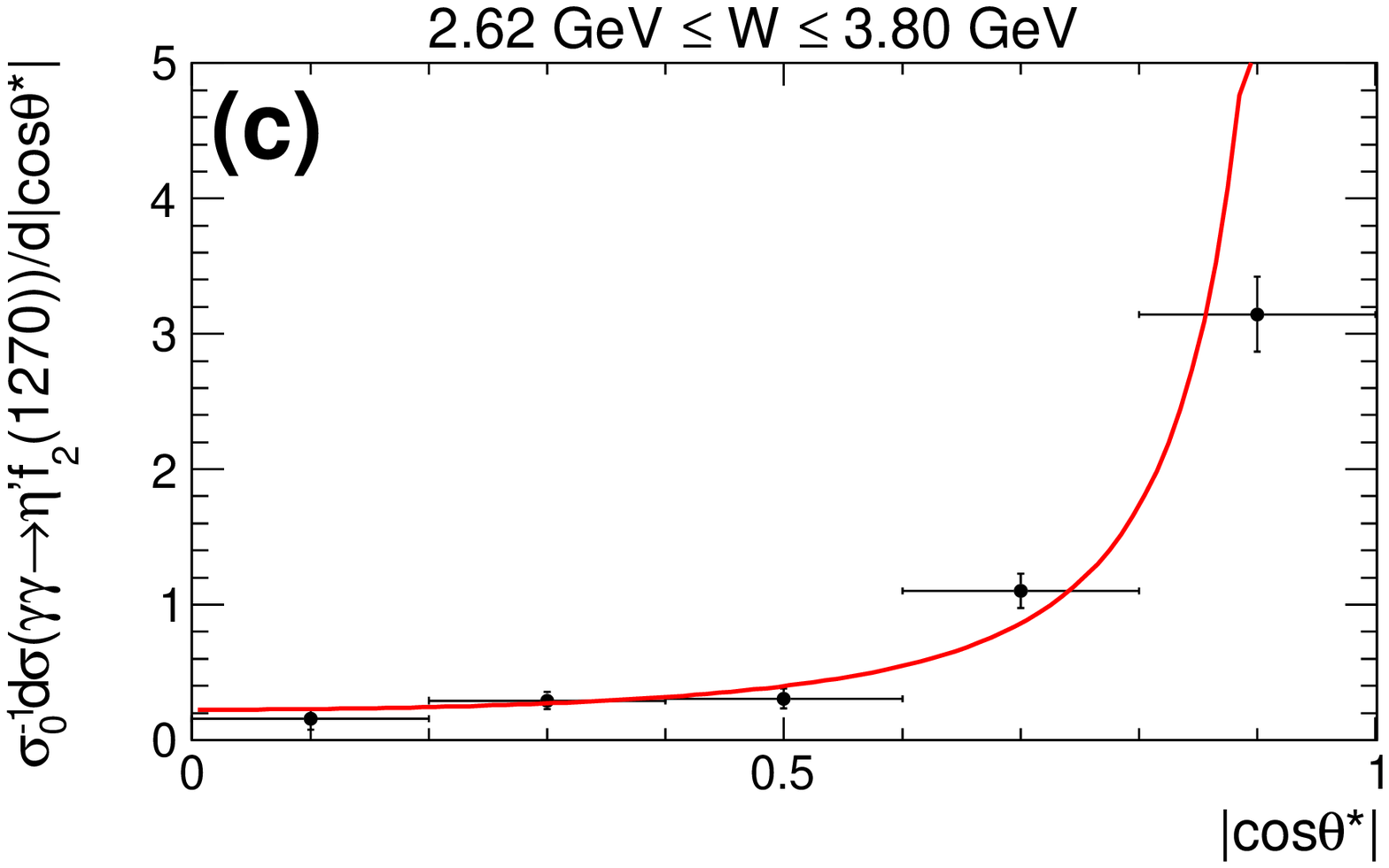}
\end{overpic}
\renewcommand{\figurename}{Fig}
\caption{(Color online) Cross sections of $\gamma\gamma\rightarrow\eta'f_{2}(1270)$ in $|\rm{cos}\theta^{*}|$ in three $W$ regions from 2.26 to 3.80 GeV. The normalizer $\sigma_0$ is the total cross section in the $\mathrm{|cos\theta^*|<0.8}$ region.
The black solid points are the data with statistical errors. The red solid  line,  normalized to the data in the same angular range follows a $1/\sin^4\theta^*$ behavior.}
\label{cr_f1270_cost}
\end{figure*}

The $W$-dependent cross section for $\gamma\gamma\rightarrow\eta'f_{2}(1270)$ in the $\eta\pi\pi$ mode, calculated with Eq.~\eqref{crs_def2}, is shown in Fig.~\ref{cr_f1270} and listed in Table~\ref{cr_f1270_sta_sys_epp}. 
The differential cross sections in $|$cos$\theta^{*}|$, averaged over $W$ bins in the three ranges $\mathrm{\mathit{W}\in [2.26, 2.50), [2.50,2.62), [2.62,3.80]}$ GeV, are given in Fig.~\ref{cr_f1270_cost}. 

We assume that the $W$ and $\theta^{*}$  dependencies of the differential cross section
follow the power law $\sigma \propto 1/W^{n}\cdot \rm{sin}^{\alpha}\theta^{*}$, which is the same as that for pseudoscalar meson pairs in the Belle data and the QCD predictions \cite{belle_result_cr}.
In a fit to the measured cross sections for $\gamma\gamma\to\eta'f_{2}(1270)$ in the range of $\mathrm{\mathit{W} \in [2.5,3.8]}$ GeV,
the resulting $W$ power-law exponent is $\mathrm{\mathit{n}=7.7\pm 1.5}$ ($\mathrm{7.5\pm 2.0}$) for $\mathrm{|\rm{cos}\theta^{*}|\in}~[0.0,0.8]$ ($\in [0.0,0.6]$).  
The differential cross sections in $|$cos$\theta^{*}|$ show an ascending trend in all three $W$ ranges, 
and its rate of increase is greater for events in the larger $W$ ranges. 
The complicated  behavior for the angular dependence of the cross sections is seen in the range of 
$\mathrm{\mathit{W}<2.50}$ GeV with markedly lower power for $\sin\theta^*$ of  $\mathrm{\alpha < 4}$, while it tends to match with the power law for the ranges of $\mathrm{\mathit{W}\in [2.50,2.62]}$ and $\mathrm{[2.62,3.80]}$ GeV.

\subsection{\texorpdfstring{Result for the $\gamma\gamma\rightarrow\eta'\pi^{+}\pi^{-}$ (excluding $\eta'f_{2}(1270)$) cross sections} {sigma(gammagamma->eta'pi+pi-) after subtraction of eta'f2(1270)}}

\begin{figure*}[!htpb]
\begin{overpic}[width=7cm]{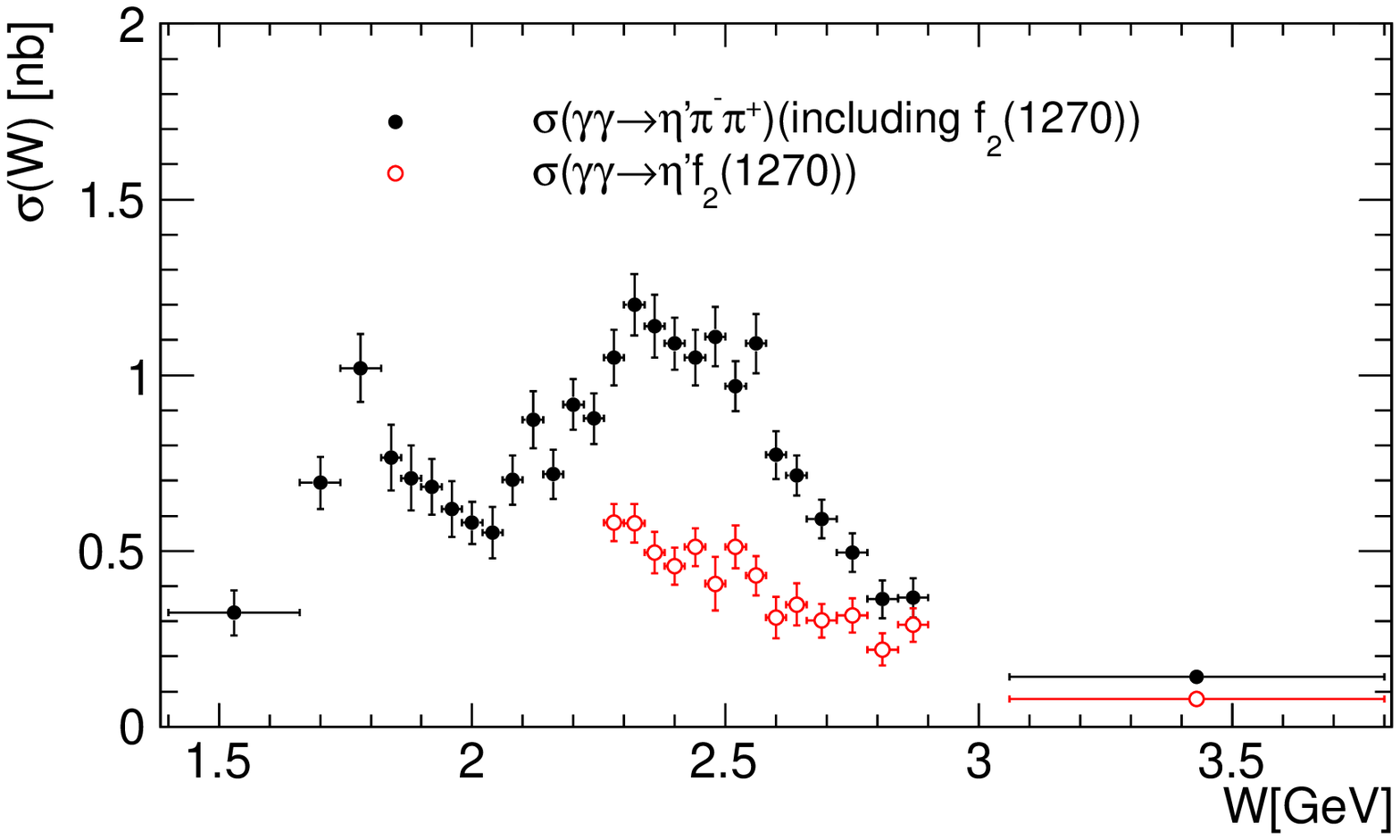}
\end{overpic}
\begin{overpic}[width=7cm]{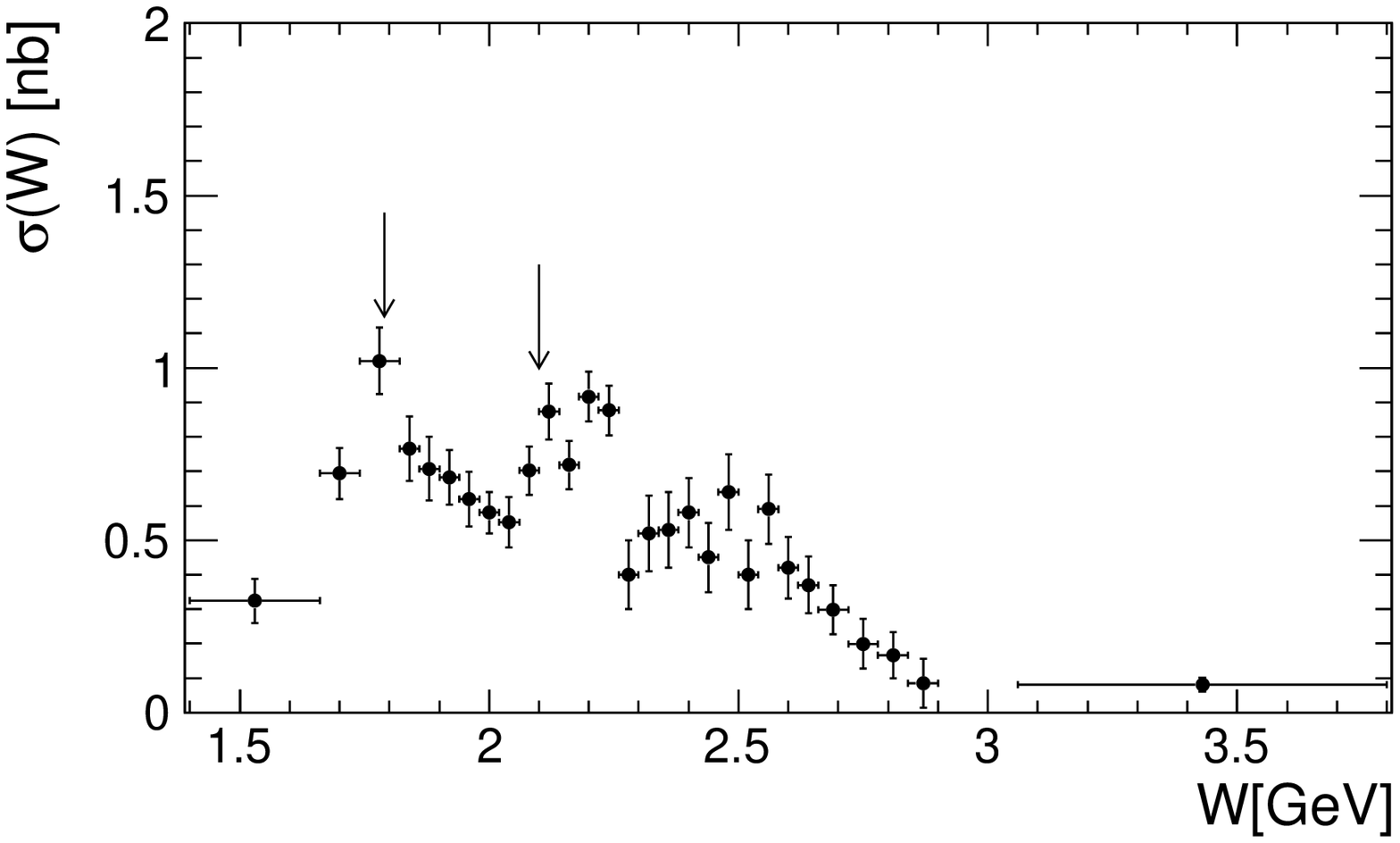}
\put(45,87){\fontsize{3pt}{3pt} $(a)$}
\put(65,80){\fontsize{3pt}{3pt} $(b)$}
\end{overpic}
\renewcommand{\figurename}{Fig}
\caption{(Color online) Left panel: cross sections of $\gamma\gamma\rightarrow\eta'\pi^{+}\pi^{-}$ [including $\eta'f_{2}(1270)]$ (black solid dots) and $\gamma\gamma\rightarrow\eta'f_{2}(1270)$ (red open dots). Right panel: cross sections of $\gamma\gamma\rightarrow\eta'\pi^{+}\pi^{-}$ [excluding $\gamma\gamma\rightarrow\eta'f_{2}(1270)$] in the $W$ range above 2.26 GeV. 
The structure (a) near 1.8 GeV arises from $X$(1835) and $\eta(1760)$; the structure (b) near 2.1 GeV is perhaps from $\gamma\gamma\rightarrow\eta'f_{0}(980)$ production. In both panels, the error bars are statistical.}
\label{cr_epp_sub1270}
\end{figure*}

In the left plot of Fig.~\ref{cr_epp_sub1270}, the measured $W$-dependent cross sections of $\gamma\gamma\rightarrow\eta^\prime f_{2}(1270)$ and $\gamma\gamma\rightarrow\eta'\pi^{+}\pi^{-}$ [including $\eta'f_{2}(1270)]$ production are shown. The former is obtained by fitting the $\pi^+\pi^-$ invariant mass spectrum for the $f_{2}(1270)$ signal and the latter is extracted in fitting the $|\Sigma p^*_{t}|$ distribution for the $\eta'\pi^{+}\pi^{-}$ signal. 
Taking the difference between the two yields in each 2D bin in data as input, the cross sections of 
$\gamma\gamma\rightarrow\eta'\pi^{+}\pi^{-}$ production without the $\eta^\prime f_{2}(1270)$ 
contribution for the $\eta\pi\pi$ mode are calculated and shown in the right plot of Fig.~\ref{cr_epp_sub1270} and summarized in Table~\ref{cr_sta_sys_epp}.
Two peaking structures are evident. The one around 1.8 GeV likely arises from the 
$\eta(1760)$ and $X(1835)$ decays to $\eta^\prime\pi^+\pi^-$~\cite{etac1S_zcc} and the other around 2.15 GeV is possibly due to $\gamma\gamma\to\eta'f_{0}(980)$ production. The $\eta_{c}(1S)$ contribution near 2.98 GeV has been subtracted. 
A larger data sample is necessary in order to understand these two structures in more detail.

\begin{figure*}[!hbpt]
\begin{overpic}[width=7cm]{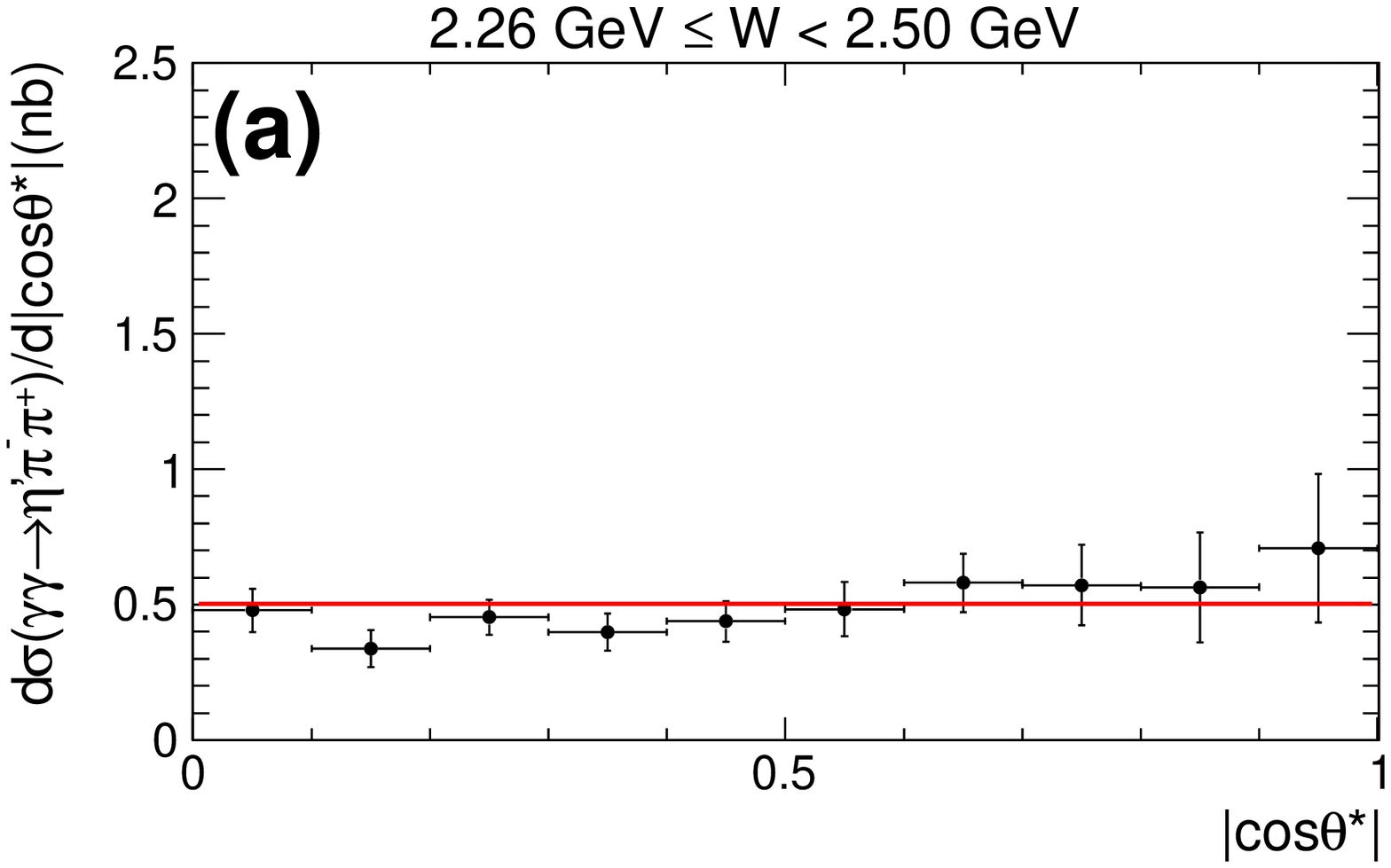}
\end{overpic}
\begin{overpic}[width=7cm]{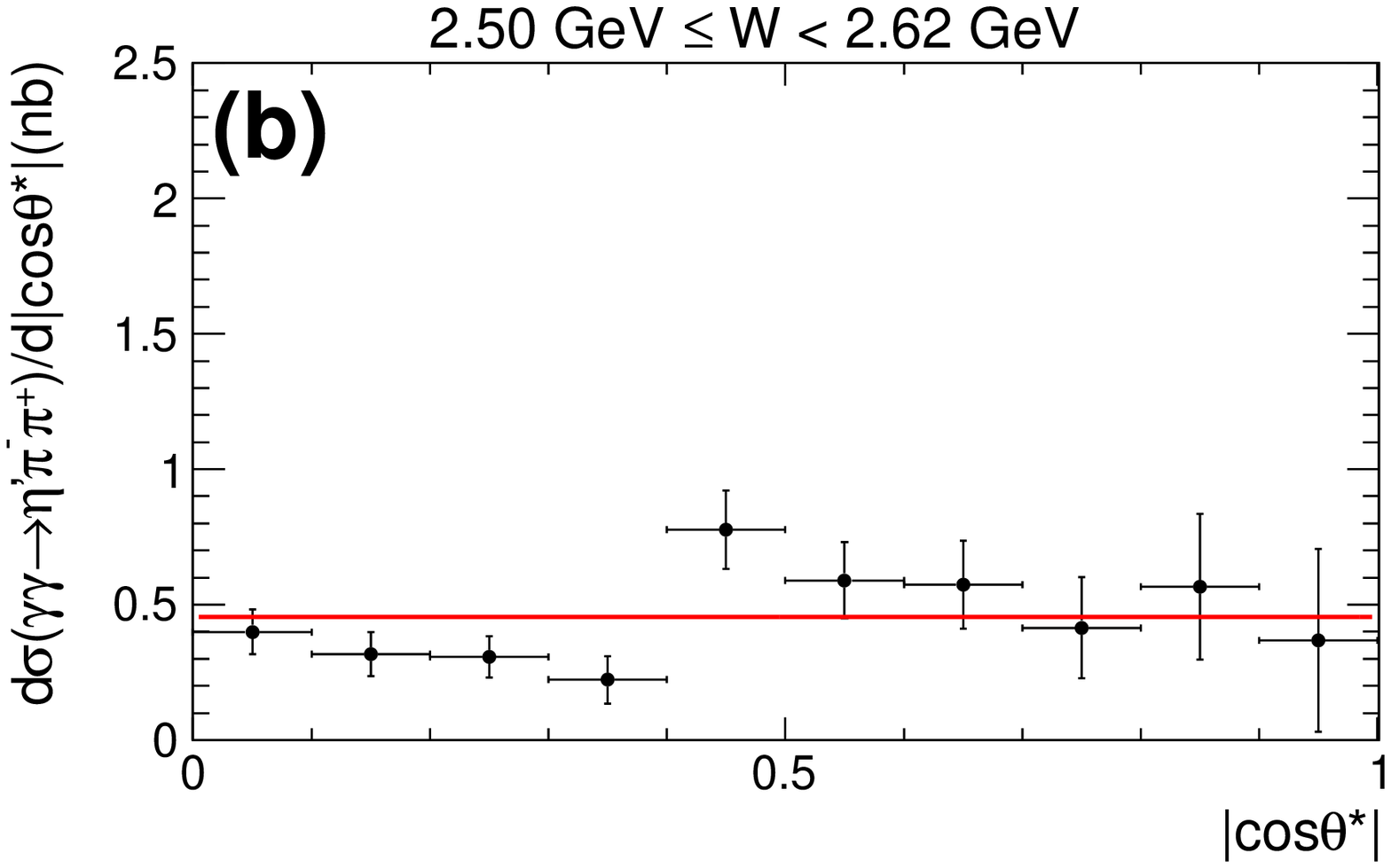}
\end{overpic}

\begin{overpic}[width=7cm]{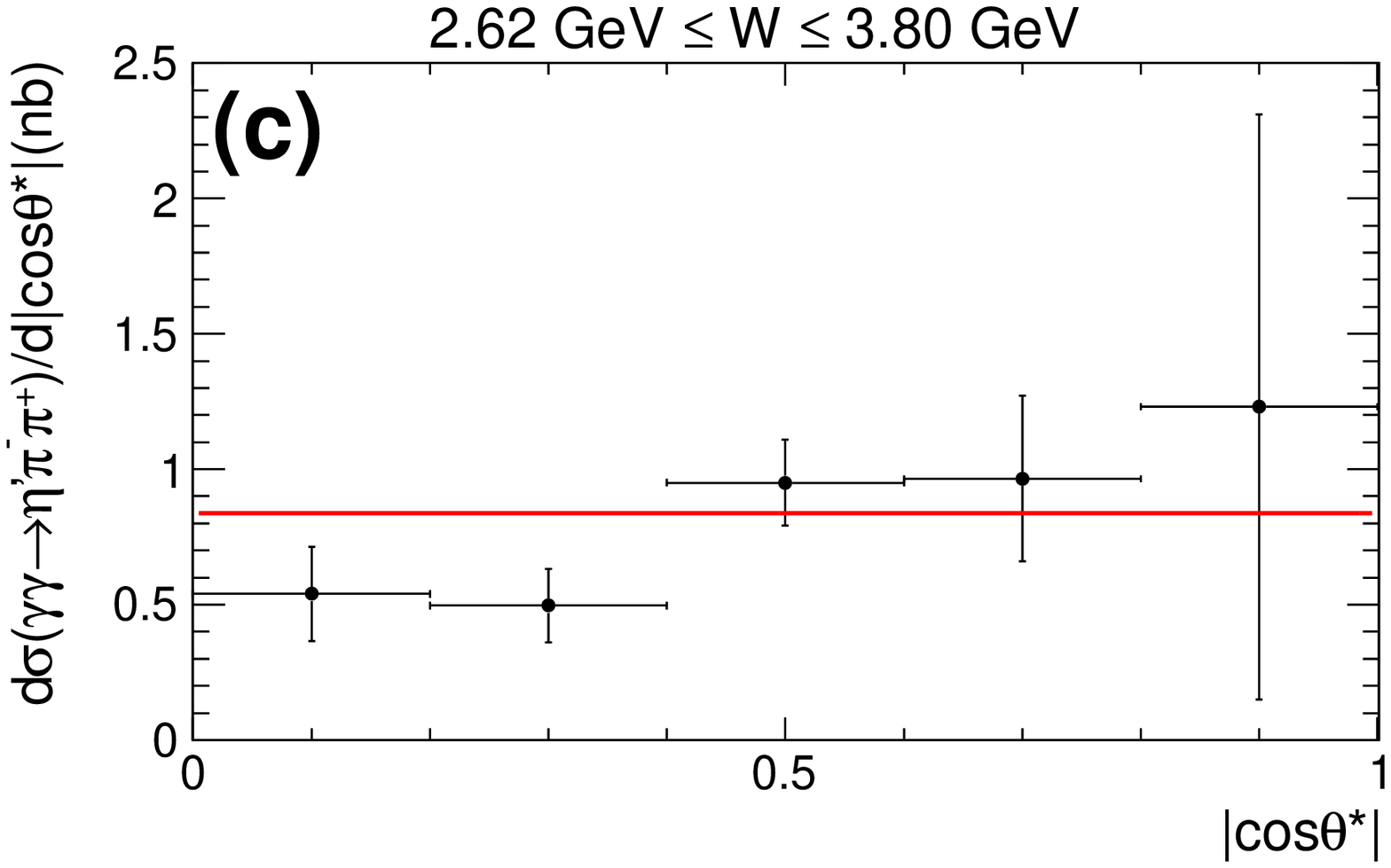}
\end{overpic}
\caption{(Color online) Differential cross sections of $\gamma\gamma\rightarrow\eta'\pi^{+}\pi^{-}$ [excluding $\eta' f_{2}(1270)$] in $|\rm{cos}\theta^{*}|$ in three $W$ regions from 2.26 to 3.80 GeV. The red solid line is a uniform distribution normalized to the data. In all panels, the error bars are statistical.}
\label{cr_epp_cost_sub1270}
\end{figure*}

The differential cross section in $|$cos$\theta^*|$ for $\gamma\gamma\rightarrow\eta'\pi^{+}\pi^{-}$  production after subtracting both contributions from $\gamma\gamma\rightarrow\eta'f_{2}(1270)$ in the $W$ region above 2.26 GeV and $\eta_c(1S)$ in the region of $\mathrm{\mathit{W} \in [2.62,3.06]}$ GeV is shown in 
Fig.~\ref{cr_epp_cost_sub1270}.  Nearly flat distributions of the cross sections in the three regions of 
$\mathrm{\mathit{W} \in [2.26,2.50]}$, $\mathrm{[2.50,2.62]}$ and $\mathrm{[2.62,3.06]}$ GeV  are consistent with the expectations from three-body final-state production via two-photon collisions.
Both the peaking structures [$\gamma\gamma\rightarrow\eta(1760)$ or $X(1835)\rightarrow \eta'\pi^+\pi^-$  and $\gamma\gamma\rightarrow\eta'f_0(980)\rightarrow\eta'\pi^+\pi^-]$ follow a  uniform angular distribution; thus, there is no distortion with or without their contribution in the resulting angular distribution in Fig.~\ref{cr_epp_cost_sub1270}.

\begin{table*}[!hbpt]
\caption{Measured cross sections for $\gamma\gamma\rightarrow\eta'\pi^{+}\pi^{-}$ after subtracting  contributions from $\gamma\gamma\rightarrow\eta'f_{2}(1270)$ in the $W$ region above 2.26 GeV and $\eta_c(1S)$ in the $W$ region of [2.62,3.06] GeV. The first error is statistical and the second is systematic.}
\begin{center}
\begin{tabular}{cccc}
\hline
\hline
\multicolumn{1}{c}{ $W$(GeV)} & \multicolumn{1}{c}{$\sigma(\gamma\gamma\rightarrow\eta'\pi^{+}\pi^{-})$ $(\rm nb)$} & \multicolumn{1}{c}{$W$(GeV)} & \multicolumn{1}{c}{$\sigma(\gamma\gamma\rightarrow\eta'\pi^{+}\pi^{-})$ $(\rm nb)$} \\
\hline
1.40 -- 1.66 & $ \mathrm{0.315\pm 0.064^{+0.046}_{-0.046}} $ &  2.30 -- 2.34 &    $  \mathrm{0.52\pm 0.11 ^{+0.10}_{-0.10}} $ \\
1.66 -- 1.74 & $ \mathrm{0.689\pm 0.074^{+0.084}_{-0.088}} $ &  2.34 -- 2.38 &    $  \mathrm{0.53\pm 0.11 ^{+0.10}_{-0.10}} $ \\
1.74 -- 1.82 & $ \mathrm{1.01 \pm 0.10 ^{+0.11 }_{-0.17 }} $ &  2.38 -- 2.42 &    $  \mathrm{0.58\pm 0.10 ^{+0.11}_{-0.11}} $ \\
1.82 -- 1.86 & $ \mathrm{0.77 \pm 0.09 ^{+0.09 }_{-0.11 }} $ &  2.42 -- 2.46 &    $  \mathrm{0.45\pm 0.10 ^{+0.09}_{-0.09}} $ \\
1.86 -- 1.90 & $ \mathrm{0.69 \pm 0.09 ^{+0.08 }_{-0.10 }} $ &  2.46 -- 2.50 &    $  \mathrm{0.64\pm 0.11 ^{+0.14}_{-0.14}} $ \\
1.90 -- 1.94 & $ \mathrm{0.661\pm 0.082^{+0.075}_{-0.091}} $ &  2.50 -- 2.54 &    $  \mathrm{0.40\pm 0.10 ^{+0.07}_{-0.08}} $ \\
1.94 -- 1.98 & $ \mathrm{0.62 \pm 0.08 ^{+0.07 }_{-0.12 }} $ &  2.54 -- 2.58 &    $  \mathrm{0.59\pm 0.10 ^{+0.11}_{-0.11}} $ \\
1.98 -- 2.02 & $ \mathrm{0.58 \pm 0.060^{+0.065}_{-0.082}} $ &  2.58 -- 2.62 &    $  \mathrm{0.42\pm 0.09 ^{+0.09}_{-0.09}} $ \\
2.02 -- 2.06 & $ \mathrm{0.552\pm 0.072^{+0.062}_{-0.094}} $ &  2.62 -- 2.66 &    $  \mathrm{0.37\pm 0.08 ^{+0.07}_{-0.07}} $ \\
2.06 -- 2.10 & $ \mathrm{0.70 \pm 0.07 ^{+0.08 }_{-0.17 }} $ &  2.66 -- 2.72 &    $  \mathrm{0.30\pm 0.07 ^{+0.06}_{-0.06}} $ \\
2.10 -- 2.14 & $ \mathrm{0.85 \pm 0.08 ^{+0.09 }_{-0.16 }} $ &  2.72 -- 2.78 &    $  \mathrm{0.20\pm 0.07 ^{+0.03}_{-0.04}} $ \\
2.14 -- 2.18 & $ \mathrm{0.71 \pm 0.07 ^{+0.08 }_{-0.12 }} $ &  2.78 -- 2.84 &    $  \mathrm{0.17\pm 0.07 ^{+0.03}_{-0.03}} $ \\
2.18 -- 2.22 & $ \mathrm{0.92 \pm 0.07 ^{+0.10 }_{-0.11 }} $ &  2.84 -- 2.90 &   $  \mathrm{0.085\pm 0.071^{+0.015}_{-0.015}} $ \\
2.22 -- 2.26 & $ \mathrm{0.86 \pm 0.07 ^{+0.10 }_{-0.11 }} $ &  3.06 -- 3.80 &   $  \mathrm{0.081\pm 0.021^{+0.021}_{-0.022}} $ \\
2.26 -- 2.30 & $ \mathrm{0.40 \pm 0.10 ^{+0.08 }_{-0.08 }} $ &       \\
\hline
\hline
\end{tabular}
\end{center}
\label{cr_sta_sys_epp}
\end{table*}

\subsection{Systematic uncertainty}

Systematic uncertainties arising from the pion identification, $\pi^{0}$-veto and $\eta'$-$sdb$ background in measurements of the cross sections for both $\gamma\gamma\rightarrow\eta'\pi^{+}\pi^{-}$ and $\gamma\gamma\rightarrow\eta'f_{2}(1270)$ production are estimated in each 2D bin, using a method similar to that in the determination of the product of two-photon width and branching fraction for the final state, $\Gamma_{\gamma\gamma}{\cal B}$. 
The uncertainty in the trigger efficiency is calculated to be 1.2--6.7\% for the $\eta\pi\pi$ mode. 
The uncertainty in the determination of the $b_{\rm any}$ background shape is estimated by changing each parameter by $\pm 1\sigma$ in the fit, and the difference in yields with and without this change in  each parameter, added in quadrature, is taken as its contribution to the systematic uncertainty. 
We study the non-$\eta'$ events with the same final state of $\gamma\gamma\rightarrow\gamma\pi\pi\pi\pi$ in MC.
We see that these non-$\eta'$ events with a wrong combination of $\gamma\pi\pi$, surviving the  $\eta'\pi\pi$ selection criteria,
have a peaking feature in the $|\Sigma p_{t}^{*}|$ distribution in the $\eta'$ signal window.  The contribution from non-$\eta'$ is regarded as a lower systematic uncertainty of the cross section.
The systematic uncertainties in the measurements of the cross sections are summarized in Table~\ref{cr_sys_sum}.

\begin{table}[!htp]
\caption{Summary of systematic uncertainties in the differential cross section measurement.}
\begin{center}
\begin{tabular}{c|cc}
\hline
\hline
Source                    & $\eta'\pi\pi$ (\%) & $\eta'f_2(1270)$  (\%) \\
\hline
Trigger efficiency            & 1.2-6.7         & 1.2-1.4  \\
Background shape              & 0.6-6.5     & 12-21     \\
$\eta'$-$sdb$ and $b_{\rm any}$   & 0.6-6.6     & 1.6-2.1       \\
$\pi^{0}$-veto                & 2.7-4.4     & 2.9-3.7       \\
$\pi^{\pm}$ identification efficiency         & 0.6-1.9     & 0.8-1.8       \\
non-$\eta'$                & 2.0-21    & --           \\
$\eta$ reconstruction efficiency     &\multicolumn{2}{c}{4.9} \\
Track reconstruction efficiency      & \multicolumn{2}{c}{5.5}  \\
Two-photon luminosity      & \multicolumn{2}{c}{5}    \\
Run dependence          & \multicolumn{2}{c}{3}    \\
\hline
\hline
\end{tabular}
\end{center}
\label{cr_sys_sum}
\end{table}

\section{summary and discussions}
The $\eta_c(1S)$, $\eta_c(2S)$, and non-resonant production of the $\eta'\pi^+\pi^-$ final state via two-photon collisions are measured.
The results for the yields, masses, and widths, as well as the product decay widths are summarized in Table~\ref{etac_resl} for the  $\eta_c(1S)$ and $\eta_c(2S)$. The differential cross sections for the non-resonant states of two-body $\eta'f_2(1270)$ with $f_2(1270)\rightarrow\pi^+\pi^-$ and  three-body $\eta'\pi^+\pi^-$ [excluding $\eta' f_{2}(1270)$] in the $\eta\pi\pi$ mode are shown in Tables~\ref{cr_f1270_sta_sys_epp} and \ref{cr_sta_sys_epp} and Figs.~\ref{cr_f1270}--\ref{cr_epp_cost_sub1270}.
  
The $\eta_c(1S)$ mass and width are measured to be 
$M$ = [2984.6 $\pm$ 0.7 (stat) $\pm$ 2.2 (syst) $\pm$ 0.3 (model)] MeV/$c^2$ 
and $\Gamma$ = $[30.8^{+2.3}_{-2.2}$ (stat) $\pm$ 2.5 (syst) $\pm$ 1.4 (model)] MeV, 
and are consistent with the world-average values \cite{PDG_2016}.  Here, the differences in the $\eta_c(1S)$ mass and width with and without interference between $\eta_c(1S)$ and non-resonant component, $\Delta M$ = 0.3 MeV/$c^2$ and $\Delta \Gamma$ = 1.4 MeV, are taken as model-dependent uncertainties in the determination of the mass and width  \cite{etac1S_zcc}. 
The directly measured product of the two-photon width and branching fraction for  $\eta_c(1S)$ decay to  $\eta'\pi^+\pi^-$ is determined to be $\Gamma_{\gamma\gamma}{\cal B}(\eta_c(1S)\rightarrow\eta'\pi^+\pi^-)$ $\mathrm{ = (65.4\pm2.6\pm 7.8)}$ eV. 
By employing the full $\Upsilon(4S)$ and $\Upsilon(5S)$ data samples (941 fb$^{-1}$) and 
an additional decay mode for the $\eta^\prime\rightarrow\gamma\rho$, the results for the  
$\eta_c(1S)$ mass, width and product of its decay width in this measurement are obtained with improved statistical 
errors, and thus supersede our previous measurement using a 673 fb$^{-1}$ data sample~\cite{etac1S_zcc}.
With the world-average value of $\Gamma_{\gamma\gamma}(\eta_c(1S))\mathrm{ = (5.1\pm 0.4)}$ keV \cite{PDG_2016} as input, the branching fraction is calculated to be ${\cal{B}}(\eta_{c}(1S)\rightarrow\eta'\pi^+\pi^-)\mathrm{ = [12.8\pm0.5~(stat)\pm1.4~(syst)\pm1.0~(PDG)]}$ $\times 10^{-3}$, where the third error is due to the $\eta_c(1S)$ two-photon decay width.

\begin{table*}[!hbpt]
\caption{Comparison of the $\Gamma_{\gamma\gamma}{\cal B}$ for $\eta_{c}(1S)$ and $\eta_{c}(2S)$ decays by CLEO, Belle, and BaBar, along with the ratio  ${\cal R}(\eta_{c}(2S)/\eta_{c}(1S))$ = $(\Gamma_{\gamma\gamma}(\eta_{c}(2S)){\cal B}(\eta_{c}(2S)))/(\Gamma_{\gamma\gamma}(\eta_{c}(1S)){\cal B}(\eta_{c}(1S)))$. The two-photon decay width $\Gamma_{\gamma\gamma}(\eta_{c}(2S)$ is estimated using the world-average value of $\Gamma_{\gamma\gamma}(\eta_{c}(1S))$ = (5.1 $\pm$ 0.4) keV  as input under the assumption of equal ${\cal B}$ for $\eta_{c}(1S)$ and $\eta_{c}(2S)$ decays.}
\begin{center}
\begin{tabular}{cccccl}
\hline
\hline
\multicolumn{1}{c}{Final state} & $\Gamma_{\gamma\gamma}{\cal B}$ for $\eta_{c}(1S)$  &  $\Gamma_{\gamma\gamma}{\cal B}$ for $\eta_{c}(2S)$  & ${\cal R}(\eta_{c}(2S)/\eta_{c}(1S))$ & $\Gamma_{\gamma\gamma}(\eta_{c}(2S))$ & Reference \\ 
  & (eV) & (eV) & ($\times10^{-2}$) & (keV) &  \\
\hline
$K^0_SK^+\pi^-$ & -- & -- & 18 $\pm$ 5 $\pm$ 2 & 0.92 $\pm$ 0.28 & \cite{etac2s_epp} CLEO 2004 \\ 
$K\bar{K}\pi$ & 386 $\pm$ 8 $\pm$ 21 & 41 $\pm$ 4 $\pm$ 6 & 10.6 $\pm$ 2.0 & 0.54 $\pm$ 0.11 & \cite{etac_babar} BaBar 2011  \\
$\eta'\pi^+\pi^-$ & 65.4 $\pm$ 2.6 $\pm$ 7.8  & 5.6 $\pm$ 1.2 $\pm$ 1.1 & 8.6 $\pm$ 2.7 & 0.44 $\pm$ 0.14 & This, Belle \\
\hline
\multicolumn{4}{c}{QCD} & 1.8 - 5.7 & \cite{tp_wid_QCD,tp_wid_QCD_1,tp_wid_QCD_2,tp_wid_QCD_3,tp_wid_QCD_4,tp_wid_QCD_5} 1992 - 2005 \\
\multicolumn{5}{c}{ } & \cite{rev_eqbr} 2008 \\
                           
\hline
\hline
\end{tabular}
\end{center}
\label{com_exp_thr}
\end{table*}

We report the first observation of $\eta_c(2S)\rightarrow\eta'\pi^+\pi^-$, with a significance of 5.5$\sigma$ including the systematic error. 
We measure the mass of the $\eta_c(2S)$ to be 
$M$ = [3635.1 $\pm$ 3.7 (stat) $\pm$ 2.9 (syst) $\pm$ 0.4 (model)] MeV/$c^2$, 
which is consistent with the world-average value~\cite{PDG_2016}, and the product of two-photon width and branching fraction to $\eta'\pi^+\pi^-$ to be $\mathrm{\Gamma_{\gamma\gamma}{\cal B}(\eta_c(2S)\rightarrow\eta'\pi^+\pi^-) = (5.6^{+1.2}_{-1.1}\pm1.1)}$ eV. 

In fact, the ratio of the two products of two-photon decay width and branching fraction for the $\eta_c(1S)$ and $\eta_c(2S)$,
\begin{eqnarray} \label{ratio}
\mathcal{R} = \frac{\Gamma_{\gamma\gamma}(\eta_c(2S)) {\cal B}(\eta_c(2S))}{\Gamma_{\gamma\gamma}(\eta_c(1S)) {\cal B}(\eta_c(1S))},
\end{eqnarray}
is a quantity directly measured in experiments.  
The $\eta_c(1S)$ and $\eta_c(2S)$ mesons in the measurements are all produced via two-photon process, and the dominant contributions to the systematic uncertainty in either product alone, such as those for the two-photon luminosity and reconstruction efficiencies of $\eta$ and charged pion tracks, cancel almost completely in this ratio. 
As shown in Table~\ref{com_exp_thr}, the ${\cal R}$ values from the two observations---one by BaBar \cite{etac_babar} with $K\bar{K}\pi$ and the other by this analysis with $\eta'\pi^+\pi^-$---are measured to be $\mathcal{R} = (10.6 \pm 2.0)\times 10^{-2}$ and $(8.6 \pm 2.7)\times 10^{-2}$, respectively.
They are consistent with each other, while a third measurement  with large uncertainty by CLEO \cite{etac2s_epp} is compatible with the former. 
It implies that the assumption of approximate 
equality of the branching fractions for $\eta_c(1S)$ and $\eta_c(2S)$ to a specific final state,
\begin{eqnarray}
\label{equcal}
&&\frac{{\cal B}(\eta_c(2S)\rightarrow\eta'\pi^+\pi^-)}{{\cal B}(\eta_c(1S)\rightarrow\eta'\pi^+\pi^-)} \nonumber\\[1mm]
&&\cong \frac{{\cal B}(\eta_c(2S)\rightarrow K\bar{K}\pi)}{{\cal B}(\eta_c(1S)\rightarrow K\bar{K}\pi)},
\end{eqnarray}
is reasonable within the errors. 
Here, the systematic uncertainty contributions in the ${\cal R}$ values [and thus the ratio of branching fractions for $\eta_c(1S)$  and $\eta_c(2S)$  decays in Eq.~\eqref{equcal}] are conservatively estimated, since their cancellation effect in determination of the ratio ${\cal R}$ errors is not subtracted yet. 

Under the assumption of equal branching fractions for $\eta_c(1S)$ and $\eta_c(2S)$ decay, the two-photon decay width for $\eta_c(2S)$ is determined to be $\Gamma_{\gamma\gamma}(\eta_c(2S)) \mathrm{=(1.3\pm 0.6)}$ keV by CLEO~\cite{etac2s_epp}, which lies at the lower bound of the QCD predictions~\cite{tp_wid_QCD,tp_wid_QCD_1,tp_wid_QCD_2,tp_wid_QCD_3,tp_wid_QCD_4,tp_wid_QCD_5}. 
The resulting $\Gamma_{\gamma\gamma}(\eta_c(2S))$ value, derived from this work, is less than half of  CLEO's (see Table~\ref{com_exp_thr}). On the other hand, the measured  unequal branching fractions for $\eta_c(1S)$ and $\eta_c(2S)$ decays to 
$K\bar{K}\pi$, albeit with good precision for the former \cite{PDG_2016} but large uncertainty for the latter~\cite{babar_etac2s}, indicates that an improved test of the assumption with experimental data is indeed needed. Precision measurements of the branching fraction for either $\eta_c(2S)$ decays to $K^0_SK^+\pi^-$ ($\eta\pi^+\pi^-$) or $B$ decays to $K\eta_c(2S)$ would be able to clarify the discrepancy in the two-photon decay width of $\eta_c(2S)$ between data and QCD predictions.

The cross sections of $\gamma\gamma\rightarrow\eta'f_{2}(1270)$ and $\gamma\gamma\rightarrow\eta'\pi^{+}\pi^{-}$ [excluding $\eta'f_{2}(1270)$] in $\eta\pi^+\pi^-$ mode are measured. 
Under the assumption of the power law dependence $\sigma \propto 1/(W^{n}\cdot \rm{sin}^{\alpha}\theta^{*})$ for pseudoscalar tensor meson pair production, the fitted index $n = 7.5\pm 2.0$ (for $|$cos$\theta^{*}|<0.6$) shows that the cross section of the $\gamma\gamma\rightarrow\eta'f_{2}(1270)$ production with $\eta'$ scattering at large angles in the $\gamma\gamma$ rest system behaves much steeper in its $W$ dependence than that at small angle, and that the $W$ dependence of cross section in the power law is compatible, within error, with the sharply dropping behavior  for neutral pseudoscalar meson pair production measured by Belle ($\mathit{n}$ = $7.8-11$)~\cite{belle_result_cr} and predicted by QCD ($\mathit{n}$ = 10)~\cite{qcd_pre_1,qcd_pre_2,qcd_pre_2_1,handbag}. On the other hand, the behavior  of the cross sections'  angular dependence for the ranges of  
$W \in [2.50, 2.62]$ and $\in [2.62, 3.8]$ GeV is compatible 
with that for $\pi^0\pi^0$ and $\eta\pi^0$ production as measured by Belle~\cite{belle_result_cr} and with that for pseudoscalar meson pair production predicted by the QCD calculations \cite{qcd_pre_1,qcd_pre_2,qcd_pre_2_1,handbag}.

In summary, the $\eta_{c}(1S)$, $\eta_{c}(2S)$ and non-resonant $\eta'\pi^{+}\pi^{-}$ production via two-photon collisions is measured. We report the first observation of the signal for   $\eta_c(2S)$ decays to $\eta'\pi^{+}\pi^{-}$, the measured products of the two-photon decay width and the branching fraction for the $\eta_c(1S)$ and $\eta_c(2S)$ decays to $\eta'\pi^{+}\pi^{-}$, and the measurement of  non-resonant production of two-body $\eta'f_2(1270)$ and three-body $\eta'\pi^+\pi^-$ final states via two-photon collisions.

\section*{Acknowledgments}
%
%

We  extend our special thanks to Y. H. Zheng and X. R. Lyu of the University of Chinese Academy of Sciences
for helpful discussions.
We thank the KEKB group for the excellent operation of the
accelerator; the KEK cryogenics group for the efficient
operation of the solenoid; and the KEK computer group,
the National Institute of Informatics, and the 
Pacific Northwest National Laboratory (PNNL) Environmental Molecular Sciences Laboratory (EMSL) computing group for valuable computing
and Science Information NETwork 5 (SINET5) network support.  We acknowledge support from
the Ministry of Education, Culture, Sports, Science, and
Technology (MEXT) of Japan, the Japan Society for the 
Promotion of Science (JSPS), and the Tau-Lepton Physics 
Research Center of Nagoya University; 
the Australian Research Council;
Austrian Science Fund under Grant No.~P 26794-N20;
the National Natural Science Foundation of China under Contracts
No.~11435013,  
No.~11475187,  
No.~11521505,  
No.~11575017,  
No.~11675166,  
No.~11705209;  
Key Research Program of Frontier Sciences, Chinese Academy of Sciences (CAS), Grant No.~QYZDJ-SSW-SLH011; 
the  CAS Center for Excellence in Particle Physics (CCEPP); 
Fudan University Grant No.~JIH5913023, No.~IDH5913011/003, 
No.~JIH5913024, No.~IDH5913011/002;                        
the Ministry of Education, Youth and Sports of the Czech
Republic under Contract No.~LTT17020;
the Carl Zeiss Foundation, the Deutsche Forschungsgemeinschaft, the
Excellence Cluster Universe, and the VolkswagenStiftung;
the Department of Science and Technology of India; 
the Istituto Nazionale di Fisica Nucleare of Italy; 
National Research Foundation (NRF) of Korea Grants No.~2014R1A2A2A01005286, No.2015R1A2A2A01003280,
No.~2015H1A2A1033649, No.~2016R1D1A1B01010135, No.~2016K1A3A7A09005 603, No.~2016R1D1A1B02012900; Radiation Science Research Institute, Foreign Large-size Research Facility Application Supporting project and the Global Science Experimental Data Hub Center of the Korea Institute of Science and Technology Information;
the Polish Ministry of Science and Higher Education and 
the National Science Center;
the Ministry of Higher Education and Science of the Russian Federation
under the grant 14.W03.31.0026;
the Slovenian Research Agency;
Ikerbasque, Basque Foundation for Science, Basque Government (No.~IT956-16) and
Ministry of Economy and Competitiveness (MINECO) (Juan de la Cierva), Spain;
the Swiss National Science Foundation; 
the Ministry of Education and the Ministry of Science and Technology of Taiwan;
and the United States Department of Energy and the National Science Foundation.

We thank the KEKB group for excellent operation of the
accelerator; the KEK cryogenics group for efficient solenoid
operations; and the KEK computer group, the NII, and 
PNNL/EMSL for valuable computing and SINET5 network support.  
We acknowledge support from MEXT, JSPS and Nagoya's TLPRC (Japan);
ARC (Australia); FWF (Austria); NSFC and CCEPP (China); 
MSMT (Czechia); CZF, DFG, EXC153, and VS (Germany);
DST (India); INFN (Italy); 
MOE, MSIP, NRF, RSRI, FLRFAS project and GSDC of KISTI (Korea);
MNiSW and NCN (Poland); MES and RFAAE (Russia); ARRS (Slovenia);
IKERBASQUE and MINECO (Spain); 
SNSF (Switzerland); MOE and MOST (Taiwan); and DOE and NSF (USA).



\begin{thebibliography}{99}
\bibitem{icb fun}
The original Crystal Ball (CB) function, defined in MINUIT, has a Gaussian in its central and upper-side regions and a  tail in the lower side. The improved Crystal Ball (ICB) function is defined as a CB with an additional tail in its upper side.
\end{thebibliography}
\end{document}